\documentclass[journal]{IEEEtran}

\usepackage{amsmath}
\usepackage{amsfonts}
\usepackage{amssymb}
\usepackage{balance}
\usepackage{bm}
\usepackage{graphics}
\usepackage{graphicx}
\usepackage{grffile}
\usepackage{footnote}
\usepackage{pifont}
\usepackage{epsfig}
\usepackage{xspace}
\usepackage{algpseudocode}
\usepackage{algorithm}
\usepackage{mathtools}
\usepackage{array}
\usepackage{todonotes}
\usepackage{etoolbox}
\usepackage{enumitem}
\usepackage{booktabs}
\usepackage{tablefootnote}
\usepackage{hyperref}
\hypersetup{
    colorlinks = false,
    linkbordercolor = {white},
}
\usepackage{silence}
\usepackage{multirow}
\usepackage{float}
\floatname{algorithm}{Listing}

\def\ie{\textit{i.e.}\xspace}
\def\etal{\textit{et al.}\xspace}
\def\etc{\textit{etc}\xspace}
\def\wrt{\textit{w.r.t.}\xspace}

\def\eg{\textit{e.g.}\xspace}

\algnewcommand\algorithmicforeach{\textbf{for each}}
\algdef{S}[FOR]{ForEach}[1]{\algorithmicforeach\ #1\ \algorithmicdo}
\graphicspath{{./figures/}}

\title{Eliminating the Barriers: Demystifying Wi-Fi Baseband Design and Introducing the PicoScenes Wi-Fi Sensing Platform}

\begin{document}

\author{Zhiping~Jiang,~Tom~H.~Luan,~Xincheng~Ren,~Dongtao~Lv,~Han~Hao,~Jing~Wang,~Kun~Zhao,~Wei~Xi,~Yueshen~Xu,~and~Rui~Li
\IEEEcompsocitemizethanks{
\IEEEcompsocthanksitem This work is supported by the National Key R\&D Program of China~(No. 2020YFC1523300), Youth Program of 
Natural Science Foundation of China (NSFC)~(No. 61802291), General Program of NSFC~(No. U1808207, 61772413) and 
National Postdoctoral Program for Innovative Talent of China~(No. BX20180235). 
(Corresponding author: Rui Li)
\IEEEcompsocthanksitem Zhiping Jiang,  Xincheng Ren, Dongtao Lv, Jing Wang, Yueshen Xu and Rui Li are with 
the School of Computer Science and Technology, 
Xidian University, Xi'an, China. 
Email:
\{\href{mailto:zpj@xidian.edu.cn}{zpj},
\href{mailto:ysxu@xidian.edu.cn}{ysxu},
\href{mailto:rli@xidian.edu.cn}{rli}\}@xidian.edu.cn,
\{\href{mailto:xcren@stu.xidian.edu.cn}{xcren},
\href{mailto:tdlv@stu.xidian.edu.cn}{tdlv}
\href{mailto:jingwang_buer@stu.xidian.edu.cn}{jingwang\_buer}
\}@stu.xidian.edu.cn.
\IEEEcompsocthanksitem Tom H. Luan is with the School of Cyber Engineering, Xidian University, Xi'an, China.
Email: \href{mailto:tom.luan@xidian.edu.cn}{tom.luan}@xidian.edu.cn.

Han Hao, Kun Zhao and Wei Xi are with 
the School of Computer Science and Technology, 
Xi'an Jiaotong University, Xi'an, China. 
Email:
\{\href{mailto:kunzhao@xjtu.edu.cn}{kunzhao},
\href{mailto:weixi@xjtu.edu.cn}{xiwei}\}@xjtu.edu.cn,
\{\href{mailto:haohan9717@stu.xjtu.edu.cn}{haohan9717}\}@stu.xjtu.edu.cn.
\IEEEcompsocthanksitem Copyright (c) 20xx IEEE. Personal use of this material is permitted. However, permission to use this material for any other purposes must be obtained from the IEEE by sending a request to pubs-permissions@ieee.org.
}
}

\maketitle

\begin{abstract}
The research on Wi-Fi sensing has been thriving over the past decade but the process has not been smooth. Three barriers always hamper the research: unknown baseband design and its influence, inadequate hardware, and the lack of versatile and flexible measurement software.
This paper tries to eliminate these barriers through the following work.
\textit{First}, we present an in-depth study of the baseband design of the Qualcomm Atheros AR9300 (QCA9300) NIC. We identify a missing item of the existing CSI model, namely, the CSI distortion, and identify the baseband filter as its origin. We also propose a distortion removal method.
\textit{Second}, we reintroduce both the QCA9300 and software-defined radio (SDR) as powerful hardware for research. For the QCA9300, we unlock the arbitrary tuning of both the carrier frequency and bandwidth. For SDR, we develop a high-performance software implementation of the 802.11a/g/n/ac/ax baseband, allowing users to fully control the baseband and access the complete physical-layer information.
\textit{Third}, we release the PicoScenes software, which supports concurrent CSI measurement from multiple QCA9300, Intel Wireless Link (IWL5300) and SDR hardware. PicoScenes features rich low-level controls, packet injection and software baseband implementation. It also allows users to develop their own measurement plugins.
\textit{Finally}, we report state-of-the-art results in the extensive evaluations of the PicoScenes system, such as the $>$2 GHz available spectrum on the QCA9300, concurrent CSI measurement, and up to 40 kHz and 1 kHz CSI measurement rates achieved by the QCA9300 and SDR. PicoScenes is available at \url{https://ps.zpj.io}.
\end{abstract}


\IEEEpeerreviewmaketitle

\section{Introduction} 
\label{sec:introduction}

\IEEEPARstart{A}{fter} a decade of advancement, channel state information (CSI)-based Wi-Fi sensing has grown into a thriving and fruitful research field and has led to new areas of sensing research, such as gesture recognition~\cite{WFID}~\cite{WiTrace}~\cite{WiDrive}, motion tracking~\cite{Widar2.0}, respiration detection~\cite{UbiBreathe}~\cite{BreathTrack}, through-wall detection~\cite{WiSpy}, sign language recognition~\cite{SignFi}~\cite{WiFinger}, \etc.
However, the 10-year evolution of the  research were not smooth and has been severely hampered by the following three barriers:

\textbf{Barrier 1: the unknown baseband design and its influence on CSI}.
The widely adopted CSI model~\cite{sensesurvey1,Tadayon:2019cj} can be simplified as follows.
\begin{align}
	\mathbf{Y} = & \mathbf{H}_{bb} \cdot \mathbf{X} + \mathbf{N} \label{eq:basic_signal_model}, \\
	& \mathbf{H}_{bb} =  \mathbf{H}_{air}  \cdot \mathbf{H}_{\theta}\label{eq:csi_basic_form}
\end{align}
Here, (\ref{eq:basic_signal_model}) describes how the baseband channel frequency response $\mathbf{H}_{bb}$, or the typically obtained CSI, transforms the baseband signal $\mathbf{X}$ into the received form $\mathbf{Y}$ with some acceptable noise $\mathbf{N}$.
(\ref{eq:csi_basic_form})
describes how, in the context of Wi-Fi sensing research, $\mathbf{H}_{bb}$ is further decomposed into two terms, the pure \textit{in-air} channel response, $\mathbf{H}_{air}$, and the sum of all linear phase errors, $\mathbf{H}_{\theta}$.
$\mathbf{H}_{air}$ characterizes the channel response of the in-air propagation, including the distance fading, multipath, Doppler effect, \etc.
$\mathbf{H}_{\theta}$ includes the carrier frequency offset (CFO), sampling frequency offset (SFO), symbol timing offset (STO), sampling clock offset (SCO), \etc.
%


%
%
\begin{figure}[t]
	\begin{center}
		\begin{tabular}{cc}
			\hspace{-0.13in}
			\includegraphics[width=0.5\columnwidth]{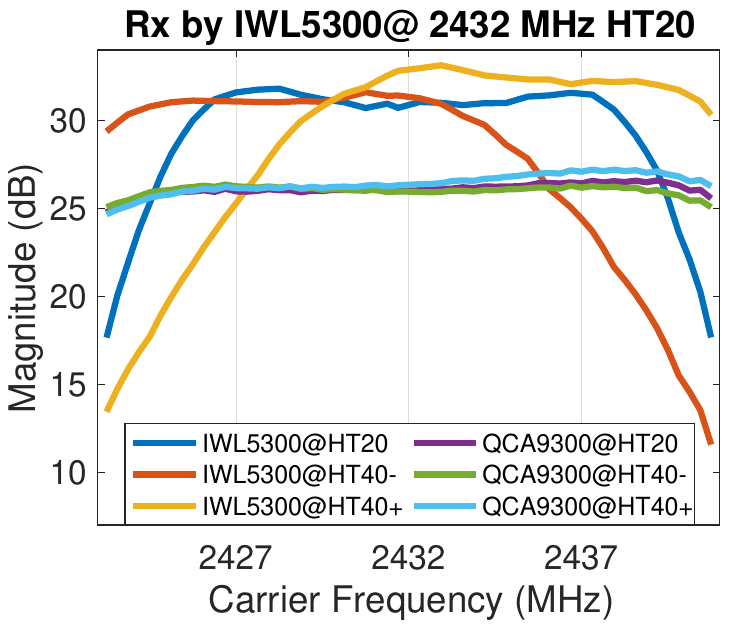} & 
			\hspace{-0.2in}
			\includegraphics[width=0.5\columnwidth]{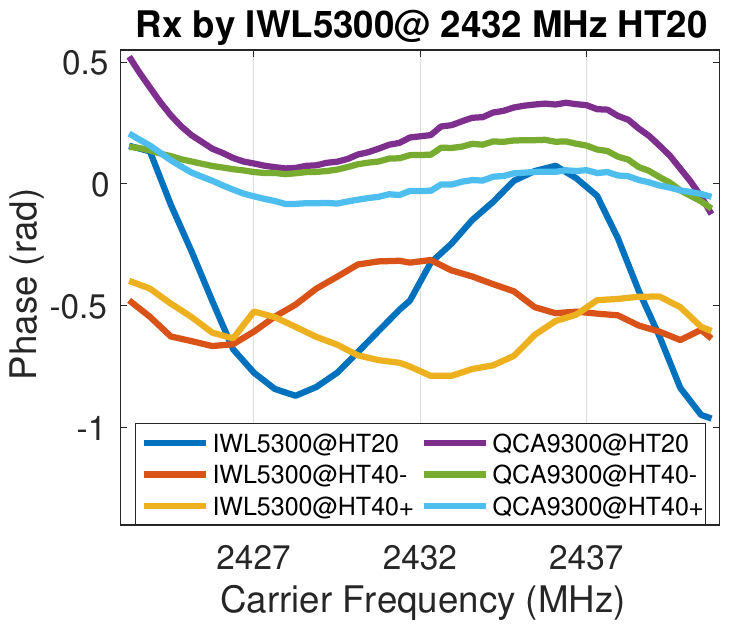}
		\end{tabular}
	\end{center}
	\caption{CSI measurement (magnitude and phase response), conducted in a radioanechoic chamber. Two Tx ends, namely, the QCA9300 and IWL5300 NICs, injected packets at 2432 MHz with the HT20 and HT40+/- channel modes. The Rx end, an IWL5300 NIC, operated at 2432 MHz in the HT20 channel mode. The magnitude shows an M-shaped and a half-M-shaped response, and the phase shows a horizontal S-shaped response.}
	\label{fig:csi_distortion}
\end{figure}

The CSI model shown in (\ref{eq:csi_basic_form}) ignores the influence of the hardware system. Researchers used to believe that the hardware imperfection is weak and flat. This assumption is reasonable because the Wi-Fi standard imposes stringent restrictions on the spectral flatness and spectral masking~\cite{Association:ttTzA7cz}.

However, our \textit{real-world evaluations indicate a very different situation}. We placed the IWL5300 NIC and the QCA9300 NIC in a radioanechoic chamber and measured the CSI between them. Since the chamber suppresses the multipath effect, we anticipated smooth and flat CSI measurements.
However, as shown in Fig.~\ref{fig:csi_distortion}, the results exhibit very strong nonflatness in both magnitude and phase. Especially in the HT40+/- cases, the differences in magnitude among the subcarriers were greater than 15 dB.
%
We repeated the measurements under a wide variety of position, carrier frequency and bandwidth configurations
and obtained the same results.
This strong nonflatness is clearly beyond the scope of hardware imperfection.
The only possible explanation is that \textbf{an unknown CSI distortion exists in the transmitter (Tx) and receiver (Rx) sides}.
This distortion, when properly considered, may affect all previous Wi-Fi sensing research,
and even \textit{challenge the correctness of the CSI model} shown in (\ref{eq:csi_basic_form}).
To address this issue, we must answer three questions:
\begin{itemize}[leftmargin=*]
	\item What is the distortion, and where does it come from?
	\item Can we alter the existing model to fit this distortion?
	\item How can we eliminate this distortion?
\end{itemize}

\textbf{Barrier 2: the inadequate hardware}. 
The 10-year development of Wi-Fi sensing research has always been impeded by inadequate hardware, reflected in three aspects.

First, \textit{there is a lack of advanced hardware features}. More advanced research demands more powerful hardware features, for example, support for an external clock source, cross-antenna phase synchronization, cross-NIC symbol synchronization, and $>$3 radio chains. However, these features have never been fulfilled, and their lack has become a significant obstacle to research advancement.

Second, \textit{the hardware low-level control is limited}. The Wi-Fi sensing research could be further developed if some low-level hardware controls were available, such as the tuning of the carrier frequency and sampling rate, radio-frequency (RF) calibration, in-phase/quadrature (I/Q) mismatch~\cite{Zhu:2018fca}, Tx/Rx gain control, beamforming, \etc. Unfortunately, these low-level hardware controls have never appeared, too.

Third, \textit{the complete physical (PHY)-layer information is inaccessible}.
CSI is just a small part of the complete PHY-layer information that the Rx baseband measures during the whole decoding process. The Rx baseband also measures the CSI based on the legacy long training field (L-LTF), denoted as ``Legacy CSI'', CSI based on the pilot subcarriers of all data symbols (``Pilot CSI''), error vector magnitudes (EVMs), estimations of the CFO and SFO, \etc. None of these valuable measurements is inaccessible to researchers.

As an alternative to the commercial off-the-shelf (COTS) Wi-Fi NICs, 
SDR has been used in some research~\cite{Xiong:2015dr, 2018BLoc, mDTrack, 246372}.
However, \textit{a dilemma arises in the broad adoption of SDR devices}. On the one hand, SDR is a promising tool for Wi-Fi sensing, as it provides full access to the raw baseband signals and complete control over the hardware.
On the other hand, the lack of a publicly available baseband implementation makes SDR practically infeasible in research. Few teams have developed their implementations, as it is an even more challenging project than the Wi-Fi sensing research.

\textbf{Barrier 3: the lack of versatile and flexible measurement software}. 
Incapable measurement software also poses a series challenges to research advancement.
The lack of support for multi-NIC concurrent CSI measurement is a very large obstacle for building a COTS NIC-based phased array~\cite{Gjengset2014Phaser}.
The lack of measurement metadata also causes trouble in CSI data alignment and preprocessing.
In addition to the insufficient functionalities,
the lack of versatility hinders the realization of complex and interactive CSI measurements, 
such as the multi-NIC collaborative measurement, round-trip measurement and synchronized channel hopping~\cite{Kumar2016Decimeter}.
To conduct these measurements, we need a Wi-Fi sensing middleware which integrates some functionalities, such as packet injection and in situ CSI data parsing, while providing APIs allowing researchers to develop their measurement-specific plugins.
Nevertheless, the existing CSI tools are merely CSI collectors and are not architecturally flexible to meet these goals.

\begin{figure*}[t]
    \centering
    \includegraphics[width=\textwidth]{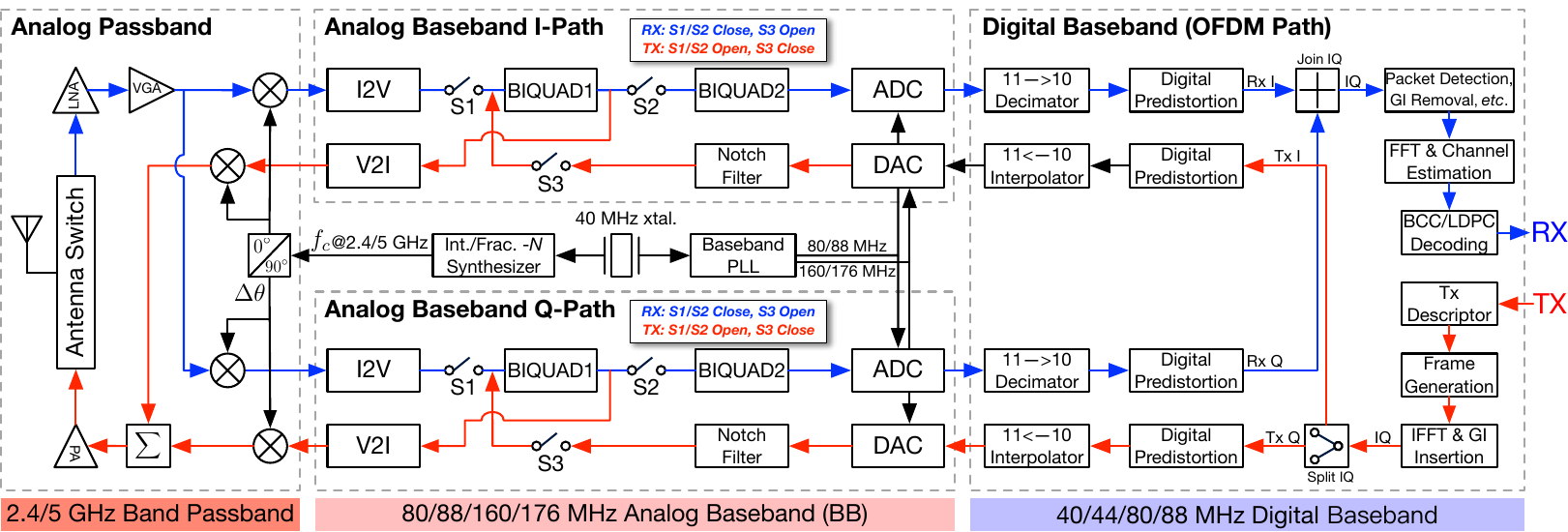}
    \caption{QCA9300 three-stage transceiver architecture. The three stages operate at different frequencies. The TX and Rx processing flows are highlighted in red and blue, respectively. For simplicity, this diagram shows only one of the three transceiver chains.}
    \label{fig:architecture}
\end{figure*}

This paper aims to address these barriers from three aspects.

To eliminate Barrier 1, we piece together a detailed architecture for the baseband design of the QCA9300 NIC. The architecture covers the RF frontend, analog baseband, and digital baseband. We study in detail three of the most critical components: the register-based control mechanism, dual-band carrier frequency synthesis and baseband multirate clocking. Then, we evaluate the impact of different baseband configurations on the CSI distortion. The evaluations show that the baseband, especially the baseband filtering, is the most influential factor in CSI distortion. The evaluations also indicate that CSI distortion is pervasive in all Wi-Fi NICs, including the QCA9300, IWL5300 and SDR devices. Finally, we propose a trivial method to eliminate the CSI distortion.

To eliminate Barrier 2,
we achieve breakthroughs in hardware capabilities for both the COTS Wi-Fi NIC and SDR devices.
For COTS NICs, as a benefit of QCA9300 baseband exploration,
we successfully unlock some of the most valuable low-level controls for the QCA9300 NIC, such as arbitrary tuning of both the carrier frequency and the baseband sampling rate, per-packet multi-CSI measurement and manual Rx gain control.
For SDR, we solve the dilemma mentioned above by developing a high-performance software baseband implementation for the 802.11a/g/n/ac/ax protocols. We provide rich APIs to control a wide range of low-level details, such as the bandwidth, packet timing, CFO, SFO, spatial mapping, Tx/Rx gain, resampling, and even I/Q mismatch.
We also provide \textit{the complete PHY-layer information}. In addition to the HT-LTF based CSI, the baseband returns the Legacy CSI, Pilot CSI, CFO and SFO estimations, packet content and fundamental per-packet raw baseband signals.

To eliminate Barrier 3, we develop and release PicoScenes, a versatile and flexible Wi-Fi sensing platform software that integrates the packet injection, in situ CSI data parsing, low-level hardware controls, multi-NIC concurrent operation and the plugin mechanism. 
\textit{PicoScenes on SDR}, the dedicated support for SDR devices, integrates our baseband implementation into PicoScenes. It enables researchers to transmit and receive 802.11a/g/n/ac/ax Wi-Fi packets in real time using SDR devices similar to ordinary COTS Wi-Fi NICs.
Regarding the architectural flexibility,
we provide a rich set of hardware-unified and high-level APIs enabling researchers to prototype their measurement plugins easily.
PicoScenes currently supports two COTS NICs, the QCA9300 and IWL5300 NICs, and \textit{all} Universal Software Radio Peripheral (USRP) models.

%




The contributions of this paper are summarized as follows.


First, to the best of our knowledge, 
this is the first in-depth study of hardware baseband design for COTS Wi-Fi NICs.
We identify the pervasiveness of CSI distortion and propose a reliable solution to eliminate it.
The lessons learned from the QCA9300 NIC are of great guiding significance for inferring how other confidential platforms operate, such as the IWL5300 and Broadcom 43xx (BCM43xx)~\cite{Nexmon} models.


Second, we \textit{reintroduce} both the QCA9300 and SDR as powerful hardware options for Wi-Fi sensing research.
For the QCA9300 NIC, we, for the first time, unlock the arbitrary access to the $>$2 GHz spectrum, up to 80 MHz bandwidth and manual Rx gain control, which enable tremendous potential for COTS NIC-based Wi-Fi sensing research.
For SDR devices, we resolve the dilemma of SDR. With the high-performance baseband implementation, SDR devices can inject and receive Wi-Fi packets in real time similar to an ordinary Wi-Fi NIC. The baseband implementation also provides complete hardware control and returns much richer PHY-layer information.


Third, we release PicoScenes, the first middleware platform for Wi-Fi sensing research and application. 
It is the first platform that supports multi-NIC concurrent CSI measurement for the QCA9300 and IWL5300 NICs. 
Researchers can build a large multi-NIC/SDR Wi-Fi sensing array and connect it to only one computer. We even release the reference design for a 27-NIC array.
PicoScenes is architecturally flexible such that researchers can quickly realize a complex and interactive CSI measurement protocol via the plugin mechanism.

Last, comprehensive evaluations demonstrate the state-of-the-art performance of the PicoScenes platform, \eg, $>$2 GHz wide full-spectrum availability for the QCA9300 NIC, an up to 8 kHz CSI measurement rate by the 27-NIC array, an up to 40 kHz packet injection rate on SDR, an up to 1 kHz packet decoding rate on SDR and a up to 200 MHz bandwidth CSI measurement on SDR.

The rest of this paper is organized as follows. Section~\ref{sec:architecture} presents an in-depth study of the baseband design of the QCA9300. Section~\ref{sec:distortion} investigates CSI distortion. Section~\ref{sec:picoscenes} introduces the PicoScenes platform. Section~\ref{sec:evaluation} reports extensive evaluations. Section~\ref{sec:related_work} briefly reviews recent works on Wi-Fi sensing. Finally, Section~\ref{sec:conclusion} concludes the paper.

\section{QCA9300 Hardware Architecture} 
\label{sec:architecture}

In this section, we present the baseband architecture of the QCA9300 NIC. We focus on three core designs that are most relevant to Wi-Fi sensing research. Then, we conclude what we learn from the QCA9300 NIC and discuss how other Wi-Fi NICs might operate.

\subsection{Architecture of the QCA9300 NIC}

Detailed baseband design is the key to unveiling the signal processing flow. However, Wi-Fi NIC vendors keep their designs confidential and release them only to NDA-signed partners. 


Fortunately, Atheros has published papers about the design of the QCA9300 NIC~\cite{4684645, 4907420, AbdollahiAlibeik:2011um, sequencing}. 
These works give us a change to glimpse the actual design of modern Wi-Fi NICs. 
Compared with the firmware-controlled IWL5300 and BCM43xx, the QCA9300 NIC is fully controlled by an open-source kernel driver, codenamed \textsf{ath9k}~\cite{ath9k_driver}. This firmware-free architecture allows us to investigate the baseband design interactively.
Based on the in-depth study of Atheros's publications and the \textsf{ath9k} driver, we piece together a detailed architecture of the QCA9300 baseband, as shown in Fig.~\ref{fig:architecture}. 

The QCA9300 adopts a three-stag design: \textit{digital baseband}, \textit{analog baseband} and \textit{analog passband}.


The \textit{digital baseband} stage deals with the discrete complex baseband signals. It is composed of the fast Fourier transform (FFT) and inverse FFT (IFFT) pairs, 802.11 packet generation and detection, channel estimation, the Rx state machine, forward error correction (FEC) codecs, \etc.

The \textit{analog baseband} stage performs signal sampling and filtering. 
The 40 MHz crystal local oscillator (LO), denoted by ``40 MHz xtal.'', drives the baseband phase-locked loop (PLL), which then feeds a group of synchronized timing signals to the digital-to-analog converters (DACs), analog-to-digital converters (ADCs) and the digital baseband. 
The Tx and Rx paths have different filtering settings.
In the Tx path, denoted by the red signal route in Fig.~\ref{fig:architecture}, the voltage-to-current converter (V2I), notch filter and BIQUAD1 (an active-RC based reconfigurable filter) form a second-order Butterworth filter, which serves as the DAC reconstruction filter.
In the Rx path, denoted by the blue signal route in Fig.~\ref{fig:architecture}, the current-to-voltage converter (I2V), BIQUAD1 and BIQUAD2 (another reconfigurable filter) form a fifth-order Butterworth filter, which serves as the ADC anti-aliasing filter and the adjacent channel rejection (ACR) filter.
It is worth noting that the ADC/DAC pairs sample the in-phase and quadrature (I/Q) signals separately; therefore, all filters in the I-path have their mirrors in the Q-path.

The \textit{analog passband} stage is a rather conventional RF frontend circuit. It performs the carrier frequency synthesis, up/down-conversion, power amplification and antenna switch control. The QCA9300 uses a single frequency synthesizer to generate the carrier frequencies of both the 2.4 and 5 GHz Wi-Fi bands.

\subsection{Three core design elements of the QCA9300}

In this section, we discuss in detail the design elements that are most relevant to Wi-Fi sensing.

\subsubsection{Highly configurable and firmware-free architecture}
\label{ssec:soft_mac}
The \textsf{ath9k} driver has complete control over the NIC hardware.
Specifically, the NIC exposes a large number of control registers. These registers, directly accessible by the kernel driver, become the control interface between the driver and the hardware.
For instance, the Atheros CSI Tool~\cite{Xie:2019ji} tells the hardware to report CSI measurements by specifying the 28th bit of the control register 0x8344.
Taking advantage of these registers, our proposed platform, PicoScenes, can direct the NIC hardware to operate in a much broader working range, which enables great potential for Wi-Fi sensing research.

\subsubsection{Wide-range and user-tunable baseband clocking}
\label{ssec:sf_tuning}

\begin{figure}[t]
    \centering
    \includegraphics[width=0.9\columnwidth]{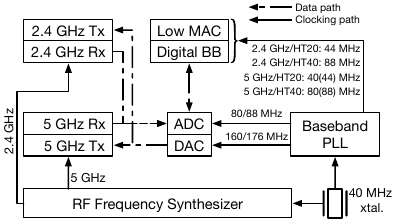}
\caption{Clocking hierarchy of the QCA9300. The 40 MHz crystal gives rise to two clocking branches. The RF Frequency Synthesizer generates the carrier frequencies, while the Baseband PLL generates multiple matched clocks for the baseband components. The values in parentheses are the frequencies when the \textit{fastclock} option is on.}
    \label{fig:clocking}
\end{figure}

The clocking architecture, as shown in Fig.~\ref{fig:clocking}, is definitely the core design of the QCA9300.
The 40 MHz LO derives two branches. One feeds into the Baseband PLL to drive the entire baseband, and the other feeds into the RF Frequency Synthesizer to generate the carrier frequencies for both the 2.4 GHz and 5 GHz bands. We first focus on the baseband branch.

The 802.11n protocol has a 20 or 40 MHz signal bandwidth, denoted by $f_{sig\_bb}$. However, the actual baseband clocking operates at much higher frequencies.
As shown in Fig.~\ref{fig:clocking}, the Baseband PLL generates a group of matched clocks that drive different parts of the baseband circuit. More confusingly, many of the frequencies have another ``paired'' frequency of $1.1\times$ higher, such as 80/88 MHz and 160/176 MHz. After a thorough study, we believe that at least three main challenges motivate this complex clocking design.

The primary challenge is to overcome the imperfections of the DAC. The DAC, modeled as a zero-order hold (ZOH) process, is fundamentally unable to reconstruct the analog signals perfectly. 
The output of the ZOH process is not smooth but rather a stair-stepped pattern. This pattern distorts the ideal output from two aspects: in-band \textit{sinc} fading and out-of-band spectral leakage~\cite{Sklar2006Digital}. The spectrum leakage repeats the spectrum at every multiple of the sampling rate in the frequency domain. These two types of distortion violate both the spectral flatness and spectral masking requirements specified by the 802.11 standard~\cite{80211n_standard}.
The conventional approach for suppressing the distortion is to attach a reconstruction filter after the DAC.
However, if the DAC sampling rate is just above $f_{sig\_bb}$, the transition zone of the reconstruction filter will be too narrow to design a filter in the IC.

Atheros addresses this problem by oversampling. The DAC of the QCA9300 operates at 160 MHz, 8$\times$ the 20 MHz signal bandwidth. 
According to sampling theory~\cite{Sklar2006Digital},
the 8$\times$ oversampling stretches both the \textit{sinc} fading and the transition zone by a factor of 8. The stretched transition zone simplifies the IC design for the reconstruction filter. 
As a result, the combination of a second-order LPF and a notch filter is sufficient to suppress the spectral images, as shown in Fig.~\ref{fig:architecture}.

\begin{table}[t!]
    \centering
    \caption{QCA9300 bandwidths calculated from parameter quadruples}
    \renewcommand{\arraystretch}{1.1} 
    \begin{tabular}{|c|c|c|} \hline
       (\textsf{DIV\_INT}, \textsf{REF\_DIV}, \textsf{CLK\_SEL}) & \textsf{HT\_2040} = 0 & \textsf{HT\_2040} = 1\\ \hline
       (22, 10, 1) & 2.5 MHz & 5 MHz \\ \hline
       (22, 10, 0) & 5 MHz & 10 MHz  \\ \hline
       (22, 5, 1) & 5 MHz & 10 MHz   \\ \hline
       (22, 5, 0) & 10 MHz & 20 MHz  \\ \hline
       (33, 5, 0) & 15 MHz & 30 MHz  \\ \hline
       (44, 5, 0) & 20 MHz & 40 MHz  \\ \hline
       (55, 5, 0) & 25 MHz & 50 MHz  \\ \hline
       ... & ... & ...  \\ \hline
       (88, 5, 0) & 40 MHz & 80 MHz  \\ \hline
    \end{tabular}
    \label{tab:pll_tuning}
\end{table}

The second challenge is power consumption. Running the entire baseband circuit at 160 MHz is unnecessary and power-inefficient.
The Baseband PLL generates three levels of subfrequencies to achieve a trade-off between power efficiency and temporal resolution.
The Tx DACs run at the highest frequency, \ie, $f_{tx\_dac}$ = 160 MHz; the Rx ADCs run at a lower frequency, \ie, $f_{rx\_adc}$ = 80 MHz; and the digital baseband runs at the lowest frequency, \ie, $f_{digi\_bb}$ = 40 MHz.

The third challenge is the mandatory backward compatibility with the 802.11b protocol.
In the 2.4 GHz band, the NIC must support the 802.11b/g/n protocols simultaneously. The 802.11b protocol runs at a 22 MHz signal bandwidth, \ie, $f_{sig\_bb}^{11b}=22$ MHz, which is not compatible with either the 20 or 40 MHz bandwidth of the 802.11 g/n protocols.
To solve this problem, the QCA9300 adopts a multi-rate design to support the 802.11b/g/n protocols simultaneously in the 2.4 GHz band.
Specifically, the QCA9300 boosts $f_{digi\_bb}$, $f_{rx\_adc}$ and $f_{tx\_dac}$ by 1.1 times to 44, 88 and 176 MHz, respectively. Then, the QCA9300 uses two types of rate converters to tackle the frequency gaps.
For the 802.11b 22 MHz bandwidth case,
the gap between $f_{sig\_bb}=22$ MHz and $f_{digi\_bb}=44$ MHz is bridged by
a ``$0.5 \leftrightarrow 2$'' rate converter, \ie, a $0.5\times$ frequency divider for the Rx path and a $2\times$ frequency multiplier for the Tx path.
For the 802.11n 40 MHz bandwidth case, \ie, 802.11n HT40+/- channel modes with $f_{sig\_bb}=40$ MHz,
the QCA9300 uses a ``$10\leftrightarrow 11$'' rate converter to bridge the gap between $f_{sig\_bb}=40$ MHz and $f_{digi\_bb}=44$ MHz.
For the 802.11g/n 20 MHz bandwidth case, \ie, HT20 or HT40+/- channel modes with $f_{sig\_bb}=20$ MHz, the QCA9300 still treats the signals as 40 MHz bandwidth signals but later discards half of the subcarriers.
Finally, in the 5 GHz band that the 802.11b protocol does not support, the QCA9300 bypasses the $10\leftrightarrow 11$ rate conversion pair and restores $f_{digi\_bb}$, $f_{rx\_adc}$ and $f_{tx\_dac}$ to 40, 80 and 160 MHz, respectively.
Note that in some later models of the QCA9300 series, namely, the QCA9380/9390/9590 models, there is a \textit{fastclock} option (ON by default) that specifies the 44/88/176 MHz clocks for both the 2.4 and 5 GHz bands.

Regarding the control mechanism,
$f_{digi\_bb}$, $f_{rx\_adc}$ and $f_{tx\_dac}$ are collectively derived from the core baseband PLL clock, denoted by $f_{pll}$.
The \textsf{ath9k} driver uses a parameter quadruple to tune $f_{pll}$. This parameter quadruple has the form of (\textsf{DIV\_INT}, \textsf{REF\_DIV}, \textsf{CLK\_SEL}, \textsf{HT20\_40}), representing the \textit{integer multiplier}, \textit{reference clock divider}, \textit{clock selection} and \textit{channel mode selection}, respectively.
Through reverse engineering, we have learned how to tune $f_{pll}$ and its derived clocks. Their mathematical models are as follows.
\begin{align}
     & f_{pll} = \textsf{DIV\_INT}\times\frac{f_{xtal.}}{\textsf{REF\_DIV}}\times\frac{2^{\textsf{HT20\_40}}}{2^{(2 + \textsf{CLK\_SEL})}},
     \nonumber \\
     &  \quad \quad \textsf{CLK\_SEL} \in \{0, 1, 2\}, \quad \textsf{HT20\_40}\in \{0, 1\} \label{eq:wlan_pll} \\
     &  f_{rx\_adc} = f_{pll} \\
     &  f_{tx\_dac} = 2\cdot f_{pll} \\
     &  f_{digi\_bb} = 1/ 2\cdot f_{pll} \\
     &  f_{sig\_bb} = 10/11\cdot f_{digi\_bb}
\end{align}
where $f_{xtal.}$=40 MHz is the LO frequency.
In the 2.4 GHz band, the parameter quadruples for the 802.11n HT20 and HT40 modes are (44, 5, 0, 0) and (44, 5, 0, 1), respectively.
Taking the 802.11n protocol with the HT20 mode as an example, 
by substituting the quadruple (44, 5, 0, 0) into (\ref{eq:wlan_pll}), we obtain a frequency of 88 MHz for $f_{pll}$ and frequencies of 20, 44, 88 and 176 MHz for $f_{sig\_bb}$, $f_{digi\_bb}$, $f_{rx\_adc}$ and $f_{tx\_dac}$, respectively.
As is clearly shown in the equations, $f_{pll}$ controls the pace of the entire baseband; therefore, \textit{tuning $f_{pll}$ is equivalent to tuning the channel bandwidth}. Table~\ref{tab:pll_tuning} lists the channel bandwidths supported by the PicoScenes platform and their quadruple values. 
Unexpectedly, the QCA9300 has a very wide operable bandwidth range. The bandwidth can be scaled from a minimum of 2.5 MHz up to 80 MHz.
PicoScenes integrates bandwidth tuning for the QCA9300, allowing users to specify the bandwidth by the ``\textsf{--rate}'' option.

\subsubsection{Wide-range and user-tunable carrier frequency synthesizer}
\label{ssec:cf_tuning}
\begin{figure}[t] 
    \centering
    \includegraphics[width=0.9\columnwidth]{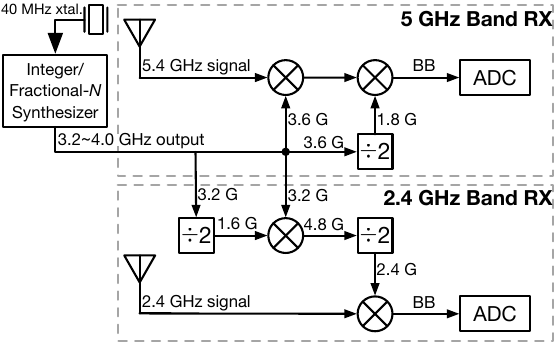}
\caption{Carrier frequency synthesis scheme of the QCA9300. An integrated integer/fractional-\textit{N} synthesizer is shared between the 2.4 and 5 GHz bands but with different frequency conversion paths. For simplicity, this diagram shows only the Rx downconversion path.}
    \label{fig:carrier}
\end{figure}

Fig.~\ref{fig:carrier} shows the detailed carrier frequency synthesizing scheme of the QCA9300.
The QCA9300 uses a shared frequency synthesizer
to generate the carrier frequency for both the 2.4 GHz and 5 GHz bands.
The frequency synthesizer is a voltage-controlled oscillator (VCO)-based synthesizer with an operating range $f_{syn}$ from 3.0 to 4.0 GHz.

The frequency synthesis scheme for the 2.4 GHz band is illustrated in the lower part of Fig.~\ref{fig:carrier}.
The synthesizer operates near 3.2 GHz, \ie, $f_{syn}=3.2$ GHz.
To generate a 2.4 GHz band carrier frequency,
$f_{syn}$ is first mixed with half of itself, \ie, 1.6 GHz, producing a 4.8 GHz signal, and the 4.8 GHz signal is further divided by 2 to obtain the 2.4 GHz frequency.

The frequency synthesis scheme for the 5 GHz band is illustrated in the upper part of Fig.~\ref{fig:carrier}.
The synthesizer operates between 3.4 and 3.9 GHz in this band.
Taking the downconversion of a 5.4 GHz Rx signal as an example,
the synthesizer operates at 3.6 GHz, \ie, $f_{syn}=3.6$ GHz.
The 5.4 GHz Rx signal is first downconverted to an intermediate frequency (IF) of 1.8 GHz by mixing it with $f_{syn}$. Then, the IF signal is downconverted again by mixing it with half of $f_{syn}$, \ie, 1.8 GHz. In this way, the signal is dual-converted to the baseband.
 
\begin{table}[t]
    \centering
    \caption{Carrier frequencies supported by the QCA9300}
    \renewcommand{\arraystretch}{1.1} 
    \begin{tabular}{|l|c|c|c|} \hline
                          & $f_{syn}$ & $f_{rf}^{2.4}$ & $f_{rf}^{5}$ \\ \hline
      Effectively supported range (GHz) & 3.0-4.0 & 2.2-2.9     & 4.4-6.1  \\ \hline
      Minimal tuning resolution (Hz) & 305.2 & 203.3         & 915.5   \\ \hline
    \end{tabular}
    \label{tab:uwb}
\end{table}

Regarding the control mechanism,
via reverse engineering,
we have learned how to tune the synthesizer, as follows:
\begin{align}
    f_{syn} & = \textsf{CHANSEL} \times\frac{f_{xtal.}}{2^{17}} \label{eq:syn} \\
    f_{rf}^{2.4} & = 3/4\cdot f_{syn}, \quad f_{rf}^{5} = 3/2\cdot f_{syn}  \label{eq:f_rf24}
\end{align}
where \textsf{CHANSEL} is an unsigned integer variable used for tuning $f_{syn}$. $f_{rf}^{2.4}$ and $f_{rf}^{5}$ are the derived carrier frequencies in the 2.4 and 5 GHz bands, respectively.
By substituting \textsf{CHANSEL}=1 into (\ref{eq:syn}), we obtain the minimal tuning step of the synthesizer: $f_{syn_{step}} \approx 305$ Hz.
In addition, bringing the operating range of the synthesizer (3.0 to 4.0 GHz) into (\ref{eq:f_rf24}),
we obtain Table~\ref{tab:uwb}, which lists the supported tuning ranges and resolutions supported by the QCA9300.
The table indicates that the QCA9300 supports a 700 MHz and a 1.7 GHz continuous spectrum in the 2.4 and 5 GHz bands. 
We believe this broad and continuously accessible spectrum can significantly benefit Wi-Fi sensing research.
The PicoScenes software also integrates the carrier frequency tuning for the QCA9300, allowing users to specify the carrier frequency by the ``\textsf{--freq}'' option.
 

\begin{figure}[t]
	\centering
	\includegraphics[width=0.9\columnwidth]{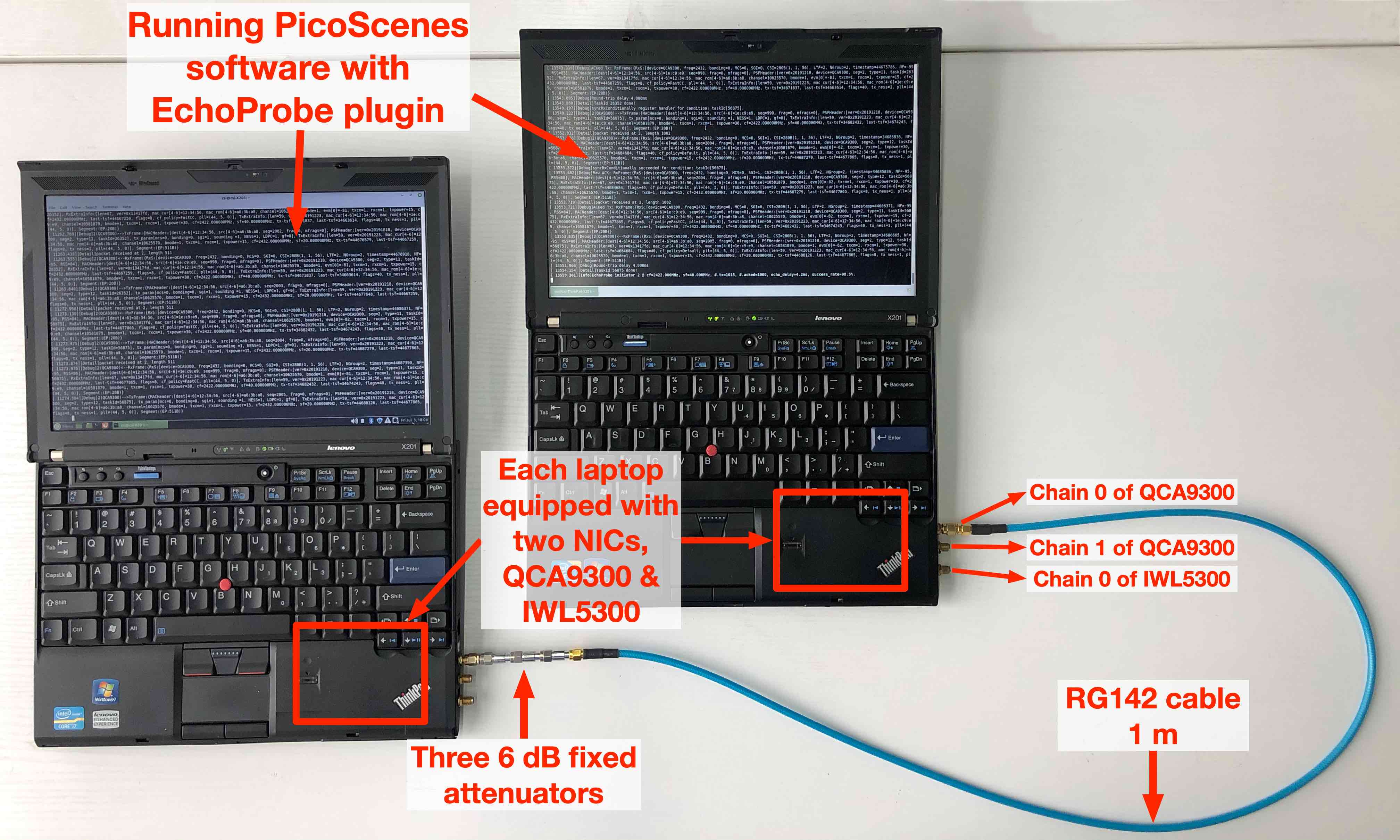}
	\caption{Hardware setup for CSI distortion evaluation.}
	\label{fig:double_laptop}
\end{figure}

\subsection{What can we learn from the QCA9300 NIC?}

The QCA9300 provides significant guidance in speculating how other Wi-Fi NICs operate. We discuss three key points below, which are common to other NICs.

The first point is related to the influence of the baseband filter.
Due to the spectral flatness and masking requirements of the 802.11 standard, all NICs have strong filters before and after the DAC/ADC pair. There is no doubt that these filters exert a certain influence on the in-band signal, including the CSI. Therefore, we may need to re-evaluate the correctness of the CSI model shown in (\ref{eq:csi_basic_form}), which assumes that the influence of the baseband hardware is merely a linear phase offset.

The second is related to the sampling time.
Phase-based Wi-Fi sensing is sensitive to the STO, which is an integer multiple of the sampling time $T_s = 1 / f_{rx\_adc}$. Prior works assumed that the hardware sampling frequency equals to the channel bandwidth~\cite{sensesurvey1, Tadayon:2019cj}, \ie, $f_{rx\_adc} = f_{sig\_bb}$.
However, this assumption is not always valid, at least for the QCA9300 and USRP cases. For the QCA9300, its Rx ADC operates at 88 and 80 MHz for the 2.4 and 5 GHz bands~\cite{4907420}, respectively. For the USRP N210 or X310, both the Tx and Rx run at a master clock rate of 100 or 200 MHz~\cite{usrp_manual}, respectively. For the IWL5300, based on our preliminary evaluations, we strongly believe that its Rx path operates at 40 MHz.

The third is related to the CFO.
Modern NICs use integer/fractional-$N$ synthesizers to generate their carrier frequencies. However, as suggested by (\ref{eq:syn}), the synthesizer has a minimum tuning resolution; thus, the synthesizer cannot precisely tune to specific frequencies. Still taking the QCA9300 as an example, if we wish to specify a frequency of 5.2 GHz, the NIC is actually specified to operate at 5199.999389 MHz or 5200.000305 MHz. This small CFO is perfectly acceptable in Wi-Fi communication and most Wi-Fi sensing applications; however, it may mix with the Doppler effect, contaminating the estimation of the Doppler frequency shift. Unfortunately, the tuning resolution depends on the hardware design, and we currently have detailed clocking architectures only for the QCA9300 and USRP models.

\section{Characterizing the \textit{Pervasive} CSI Distortion} 
\label{sec:distortion}


In this section, we examine the CSI distortion under various channel configurations and experimentally identify the most likely origin of the distortion. Then, we try to explain what this distortion is and where it comes from. Finally, we propose a trivial method to eliminate this distortion.

\subsection{Where does the CSI distortion come from?}
We conducted 5 tests (T1 to T5) to identify the origin of the distortion.

\subsubsection*{Test setup}
Two Lenovo ThinkPad X201 laptops were used in the tests. Both ran the Linux Mint 20 OS (based on Ubuntu 20.04 LTS) with kernel version 5.4. We used this old laptop model because the X201 has two mini PCI-E slots, which enable us to install both the QCA9300 and IWL5300 NICs on a single laptop.
As shown in Fig.~\ref{fig:double_laptop}, the NICs under test were connected by a double-shielded coaxial cable (RG142) and three 6 dB fixed attenuators.
During the tests, we used the PicoScenes software to control the injection-based Tx and monitor mode-based Rx. We used a low Tx power of 5 dBm to prevent Rx ADC saturation. We also shut down the Rx end's second and third radio chains to avoid the undesired maximal ratio combining (MRC).

\begin{figure}[t]
	\begin{center}
		\begin{tabular}{cc}
			\hspace{-0.13in}
			\includegraphics[width=0.5\columnwidth]{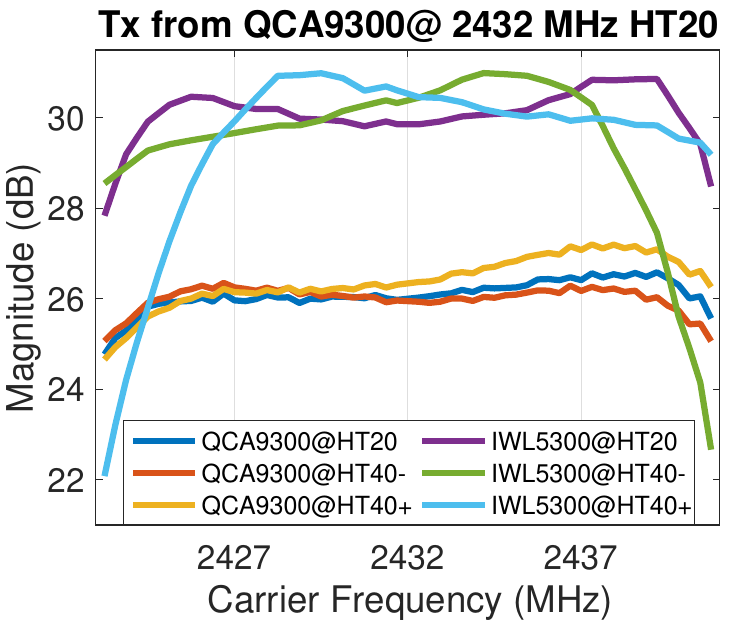} &
			\hspace{-0.2in}
			\includegraphics[width=0.5\columnwidth]{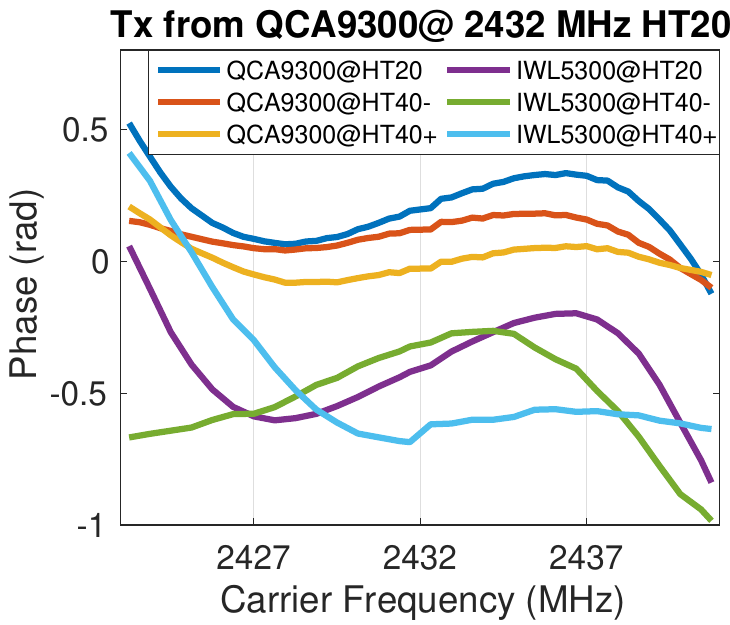}
		\end{tabular}
	\end{center}
	\caption{CSI distortion (magnitude and phase response) under different Rx NIC-channel mode configurations. The Tx end was a QCA9300 NIC continuously injecting packets in the HT20 channel mode. The QCA9300 and IWL5300 installed on the other laptop, both acting as Rx ends, switched among the HT20, HT40+ and HT40- channel modes.}
	\label{fig:rx_diversity}                                                       
\end{figure}

\subsubsection{T1, test for the influence of the Rx baseband}
\label{sssec:rx_contribution}
In this test, the QCA9300 NIC installed on laptop A continuously injected 802.11n HT20-rate packets at the 2432 MHz channel. Both the QCA9300 and IWL5300 NICs installed on laptop B received the packets with three different channel modes, namely, HT20, HT40+ and HT40-. In this way, we collected the received CSI data from a total of 6 NIC-channel mode combinations (2 NIC models $\times$ 3 channel modes). Fig.~\ref{fig:rx_diversity} shows the average magnitude and phase of the CSI.

Before the analysis, we briefly recap the 802.11n channel modes. HT20 or HT40+/- refers to the 802.11n HT-format channel with a 20 or 40 MHz bandwidth, respectively. The HT40+/- modes double the bandwidths to 40 MHz by subsuming an adjacent 20 MHz bandwidth channel with a higher or lower carrier frequency, shifting the carrier frequency toward the center of the merged channels.
An HT40+/- channel can also communicate with an HT20 channel by transmitting and receiving the signals with only the higher or lower half of its 40 MHz bandwidth.

We made three observations about the distortion from Fig.~\ref{fig:rx_diversity}.
\begin{enumerate}[leftmargin=*]
	\item The CSI difference shown in Fig.~\ref{fig:rx_diversity} can be attributed only to the Rx baseband and the NIC-channel mode configurations because the Tx end remained unchanged during the test.
	\item In the HT20 channel mode, both NICs exhibited approximately symmetrical distortion. The magnitude and phase show an M-shaped distortion and a horizontal S-shaped distortion, respectively. Both types of distortion were strong.
	\item In the HT40+/- channel modes, both the magnitude and phase distortion were heavily biased, especially in the case of the IWL5300. Interestingly, we find that the biased response is similar to a stretched version of the left or right half of the HT20 response.
\end{enumerate}
%



\begin{figure}[t]
	\begin{center}
		\begin{tabular}{cc}
			\hspace{-0.13in}
			\includegraphics[width=0.5\columnwidth]{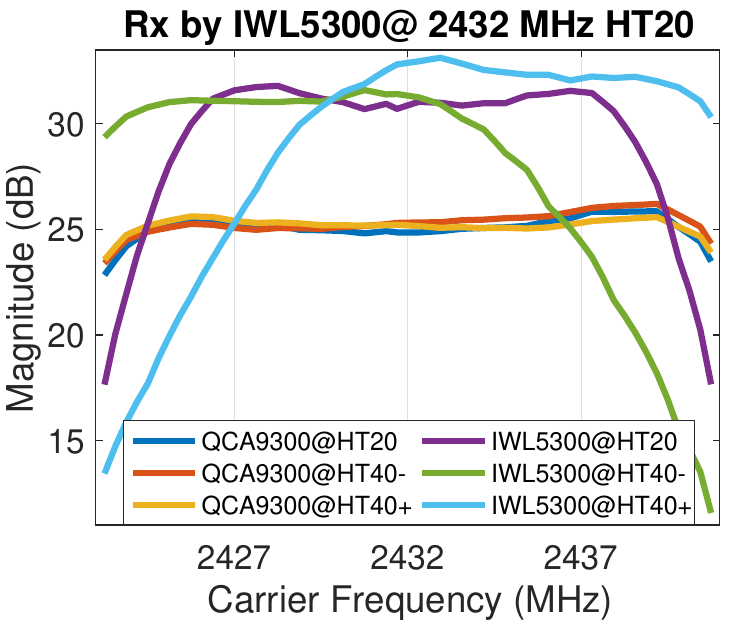} &
			\hspace{-0.2in}
			\includegraphics[width=0.5\columnwidth]{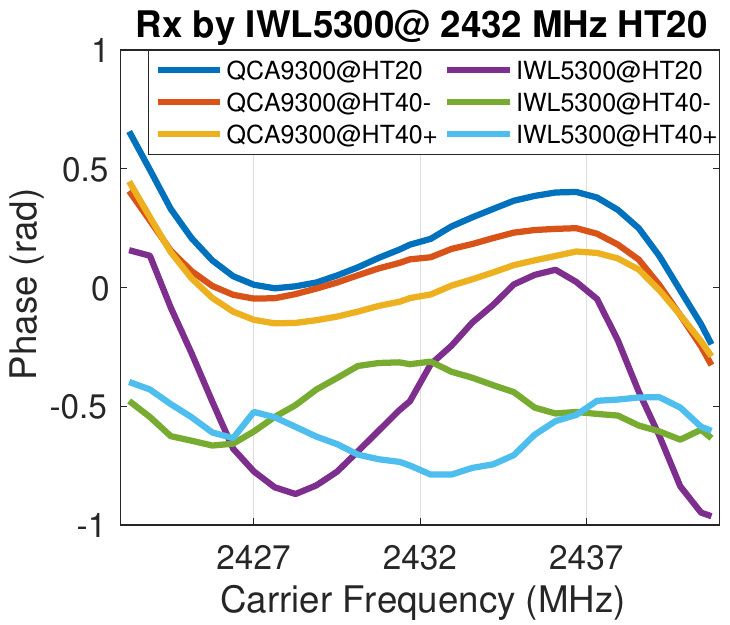}
		\end{tabular}
	\end{center}
	\caption{CSI distortion (magnitude and phase response) under different Tx NIC-channel mode configurations. The Rx end was an IWL5300 NIC operating at 2432 MHz in the HT20 channel mode. A QCA9300 and a second IWL5300, both acting as Tx ends, injected packets in the HT20, HT40+ and HT40- channel modes.}
	\label{fig:tx_diversity}                
\end{figure}

\subsubsection{T2, test of the influence of the Tx baseband}
\label{sssec:tx_contribution}
In this test, we swapped the Tx and Rx roles. Specifically, the IWL5300 NIC installed on laptop A was used as the Rx NIC\footnote{The QCA9300 NIC does not report CSI measurements for packets sent by an IWL5300 NIC, but the reverse is possible. Therefore, we used an IWL5300 NIC as the Rx NIC in this test.}. It operated at 2432 MHz in the HT20 channel mode and remained unchanged during the test.
The QCA9300 and IWL5300 NICs installed on laptop B transmitted packets in each of the three channel modes.
Fig.~\ref{fig:tx_diversity} shows the average magnitude and phase of the received CSI. We made the following key observations:
\begin{enumerate}[leftmargin=*]
	\item The CSI difference shown in Fig.~\ref{fig:tx_diversity} can be attributed only to the Tx baseband and the NIC-channel mode configurations because the Rx end remained unchanged during the test. This test reveals a long-ignored aspect of Wi-Fi sensing: \textit{the Tx signal emitted from the antenna is not flat in the spectrum, either in magnitude or phase}.
	\item Given the remarkably high similarity between Figs.~\ref{fig:rx_diversity} and \ref{fig:tx_diversity}, we suspect that in both NICs, there \textit{may} be one or more shared stages between the Tx and Rx signal processing flows. The QCA9300 baseband design supports this conjecture in that the \textsf{BIQUAD1} filter is indeed shared between the Tx and Rx flows, as shown in Fig.~\ref{fig:architecture}.
	\item The IWL5300 performs worse than the QCA9300 in terms of spectral flatness and spectral masking. In the HT40+/- cases, the difference in magnitude for the IWL5300 is greater than 10 dB; in contrast, the QCA9300 has better spectral flatness than the IWL5300.
\end{enumerate}

\begin{figure}[t]
	\begin{center}
		\begin{tabular}{cc}
			\hspace{-0.13in}
			\includegraphics[width=0.5\columnwidth]{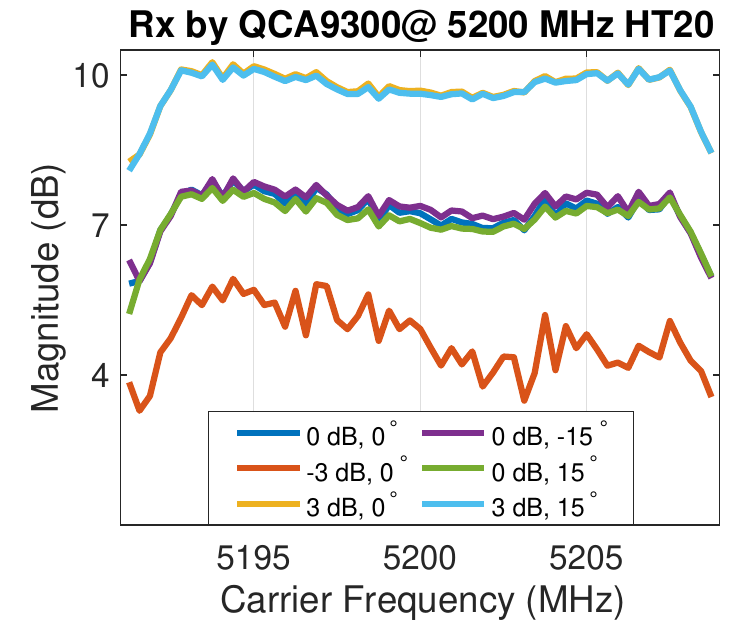} &
			\hspace{-0.2in}
			\includegraphics[width=0.5\columnwidth]{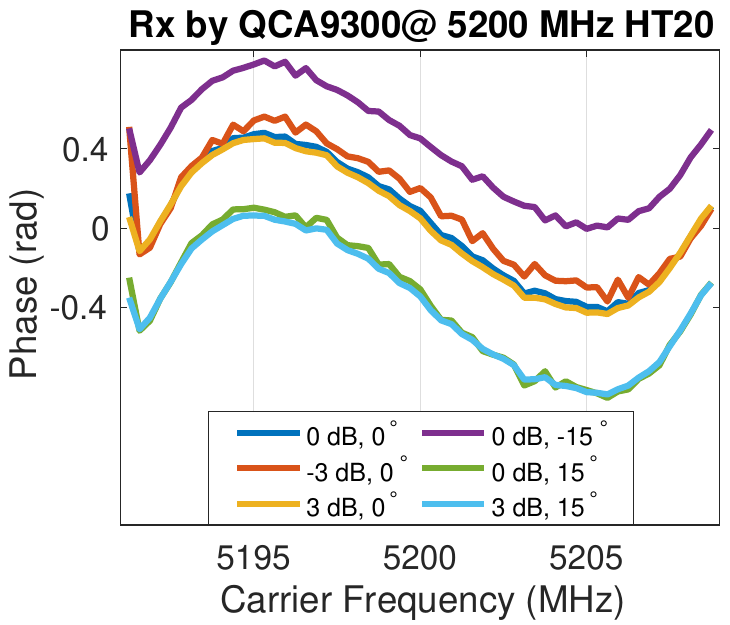}
		\end{tabular}
	\end{center}
	\caption{CSI distortion (magnitude and phase response) under different Rx I/Q mismatch configurations. Both the Tx and Rx ends were QCA9300 NICs operating at 5200 MHz in the HT20 channel mode. On the Rx end, different I/Q mismatch parameters were scanned.}
	\label{fig:iqmismatch_scan}                
\end{figure}

\subsubsection{T3, test of the influence of the I/Q mismatch}
I/Q mismatch is a common imperfection of the radio frontend, which is reflected in two mismatches, namely, \textit{magnitude inequality} and \textit{phase nonorthogonality}~\cite{Anonymous:Vp0PiVlj}. Previous work~\cite{Zhu:2018fca} claimed that the I/Q mismatch causes the phase distortion, but it was not verified.
In this test, we overrode the Rx I/Q imbalance configuration in the \textsf{ath9k} driver and scanned both the I/Q magnitude ratio and the I/Q phase offset of the Rx QCA9300, while the Tx end continuously injected packets at 2432 MHz in the HT20 channel mode. Fig.~\ref{fig:iqmismatch_scan} shows the average magnitude and phase of the received CSI.

Fig.~\ref{fig:iqmismatch_scan} shows that the I/Q mismatch created in-band CSI disturbance and an overall translation of the magnitude; however, both the M-shaped magnitude distortion and the horizontal S-shaped phase distortion remained unchanged. This test shows that I/Q mismatch may \textit{not} be associated with CSI distortion.

\begin{table*}[t]
	\centering
    \caption{Types of CSI distortion}
	\renewcommand{\arraystretch}{1.1} 
	\begin{tabular}{|c|c|c|c|} \hline
		Type No. & Visual Form & Symmetry & Trigger Condition \\ \hline
		Type-I & M-shaped magnitude, horizontal S-shaped phase & Close to symmetrical & BW $>$ 20 MHz, both Tx/Rx in the HT20 or HT40 mode\\ \hline
		Type-II & Inverted V-shaped magnitude, straight-line phase & Close to symmetrical & BW $<$ 20 MHz, both Tx/Rx in the HT20 or HT40 mode\\ \hline
		Type-III & Left or right half of Type-I & Asymmetrical & One Tx/Rx in the HT20 mode, the other in the HT40+/- mode\\ \hline
    \end{tabular}
    \label{tab:distortion_form}
\end{table*}

\begin{figure}[t]
	\begin{center}
		\begin{tabular}{cc}
			\hspace{-0.13in}
			\includegraphics[width=0.5\columnwidth]{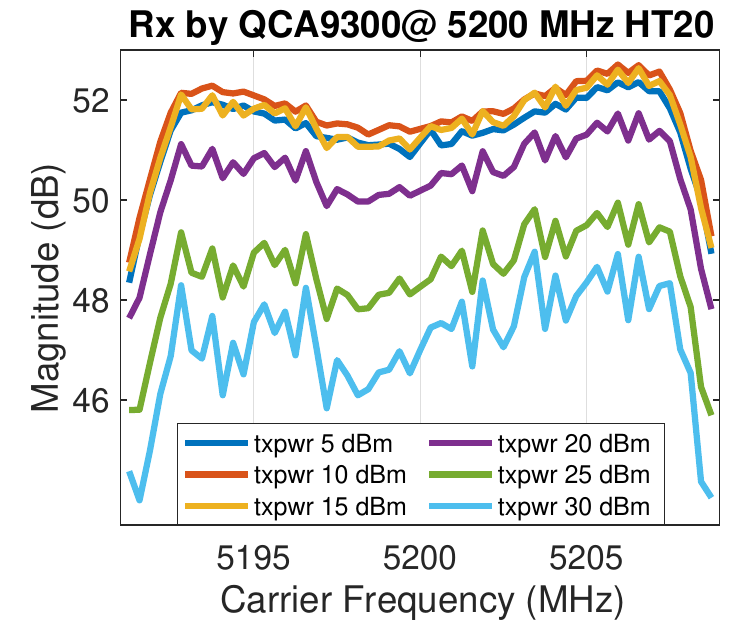} &
			\hspace{-0.2in}
			\includegraphics[width=0.5\columnwidth]{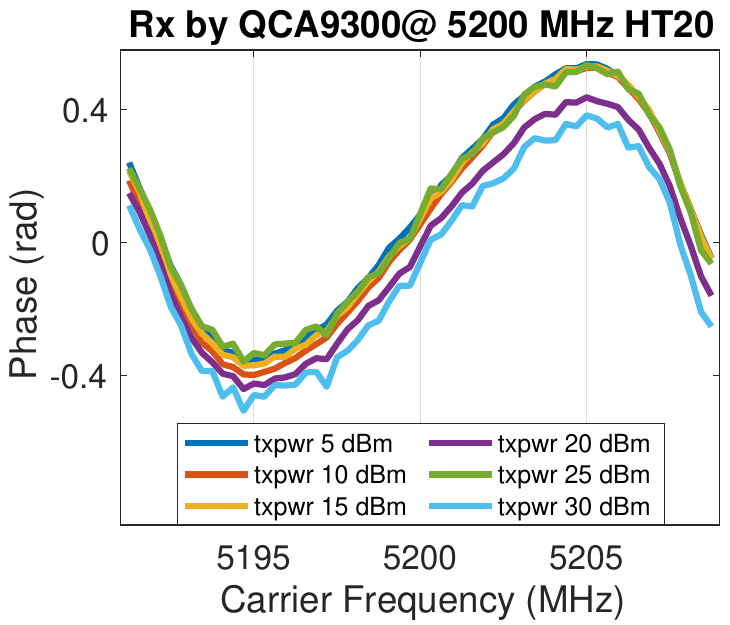}
		\end{tabular}
	\end{center}
	\caption{CSI distortion (magnitude and phase response) under different Tx power levels. Both the Tx and Rx ends were QCA9300 NICs operating at 5200 MHz in the HT20 channel mode. On the Tx end, the transmission power was scanned from 1 to 30 dBm.}
	\label{fig:txpower_scan}                
\end{figure}

\subsubsection{T4, test of the influence of the Tx power}
Similar to the previous test, we used PicoScenes to scan the transmission power of the Tx end. Fig.~\ref{fig:txpower_scan} shows the average magnitude and phase of the received CSI.

It is clear from Fig.~\ref{fig:txpower_scan} that the Tx power is also \textit{not} associated with the CSI distortion.
A strange phenomenon is that as the Tx power increases, the CSI magnitude drops. We believe this is caused by the Rx AGC, which suppresses the signal to a lower level to prevent ADC saturation.

\subsubsection{T5, test of the influence of the bandwidth}
\label{ssec:bandwidth_distortion}
\begin{figure}[t]
	\begin{center}
		\begin{tabular}{cc}
			\hspace{-0.13in}
			\includegraphics[width=0.5\columnwidth]{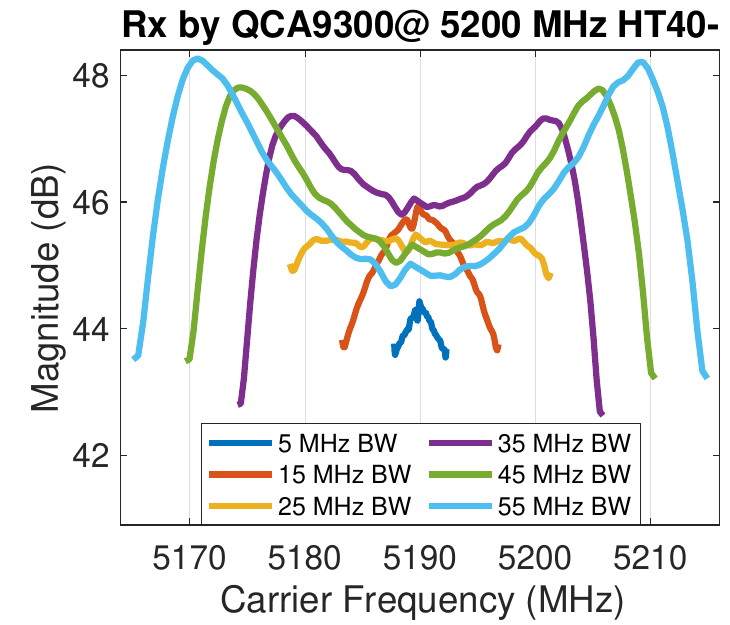} &
			\hspace{-0.2in}
			\includegraphics[width=0.5\columnwidth]{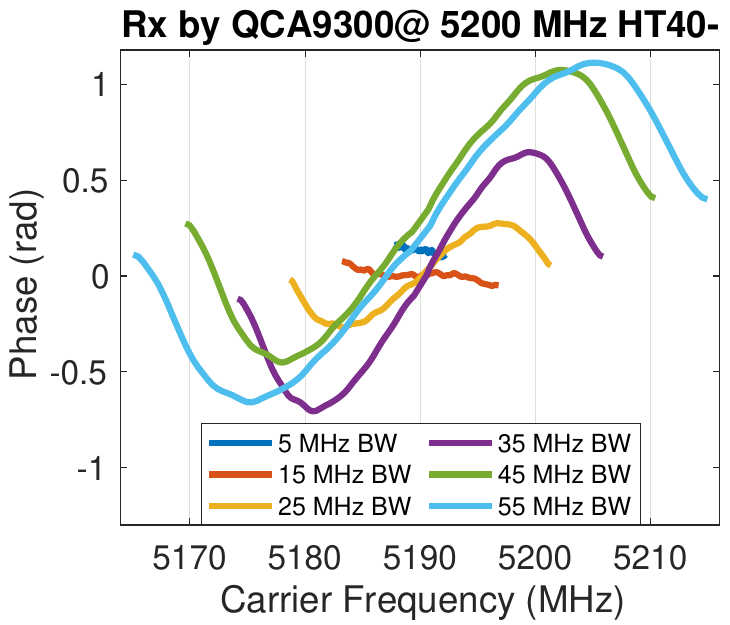} \\
		\end{tabular}
	\end{center}
	\caption{CSI distortion (magnitude and phase response) under different bandwidths. Both Tx and Rx ends were QCA9300. They operated at 5200 MHz with HT20 channel mode and varying bandwidths.}
	\label{fig:rate_scan}               
\end{figure}

In this test, we used PicoScenes to scan the bandwidth of both QCA9300 NICs from 5 to 55 MHz with a 5 MHz spacing and, at each bandwidth, performed 1000 CSI measurements.
To the best our knowledge, this is \textit{the first measurement of CSI at non-standard bandwidths on COTS NICs}.
Fig.~\ref{fig:rate_scan} shows the average magnitude and phase of the received CSI.

The results are remarkable: \textit{the distortion clearly shows a bandwidth-related shape change}. Specifically,
\begin{itemize}[leftmargin=*]
\item as the bandwidth increases, we observe an increase in the curvature of both the M-shaped magnitude distortion and the horizontal S-shaped phase distortion, and
\item as the bandwidth decreases, the magnitude response deforms from an M shape to an inverted V shape, and the phase response deforms from a horizontal S shape to an approximately straight line.
\end{itemize}

\vspace{0.1in}
\subsubsection*{Summary of the test results}
Based on the above analysis, we draw the following conclusions:
\begin{enumerate}[leftmargin=*]
\item We observed three types of distortion as listed in Table~\ref{tab:distortion_form}.
\item The bandwidth is the dominant influencing factor.
\item \textit{Both} the Tx and Rx ends contribute to the distortion, and the Rx end contributes more.
\item Distortion is \textit{pervasive} in all hardware and configurations.
\item The IWL5300 has stronger distortion than the QCA9300.
\item Tx power and I/Q imbalance have \textit{no} impact on distortion.
\end{enumerate}

\subsection{Reasonable conjectures about the cause of the distortion}

Based on some widely accepted RF design principles, we make three conjectures about the causes of the CSI distortion.

The Type-I distortion is probably caused by the combined effect of the DPD and the Rx ACR filter. The DPD \textit{intentionally} bends the signals with an \textit{inverse sinc} response to compensate for the \textit{sinc} fading caused by the Tx DAC. The ACR filter imposes strong filtering at the spectrum edges to prevent possible interference from the adjacent channels.
More specifically, the overcompensation caused by DPD maps to the central part of the M-shaped magnitude distortion and the central straight-line part of the horizontal S-shaped phase distortion. 
The  Rx-end ACR filter corresponds to the rapid fading at both ends of the spectrum.
Regarding the QCA9300, the ACR filter is a fifth-order LPF, as previously discussed.

The Type-II distortion may be caused by the combined effect of the Tx DAC fading and weak DPD.
Type-II distortion occurs only at a low bandwidth. In this case, the DAC \textit{sinc} fading shrinks to a narrower spectrum around the direct current (DC). However, the DPD, which is not associated with the bandwidth, maintains its weak response around the DC. This frequency mismatch causes the inverted V-shaped response.

The Type-III distortion may be the left or right half of the Type-I distortion.
As mentioned above, when the HT40+/- mode channel communicates with the HT20 mode channel, only half of the 40 MHz bandwidth is utilized. Therefore, the distortion corresponds to the left or right half of the HT40 mode Type-I distortion. When an HT20 channel communicates with an HT40+ channel, the CSI is the left half of the M shape; when communicating with an HT40- channel, the CSI is the right half of the M shape.

\subsection{Revised CSI model for Wi-Fi sensing}
\label{ssec:distortion_explanation}

According to tests T1, T2 and T5, both the Tx and Rx basebands introduce distortion in the CSI. We use $\mathbf{H}_{tx}$ and $\mathbf{H}_{rx}$ to denote their influences, respectively.
Substituting both terms into (\ref{eq:csi_basic_form}),
we obtain the revised CSI model:
\begin{equation}
	\mathbf{H}_{bb} = \mathbf{H}_{err}\cdot\mathbf{H}_{rx}\cdot \mathbf{H}_{air}\cdot\mathbf{H}_{tx}
	\label{eq:new_csi_form}
\end{equation}
(\ref{eq:new_csi_form}) indicates that we can isolate $\mathbf{H}_{air}$ from the raw CSI measurement if we can measure the baseband distortion.
However, it is difficult to measure $\mathbf{H}_{tx}$ and $\mathbf{H}_{rx}$ independently.
Fortunately, we can merge these two distortions\footnote{This is because $\mathbf{H}_{tx}$ and $\mathbf{H}_{rx}$ are both diagonal matrices, which allows the commutative law to be used on (\ref{eq:new_csi_form}).} and transform (\ref{eq:new_csi_form}) into a simpler version, as follows:
\begin{align}
	\mathbf{H}_{bb} = & \mathbf{H}_{dist}\cdot\mathbf{H}_{err}\cdot \mathbf{H}_{air} \nonumber \\
	& \mathbf{H}_{dist} =  \mathbf{H}_{tx}\cdot \mathbf{H}_{rx}
	\label{eq:new_csi_form_practical}
\end{align}
(\ref{eq:new_csi_form_practical}) is more practical in real-world measurement because the combined distortion, \ie, $\mathbf{H}_{dist}$, can be easily measured by connecting the Tx and Rx via a coaxial cable or placing them under the strong line-of-sight (LoS) condition.




\subsection{Why does the distortion contaminate Wi-Fi sensing but not Wi-Fi communication?}
For OFDM communication, what matters is not the specific processes or filters applied to the signals but rather that all signals undergo the same process. Taking the 802.11n protocol as an example, the HT-Data and HT-LTF symbols experience the same channel influences, including the baseband distortion. Therefore, all the channel influences imposed on the HT-Data symbols can be canceled by the channel estimation, measured by the HT-LTF symbols.
 In the Wi-Fi sensing context, unfortunately, the baseband distortion can be fatal. The distortion contaminates the CSI in a frequency-selective manner.  For model-based Wi-Fi sensing technologies, the distortion causes a phantom object that interferes with the $H_{air}$ measurement.
 
 In other words, \textbf{``channel'' is interpreted differently in the contexts of OFDM communication and Wi-Fi sensing research}. In the former context, ``channel'' refers to the combined effect of all kinds of influences, including the baseband distortion. In the latter context, the popular interpretation of ``channel'' refers to only one specific stage of the combined effect, \ie, the in-air signal propagation. However, we believe this misalignment in the interpretation is detrimental to Wi-Fi sensing research as a whole, as it ignores the substantial influences of the Wi-Fi baseband.

 \begin{figure}[t]
	\begin{center}
		\begin{tabular}{cc}
			\hspace{-0.13in}
			\includegraphics[width=0.5\columnwidth]{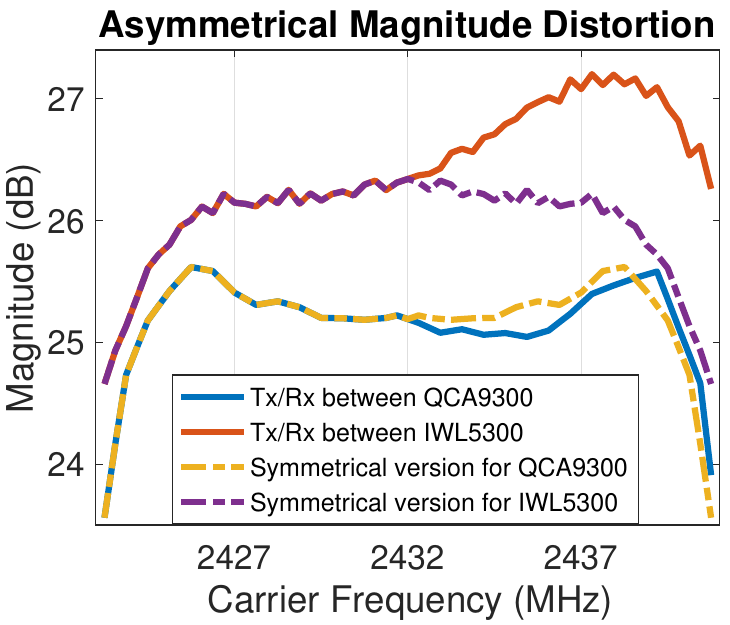} &
			\hspace{-0.2in}
			\includegraphics[width=0.5\columnwidth]{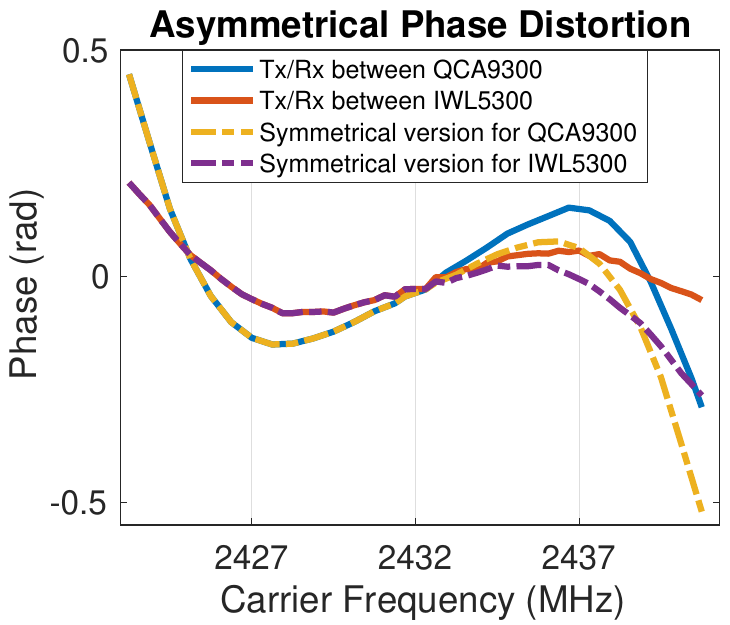} \\
		\end{tabular}
	\end{center}
	\caption{Asymmetry property of the CSI distortion. The dashed lines on the right, which mirror the left parts of the curves, show significant differences from the actual measurements.}
	\label{fig:asymmetry}                
\end{figure}

\begin{figure}
	\begin{center}
		\begin{tabular}{cc}
			\hspace{-0.13in}
			\includegraphics[width=0.5\columnwidth]{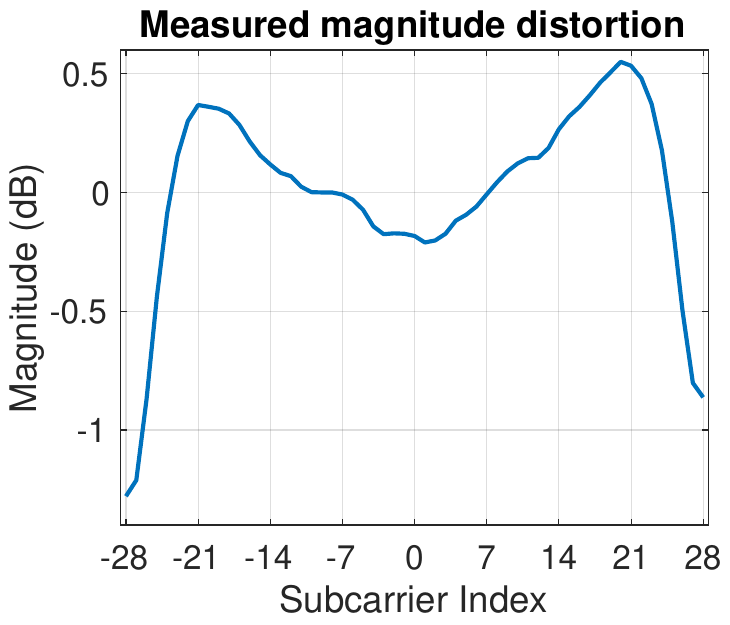} &
			\hspace{-0.2in}
			\includegraphics[width=0.5\columnwidth]{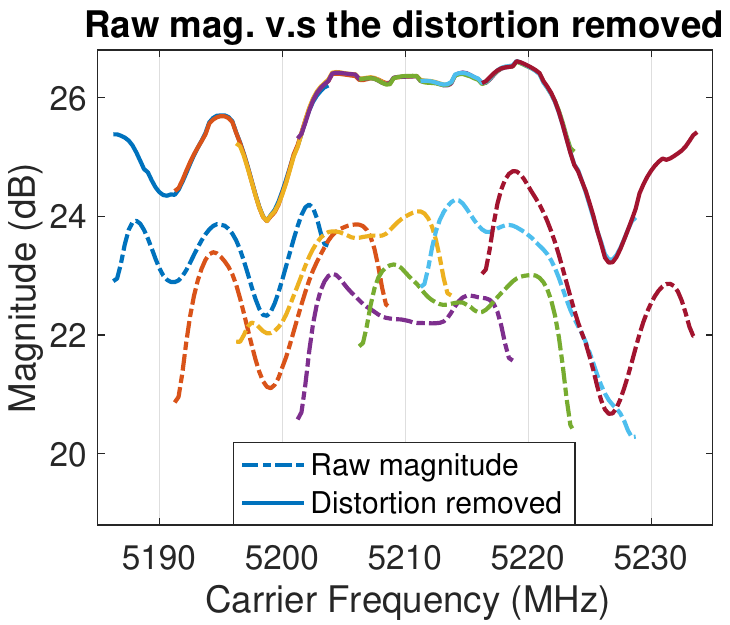} \\
			\hspace{-0.13in}
			\includegraphics[width=0.5\columnwidth]{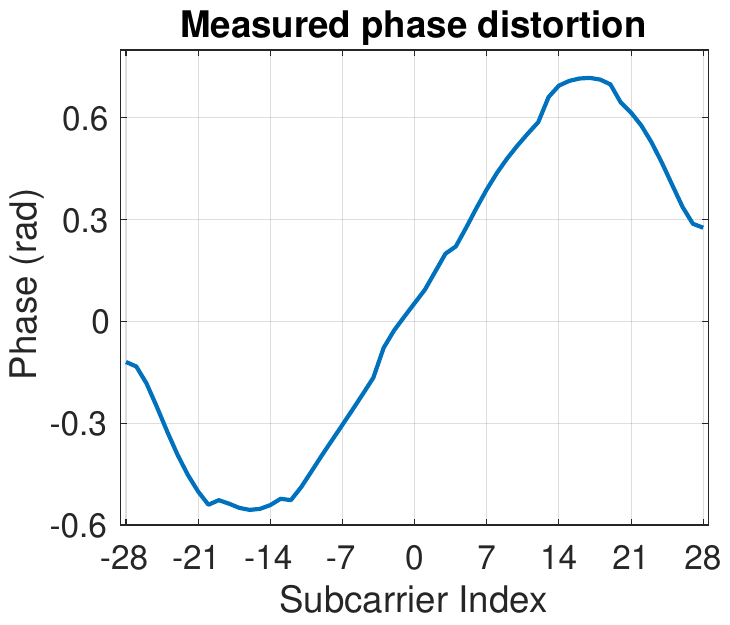} &
			\hspace{-0.2in}
			\includegraphics[width=0.5\columnwidth]{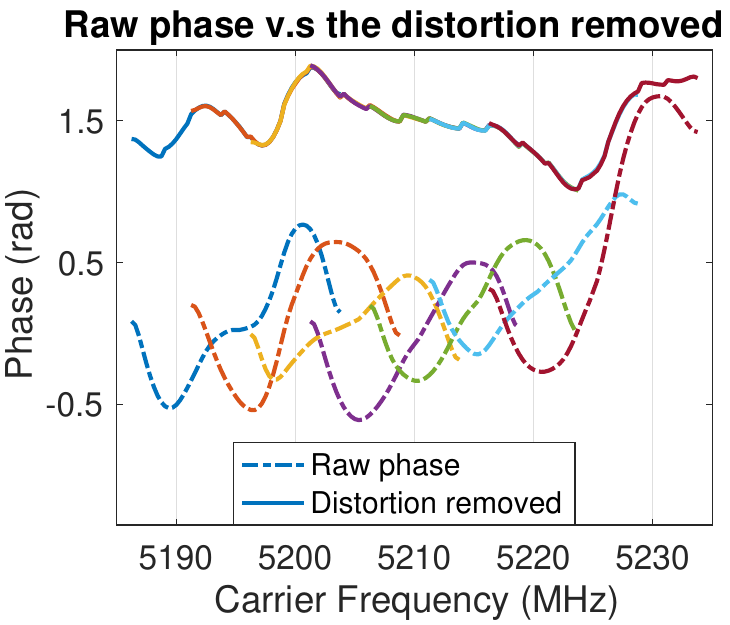} \\
		\end{tabular}
	\end{center}
	\caption{Measured CSI distortion and the results of distortion removal. The left figures show the magnitude and phase distortion measured in a strong LoS situation. The figures on the right show both the magnitude and phase measurements before and after distortion removal. The severely distorted measurements become aligned and smooth after distortion removal.}
	\label{fig:stitching}                
\end{figure}

\begin{figure*}[t]
	\centering
	\includegraphics[width=\textwidth]{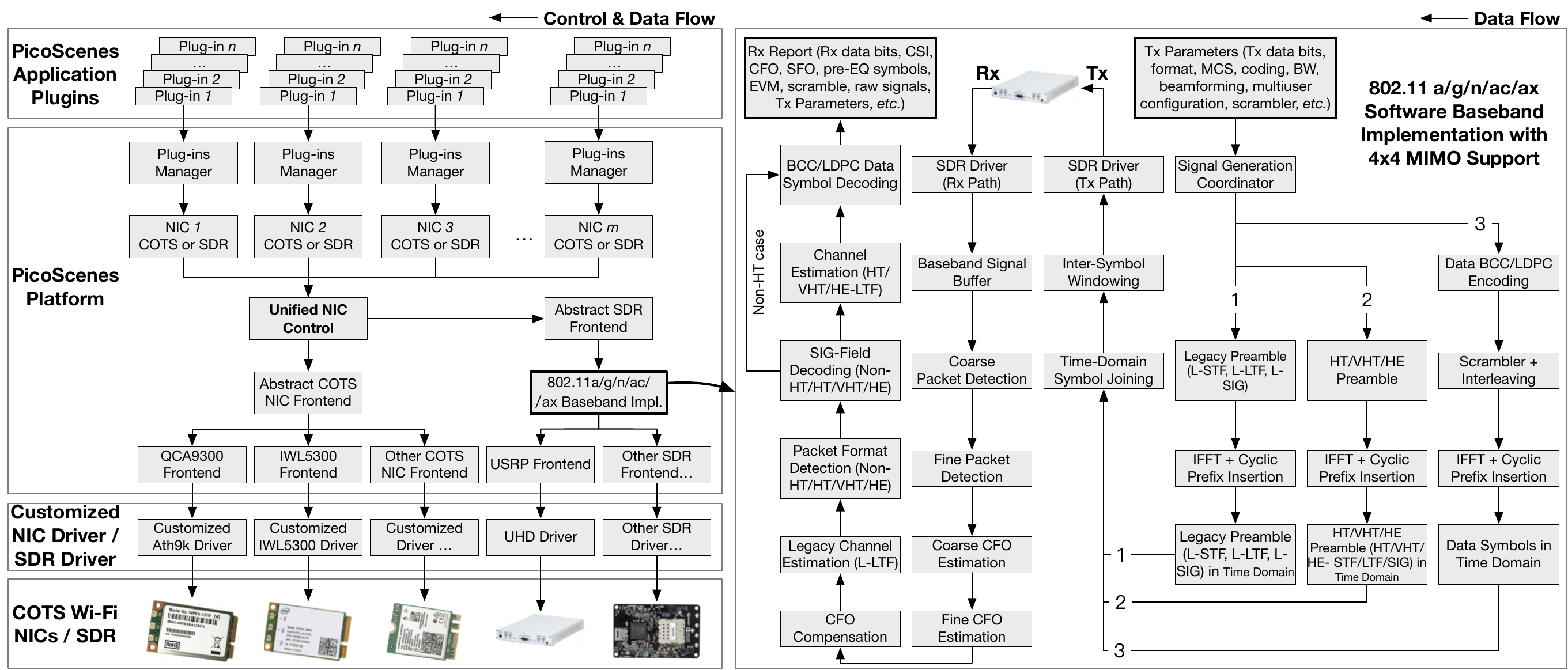} 
	\caption{Software architecture of the PicoScenes system. PicoScenes consists of 3 layers, namely, the PicoScenes drivers, the PicoScenes platform and the PicoScenes plugin subsystem. The drivers provide access to various hardware features. The platform abstracts the features into powerful APIs. The plugins perform the core measurement tasks. To support SDR-based Wi-Fi sensing, PicoScenes includes an embedded software baseband implementation, as shown on the right.}
	\label{fig:picoscenes_architecture}
  \end{figure*}
  
\subsection{How can the distortion be eliminated?}

Previous works have proposed methods~\cite{Xie:2019ji,Zhu:2018fca, Tadayon:2019cj} to eliminate the CSI distortion; however, they ignore magnitude distortion and focus only on phase distortion, particularly only Type-I phase distortion.
Xie~\etal~\cite{Xie:2019ji} prunes the severely distorted subcarriers at both ends.
Zhu~\etal~\cite{Zhu:2018fca} and Tadayon~\etal~\cite{Tadayon:2019cj} both attempt to curve-fit the Type-I phase distortion using a centrally symmetric function. However, this method may lead to additional errors because, shown in Fig.~\ref{fig:asymmetry}, the Type-I phase distortion is \textit{not} precisely symmetrical.

To eliminate the three types of magnitude/phase distortion, 
we generalize the approaches of Zhu~\etal~\cite{Zhu:2018fca} and Tadayon~\etal~\cite{Tadayon:2019cj}.
Similar to the previous works, our distortion removal method has two steps: predeployment distortion measurement and in situ distortion removal.
In the first step, 
users measure the CSI distortion between the Tx and Rx. This measurement can be done by connecting the Tx and Rx with a coaxial cable or placing them under a strong LoS condition. For each Tx/Rx link, the average magnitude and phase response are measured and stored as the distortion profile.
In the second step, the stored distortion profile is subtracted from the new CSI measurements. In this way, we remove the distortion.

The next problem is how to verify the correctness of the distortion removal.
Zhu~\etal~\cite{Zhu:2018fca} stitches the adjacent CSI measurements based on an idea that \textit{the CSI measurements from adjacent and partially overlapping channels should have identical CSI measurements for the same subcarriers}.
We adopt the same approach to verify the correctness of the distortion removal.
Fig.~\ref{fig:stitching} shows an example of distortion removal.
In the figure, the dashed lines denote the raw CSI measurements. These lines cannot be stitched, as they have different measurements for the overlapping parts. The solid lines, representing the distortion-removed measurements, are perfectly stitched with the adjacent measurements. The overlapping parts are perfectly aligned, indicating that the distortion-removed results contain only the in-air CSI portion.


\section{PicoScenes Platform} 
\label{sec:picoscenes}
In this section, we introduce our Wi-Fi sensing platform software, PicoScenes.
We herein present the architecture of PicoScenes and several design highlights.

\subsection{PicoScenes architecture}
PicoScenes consists of 3 layers from bottom to top, as shown on the left side of Fig.~\ref{fig:picoscenes_architecture}: PicoScenes drivers, PicoScenes platform and PicoScenes plugin subsystem.
                  
\subsubsection*{PicoScenes drivers} We created our modified kernel drivers for both the QCA9300 and IWL5300. These drivers extract the CSI and expose various hardware controls.
We improved the original drivers in three aspects:
First, we added the multi-NIC concurrent CSI measurement functionality by refactoring the CSI data collection onto a per-device data structure, which allows the PicoScenes software to access the CSI from multiple NICs concurrently.
Second, we provided a unified data format across the QCA9300, IWL5300 and SDR devices. All of the data components, the Rx descriptor, the CSI and the packet content are all self-descriptive and versioned \textit{segments}. All segments can be decoded only by version-matched parsers. In this way, we achieve forward compatibility for future upgrades.
Third, we simplified the installation of the PicoScenes drivers. Instead of building an old kernel, the PicoScenes drivers are released directly in a prebuilt Debian .deb package. Users can install it by double-clicking.
                    
\subsubsection*{PicoScenes platform} The PicoScenes platform is essentially the middleware for Wi-Fi sensing research. In addition to the CSI data collection functionality, the platform integrates the packet-injection-based Tx control and low-level hardware controls. It abstracts all types of frontends and presents unified and powerful APIs to the upper plugin layer. The platform is also extensible; new CSI-available hardware can be easily integrated by adding a frontend description class.

The PicoScenes platform is itself a layered architecture.
At the bottom is the frontend.
For each supported NIC and SDR device,
we have a Frontend class that encapsulates all supported controls.
Benefiting from our detailed study of the QCA9300 hardware,
PicoScenes is advantageous for the QCA9300 NICs.
In addition to the aforementioned arbitrary tuning of the carrier frequency and bandwidth,
PicoScenes enables other valuable features for the QCA9300,
such as selecting the Tx/Rx radio chain and transmitting HT-rate packets with extra spatial sounding (ESS), \etc.
On top of the frontend layer, an abstraction layer exports the unified APIs to the upper-level plugins.
To support the SDR-based frontend, we develop and embed a high-performance Wi-Fi baseband implementation, which can drive the SDR hardware to transmit and receive packets similar to a full-featured Wi-Fi NIC.
We call this mode PicoScenes on SDR, which we describe later in this section.
Above the abstraction layer, each NIC abstraction has a plugin manager that finds, installs and controls the PicoScenes plugins.

\begin{table*}[t]
  \newcommand{\tabincell}[2]{\begin{tabular}{@{}#1@{}}#2\end{tabular}}
  \caption{An incomplete list of features for comparison between PicoScenes and existing CSI extraction tools}
  \centering
  \renewcommand{\arraystretch}{1.1} 
  \begin{tabular}{|l|l|l|l|l|}\hline
    \multirow{3}*{Feature}&\multicolumn{4}{c|}{CSI Measurement/Extraction Tools}\\\cline{2-5}
    &\multirow{2}*{PicoScenes} &Intel 5300 & Atheros & Nexmon CSI \\ 
    && CSI Tool~\cite{Halperin_csitool} & CSI Tool~\cite{Xie:2019ji} & Extractor~\cite{Nexmon} \\\hline
    Supported frontend &IWL5300, QCA9300 and all USRP &IWL5300& QCA9300 & BCM43xx \\ \hline
    $>$10 kHz packet injection \& CSI measurement &  \multicolumn{1}{l|}{$\checkmark$ for all} & \multicolumn{3}{c|}{\ding{53}} \\\hline
    Concurrent multi-NIC CSI extraction & $\checkmark$ for all& \multicolumn{3}{c|}{\ding{53}}\\\hline
    In situ CSI parsing \& processing & $\checkmark$ for all& \multicolumn{3}{c|}{\ding{53}}\\\hline
    CSI measurement for the QCA9300$\rightarrow$IWL5300 & $\checkmark$ for all &\multicolumn{3}{c|}{\ding{53}}  \\\hline
    Turning on/off the selected radio chain(s) & $\checkmark$ for all & \multicolumn{3}{c|}{\ding{53}} \\\hline
    Unified, open and self-descriptive CSI data format & $\checkmark$ for all & \multicolumn{3}{c|}{\ding{53}}\\\hline
    Support for secondary development &$\checkmark$ for all & \multicolumn{3}{c|}{\ding{53}}  \\\hline
    Arbitrary bandwidth and carrier frequency tuning & $\checkmark$ for the QCA9300 and SDR & \multicolumn{3}{c|}{\ding{53}} \\\hline
    Transmission of extra sounding HT-LTF & $\checkmark$ for the QCA9300 and SDR & \multicolumn{3}{c|}{\ding{53}} \\\hline
    Turn off AGC \& manual Rx gain control & $\checkmark$ for the QCA9300 and SDR & \multicolumn{3}{c|}{\ding{53}} \\\hline
    Access to Rx EVM & $\checkmark$ for the QCA9300 and SDR & \multicolumn{3}{c|}{\ding{53}} \\\hline
    CSI measurement for all MAC addresses & $\checkmark$ for SDR &\multicolumn{3}{c|}{\ding{53}} \\\hline
    VHT/HE-rate packet injection & $\checkmark$ for SDR & \multicolumn{3}{c|}{\ding{53}} \\\hline
    Tx/Rx with 4$\times$4 MU-MIMO/OFDMA/beamforming & $\checkmark$ for SDR & \multicolumn{3}{c|}{\ding{53}} \\\hline
    Reporting of CSI for 11ax packets & $\checkmark$ for SDR & \multicolumn{3}{c|}{\ding{53}} \\\hline
    Easy installation without a kernel build & $\checkmark$ ``\textsf{apt install}'' + auto-update & \multicolumn{3}{c|}{\ding{53}} \\ \hline
    Round-trip measurement with channel scan &$\checkmark$ (by EchoProbe plugin) & \multicolumn{3}{c|}{\ding{53}}  \\\hline
    Support for latest kernel versions & $\checkmark$ currently on v5.4 LTS & \multicolumn{2}{c|}{\ding{53}} & $\checkmark$ v5.4 LTS \\\hline
    Works on latest host OSs & $\checkmark$ Ubuntu 20.04 LTS & \multicolumn{2}{c|}{\ding{53}} & $\checkmark$ 18.04 LTS\\\hline
  \end{tabular}
  \label{tab:feature_comparison}
\end{table*}

The whole platform, written in C++20, embraces a multithread design from the ground up. Performance-sensitive tasks, such as the frontend I/O, baseband decoding/encoding for SDR, per-NIC jobs, and plugin instances, all run in separate threads. In this way, a multi-NIC CSI measurement can be performed concurrently without congestion.

\subsubsection*{PicoScenes plugin subsystem} This subsystem performs application- and measurement-specific tasks. The plugins invoke the hardware-independent APIs presented by the platform to implement various Wi-Fi sensing or communication tasks. We made the PicoScenes plugin development kit (PSPDK) open-source, enabling users to develop their own measurement plugins.
As a demonstration of the PSPDK, we develop EchoProbe, a PSPDK-based plugin that can orchestrate two PicoScenes nodes performing a round-trip CSI measurement over a large spectrum.
Two roles are defined by EchoProbe: an initiator and a responder.
The initiator injects the \textsf{CSIProbeRequest} frames. The responder receives the frame, packages the measured CSI as the reply payload and transmits this payload back to the initiator. In this way, EchoProbe performs the round-trip CSI measurement. Furthermore, by specifying $f_c$ and $f_{bb}$ in the \textsf{CSIProbeRequest} frames, the initiator and responder can perform synchronized frequency and bandwidth hopping. In this way, EchoProbe achieves round-trip measurement over a large spectrum.

\subsubsection*{PicoScenes CLI}
PicoScenes provides a powerful and user-friendly command-line interface (CLI). For example, the following commands direct two NICs, \ie, \#1 on laptop A and \#2 on laptop B, to scan both the carrier frequency and bandwidth:
\begin{description}
\item \textsf{PicoScenes -i 1 --mode responder} (run on laptop A)
\item \textsf{PicoScenes -i 2 --mode initiator --cf 2.3e9:5e6:2.4e9 --sf 20e6:5e6:60e6 --repeat 200 --delay 1e3 --mcs 2 --ess 1 --txcm 4 --rxcm 7} (run on laptop B)
\end{description}
where NIC \#1, on laptop A, works in the EchoProbe \textit{responder} mode (\textsf{-i 1 --mode responder}), whereas NIC \#2, on laptop B, is the round-trip measurement \textit{initiator} (\textsf{-i 2 --mode initiator}). The initiator scans both the carrier frequency from 2300 to 2400 MHz in 5 MHz increments (\textsf{--cf 2.3e9:5e6:2.4e9}) and the bandwidth from 20 to 60 MHz in 5 MHz increments (\textsf{--sf 20e6:5e6:60e6}). For each rate combination, NIC \#2 performs 200 round-trip measurements at 1000 $\mu$s intervals (\textsf{--repeat 200 --delay 1e3}). Each packet is transmitted with MCS index 2 and with 1 ESS HT-LTF (\textsf{--mcs 2 --ess 1}). Finally, the command further specifies the Tx/Rx radio chains of NIC \#1: the Tx uses the third radio chain, and the Rx uses all three radio chains (\textsf{--txcm 4 --rxcm 7}).

\vspace{0.1in}
Table~\ref{tab:feature_comparison} lists the major advantages of the PicoScenes system over existing CSI tools.

\begin{table}[t!]
  \centering
  \caption{Extra PHY-layer information provided by PicoScenes on SDR}
  \renewcommand{\arraystretch}{1.1} 
  \begin{tabular}{|l|c|} \hline
     Feature Name & Data Type \\ \hline
     HT/VHT-SU/HE-SU CSI & complex double array \\ \hline
     VHT-MU/HE-MU per-user CSI & complex double array \\ \hline
     Legacy CSI (two sets of L-LTF CSI) & complex double array \\ \hline
     Pilot subcarrier CSI & complex double array \\ \hline
     Raw baseband signal & complex double array \\ \hline
     Pre-equalized OFDM symbols & complex double array \\ \hline
     CFO estimation (based on L-LTF) & double \\ \hline
     SFO estimation (based on pilot subcarriers) & double \\ \hline
     Rx EVM & double \\ \hline
     Noise floor & double \\ \hline
     Timestamp (by hardware baseband clock) & double \\ \hline
     Scrambler initial value  & 8-bit integer \\ \hline
  \end{tabular}
  \label{tab:extra_phy}
\end{table}

\subsection{PicoScenes on SDR}
The broad adoption of SDR in Wi-Fi sensing is severely hampered by a lack of baseband signal processing functionality.
As illustrated on the right side of Fig.~\ref{fig:picoscenes_architecture},
we address this issue by developing a high-performance 802.11a/g/n/ac/ax baseband implementation. This implementation is embedded into the PicoScenes platform, thereby transparently empowering SDR devices to function as full-featured Wi-Fi NICs.
We call this feature PicoScenes on SDR\footnote{PicoScenes on SDR contains proprietary code provided by an upstream vendor under an NDA; therefore, we do not discuss its design details here.}, and it currently supports all USRP models.
With PicoScenes on SDR, the adoption of SDR in Wi-Fi sensing is \textit{unprecedentedly simplified}. Taking the PicoScenes commands above as an example, replacing ``-i 2'' with ``-i usrp192.168.10.2'' is all that is required to switch to the USRP with the 192.168.10.2 IP address. The EchoProbe initiator performs the same measurement process, except that it returns much richer measurement results.

One of the most attractive aspects of PicoScenes on SDR is that it provides \textit{complete} control over the Wi-Fi Tx and Rx, which offers overwhelming advantages over COTS Wi-Fi NICs.
On the Tx side, the user can specify the initial scrambler value, which is crucial for Wi-Fi-based cross-technology communication (CTC); beamforming, which enables fine-grained sensing and calibration; and ESS, which enables a COTS NIC to measure additional CSI for a single 802.11n frame.
On the Rx side, as listed in Table~\ref{tab:extra_phy}, PicoScenes returns the complete PHY-layer information. We believe that this unprecedented PHY-layer information can enable more diverse and more accurate sensing applications.
In addition, the unified interface for Wi-Fi communication (from the 802.11a to 802.11ax multiuser (MU)) and high spectrum and bandwidth accessibility simplify the prototyping and development of new Wi-Fi communication and sensing applications.

\subsection{Software release}
The PicoScenes software is available at \url{https://ps.zpj.io}. We provide rich documentation, including the installation guide, CLI reference and MATLAB toolbox. We also offer technical support via GitLab issue tracker or instant messaging App.

\section{Evaluation of the PicoScenes Platform} 
\label{sec:evaluation}

In this section, we report extensive evaluations of the functionality and performance of the PicoScenes system.
We divide the evaluations into two subsections based on the hardware used: PicoScenes on COTS Wi-Fi NICs and PicoScenes on SDR. Finally, we briefly summarize the evaluations.

\subsection{Evaluation of PicoScenes on COTS Wi-Fi NICs}

In this section, we evaluate three popular features of PicoScenes: spectrum and bandwidth tuning on the QCA9300, multi-NIC CSI measurement, and high-speed packet injection.

\subsubsection{Evaluation of the channel availability of the spectrum and bandwidth supported by the QCA9300}

Sections~\ref{ssec:sf_tuning} and~\ref{ssec:cf_tuning} describe how we achieve arbitrary tuning of both the baseband bandwidth $f_{bb}$ and the carrier frequency $f_{rf}$ for the QCA9300.
However, some questions remain: \textit{Are these frequencies and bandwidths practically feasible? How good is the link quality at these frequencies and bandwidths?}

\begin{figure*}[t]
	\begin{center}
		\begin{tabular}{ccc}
			\hspace{-0.1in}
            \includegraphics[width=0.66\columnwidth]{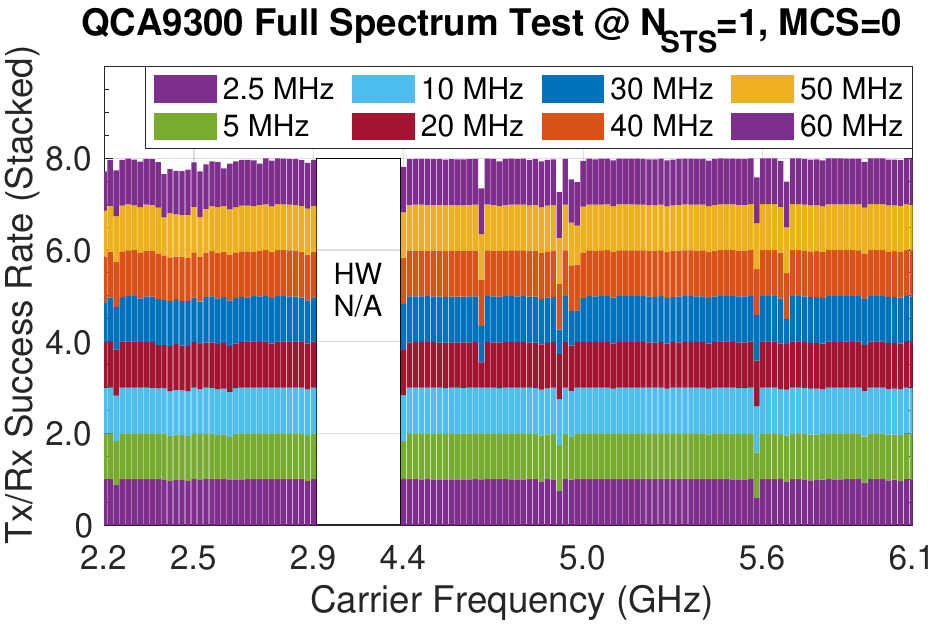} & \hspace{-0.15in}
            \includegraphics[width=0.66\columnwidth]{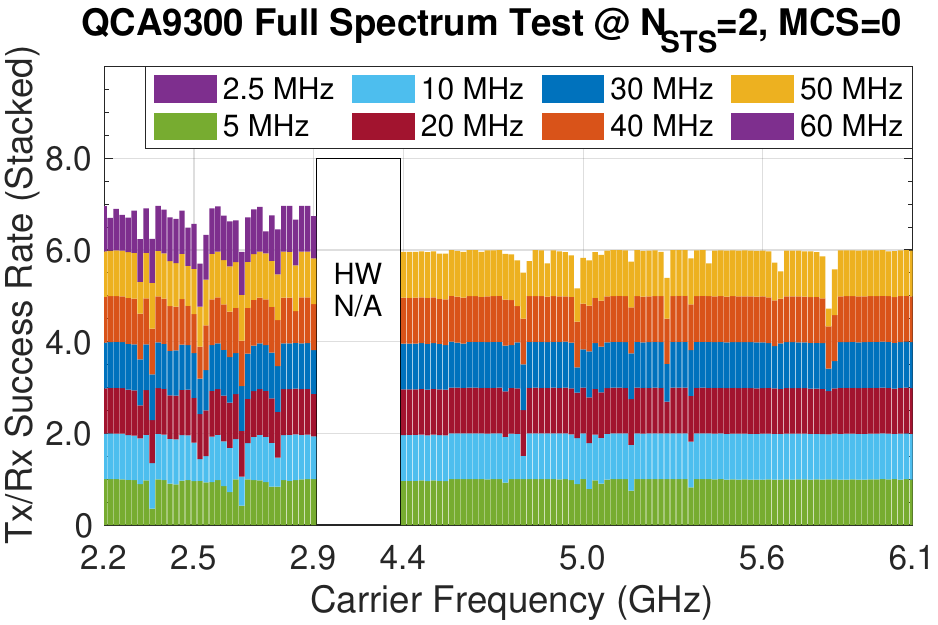} & \hspace{-0.15in}
            \includegraphics[width=0.66\columnwidth]{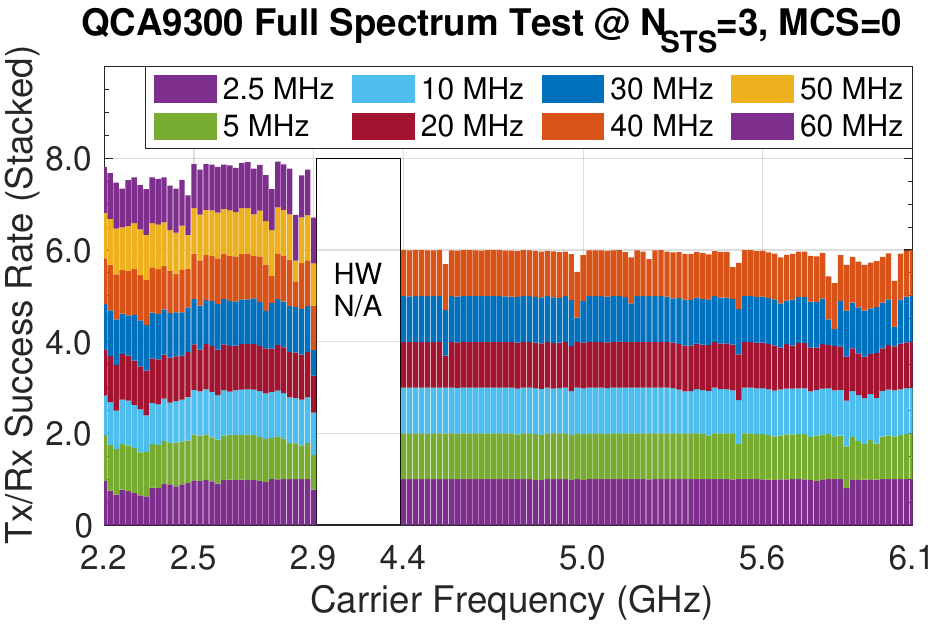} \\
            (A) & (B) & (C) \\
            \hspace{-0.1in}
            \includegraphics[width=0.66\columnwidth]{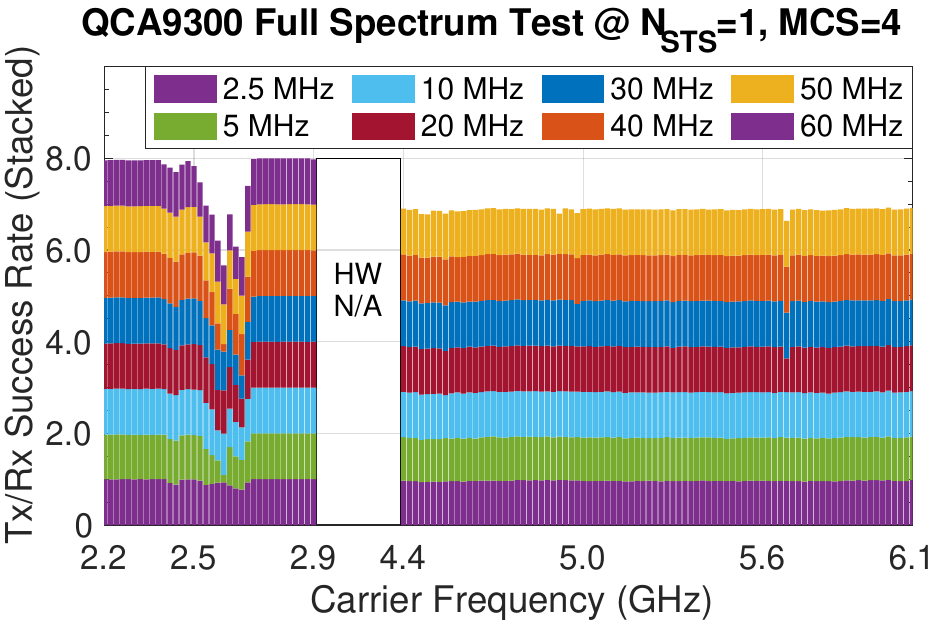} & \hspace{-0.15in}
            \includegraphics[width=0.66\columnwidth]{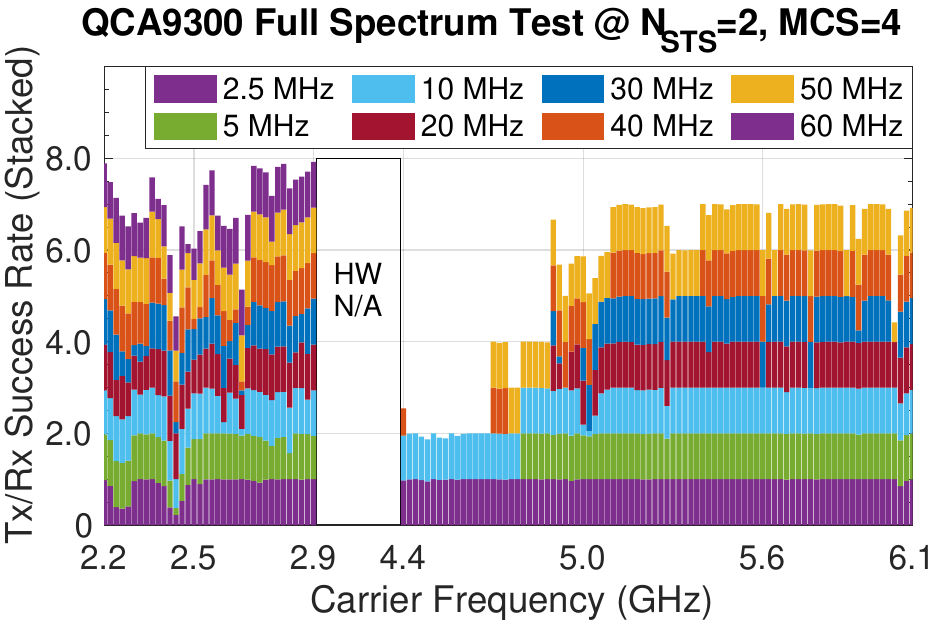} & \hspace{-0.15in}
            \includegraphics[width=0.66\columnwidth]{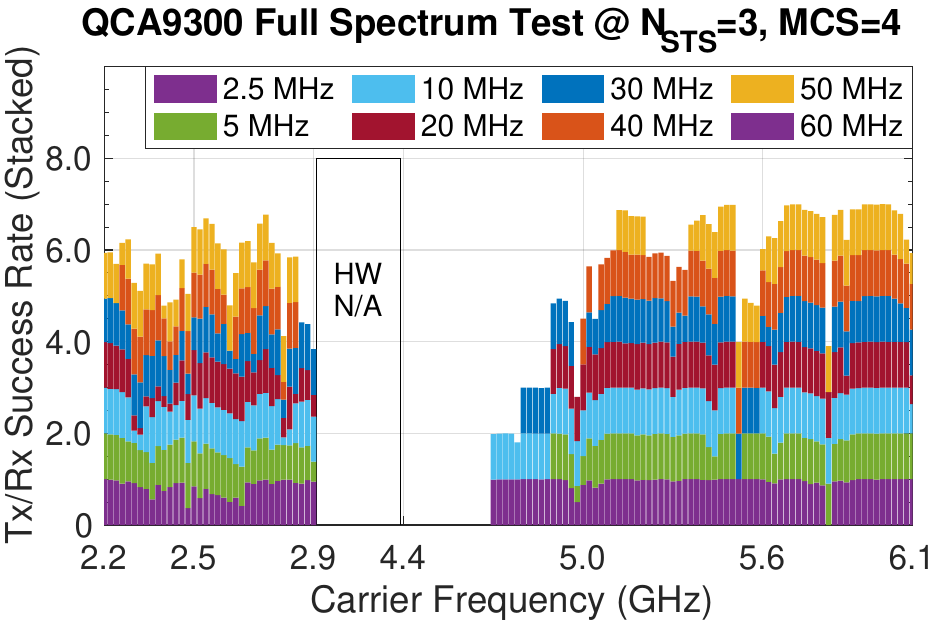} \\
            (D) & (E) & (F) \\
		\end{tabular}
	\end{center}
    \caption{Stacked Rx success rates in a full-spectrum and full-bandwidth link quality evaluation on the QCA9300. The evaluation covers 5856 configurations.}
	\label{fig:qca9300_full_spectrum}                
\end{figure*}

To answer these questions,
we conducted a comprehensive link quality evaluation that covers the full spectrum and bandwidth that are supported by the QCA9300 hardware.
In the evaluation, two ThinkPad X201 laptops were placed 3 m apart in a small room. Both laptops were equipped with the QCA9300 NICs and ran Linux Mint 20 (a variant of Ubuntu 20.04 LTS), kernel version 5.4.65.
We enumerated a total of 5856 different channel configurations (122 carrier frequencies, 8 bandwidths, 3 spatial-time stream ($N_{STS}$)\footnote{The definition of the MCS index is narrowed in 802.11ac/ax protocols, where $N_{STS}$ is decoupled from the MCS index. For consistency of description, we use the 802.11ac/ax-based definition of the MCS index throughout the evaluations.} values, and 2 MCS indices), as detailed in Listing~\ref{list:list_full_spectrum}.
In each configuration, we used PicoScenes (with the EchoProbe plugin) to perform 5000 round-trip CSI measurements between the two laptops and logged their success rates. We defined the success rate as the ratio between the target number of 5000 and the number of actual round-trip measurement attempts.
Fig.~\ref{fig:qca9300_full_spectrum} shows the stacked success rates of the evaluation.

\algrenewcommand\algorithmicindent{0.6em}%
\begin{algorithm}[t]
    \caption{QCA9300 full-spectrum evaluation procedure}
    \label{list:list_full_spectrum}
    \begin{algorithmic}[1]
        \State $F_C$= $\{2.2\sim 2.9, 4.4\sim 6.1\}$ GHz with 20 MHz intervals
        \State $F_{BB}=\{2.5, 5, 10, 20, 30, 40, 50, 60\}$ MHz, 30 MHz and below with HT20, 40 MHz and above with HT40+
        \State $N_{STS}=\{1, 2, 3\}$, $MCS=\{0, 4\}$
        \ForEach {$f_c \in F_C $}
            \ForEach {$f_{bb} \in F_{BB} $}
                \ForEach {$n_{sts} \in N_{STS} $}
                    \ForEach {$mcs \in MCS $}
                        \State Perform 5000 round-trip CSI measurements
                        \State and log the success rate.
                    \EndFor
                \EndFor
            \EndFor
        \EndFor
    \end{algorithmic}
\end{algorithm}

From Fig.~\ref{fig:qca9300_full_spectrum}, especially (A) to (C), we can see that
the QCA9300 exhibits a high level of link quality consistency across the entire 2.4 GHz wide spectrum (2.2-2.9 GHz and 4.4-6.1 GHz). In other words, we do not see differences between the standard Wi-Fi channels and the channels unlocked by PicoScenes, even if the latter lack a dedicated RF calibration.
From (B) to (E), we see that a bandwidth of 50 MHz and above seems to yield better performance in the 2.4 GHz band than in the 5 GHz band, possibly because the RF synthesizer provides a higher adjustment accuracy in the 2.4 GHz band.
(E) and (F) show a large decrease in link quality at approximately 4.4-4.9 GHz; compared to (B) and (C), their only difference is the increase in the MCS index from 0 to 4. This comparison indicates that the RF synthesizer seems to have higher error in the 4.4-4.9 GHz spectrum. To improve the performance in this short spectrum, we need to recalibrate the device; the related reverse engineering process is still under investigation.

From the bandwidth, the QCA9300 shows good link quality at bandwidths of 50 MHz and below.
The sub-20 MHz bandwidth presents superior link resilience in all test cases, including the MCS=4 cases, because the lower baseband bandwidth means higher clocking error tolerance.
If we focus on the 2.4 GHz band, the QCA9300 still performs quite well in the above-standard 30 to 60 MHz bandwidth cases.
Taking the MCS=0 cases as an example, the performance at a 60 MHz bandwidth, 1.5$\times$ the standard bandwidth, shows good link quality with 3 spatial streams.
The practical availability of higher bandwidths is quite useful in Wi-Fi sensing, as these bandwidths provide higher temporal resolution and a shorter transmission time.
We also tried to expand the evaluation to 70 MHz and above. Unfortunately, at these high bandwidths, the excessively high numbers of transmission failures led to early termination of the evaluation.

\begin{figure}[t] 
    \centering
    \includegraphics[width=1\columnwidth]{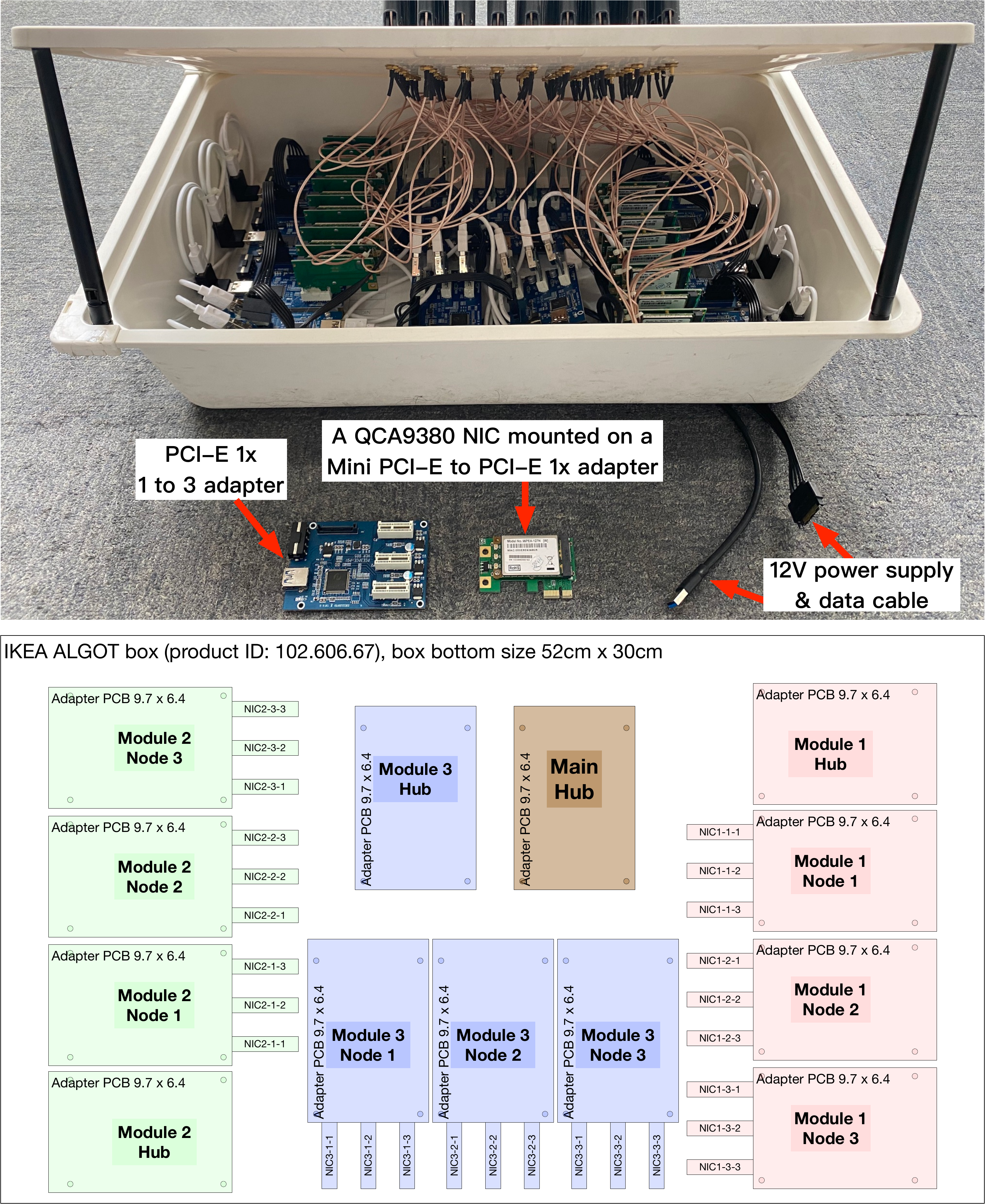}
    \caption{Photograph and design layout of a 27-NIC Wi-Fi sensing array. The 27-NIC array consists of 3 ``modules'' and a ``main hub''. Each module contains 4 PCI-E 1$\times$ 1-to-3 bridge adapters and 9 QCA9300 Wi-Fi NICs. The array is contained in an IKEA box.}
    \label{fig:nicarray}
\end{figure}

\subsubsection{Evaluation of concurrent CSI measurements for multiple COTS Wi-Fi NICs}
In this section, we present a reference design for a 27-NIC Wi-Fi sensing array. We then evaluate its concurrent CSI measurement performance.

Fig.~\ref{fig:nicarray} shows a photograph and the layout of the 27-NIC Wi-Fi sensing array. Its core architecture is a 3-layer PCI-E hierarchy whose branch nodes are PCI-E 1-to-3 bridge adapters and whose leaf nodes are the QCA9300 NICs.
For the data connections, PCI-E extension cords (actually USB 3.0 cables) are used to connect the bridge adapters to the host PC.
For the power supply, we use multiple power-splitting cords to distribute electricity to the bridge adapters and the NICs.
The array is well encapsulated in an IKEA box. The total bill of materials (BOM) for the array is less than 700 USD, \ie, merely 8.6 USD per radio chain. To the best of our knowledge, this is the largest and most cost-efficient Wi-Fi sensing array built on COTS NICs.

To evaluate the performance of the sensing array, we attached the 27-NIC array and another QCA9300 NIC to a desktop computer, \ie, a total of 28 NICs were connected. The additional QCA9300 NIC was used for packet injection. The computer was equipped with an i9-10850K CPU, 32 GB of RAM and 512 GB of SSD.
We evaluated the performance of concurrent CSI measurements for different numbers of Rx NICs ($N_{NIC}\in \{9,18,27\}$), bandwidths ($BW\in\{20,40\}$ MHz) and Tx injection speeds.
In each test case, we used PicoScenes to inject 100k packets and then log the Rx success rate (Rx rate hereafter).
We define the Rx rate as the ratio between the number of received frames and the total number of injected frames.
The injected packets were 32 B data frames. By modulating these frames with MCS$\geq$4, $N_{STS}=1$ and a 400 ns guard interval, each frame was encoded to only 864 samples, which is also the shortest 802.11n frame from which both the QCA9300 and IWL5300 can measure CSI. If transmitting at a 20 MHz bandwidth, the duration of this frame is 43.2~$\mu$s. The reason for transmitting this short frame was to maximize the injection rate; accordingly, we call this frame the ``CSI Probing Frame'' hereafter.

\begin{figure}[t] 
    \centering
    \includegraphics[width=1\columnwidth]{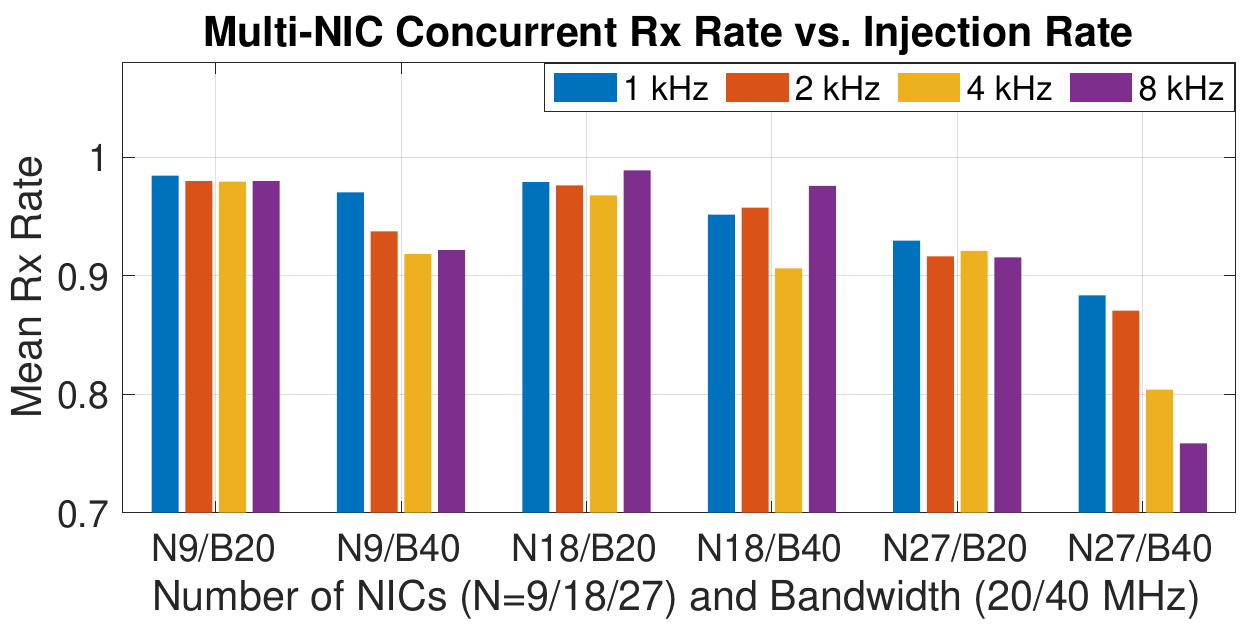}
    \caption{Rx success rates of multi-NIC concurrent CSI measurements given different numbers of NICs, bandwidths and Tx injection rates. The Y-axis represents the \textit{mean} Rx rate in multiple tests.}
    \label{fig:nicarray_performance}
\end{figure}

Fig.~\ref{fig:nicarray_performance} shows the results.
In the 20 MHz bandwidth cases, the array achieved a $>$97\% mean Rx rate when $N_{NIC}<=18$ and still had a $>$91\% Rx rate when $N_{NIC}=27$.
Moreover, in all 20 MHz bandwidth cases, an increase in the injection rate seemed to have a negligible impact on the Rx rate.
The two results above clearly show the high efficiency of the concurrent architecture of PicoScenes.

To explore the performance limits, we boosted the bandwidth from 20 to 40 MHz.
At a 40 MHz bandwidth, an approximately 7\% decline in the Rx rate was observed in the $N_{NIC}=9$ and $N_{NIC}=18$ cases.
This decline was due to the low tolerance to clocking error associated with the high bandwidth.
The expected performance drop occurred in the $N_{NIC}=27$ case.
In this case, the kernel driver seemed to become the performance bottleneck, as the QCA9300 is a soft-MAC NIC; consequently the reception of every single Wi-Fi frame required the host CPU to execute thousands to tens of thousands of instructions.
However, the \textsf{ath9k} kernel driver did not seem to be architecturally optimized for the multi-NIC communication scenario.
Therefore, the driver dispatches the massive number of kernel jobs to only a few specific CPU cores. This led to a greatly biased CPU load in the kernel space and, consequently, the decline in Rx performance.
In the 40 MHz bandwidth case, the doubled instantaneous traffic further intensified this problem.

Before ending this discussion of the multi-NIC evaluation, we answer two important questions.


\textit{Is this a phased array?} No. In the current setup, all 27 NICs are independent and therefore unsynchronized.
Phaser~\cite{Gjengset2014Phaser} proposed a method to synchronize COTS Wi-Fi NICs; however, it requires sacrificing 1/3 of the antennas to perform cross-NIC time synchronization.
We are working on an optimized solution, but it is in the preliminary stage and is beyond the scope of this paper.

\textit{Can we build an IWL5300-based multi-NIC sensing array?} Yes, of course. The kernel driver enhancement for concurrent CSI measurement is actually \textit{shared} between the QCA9300 and IWL5300 models. In addition to the driver, the PicoScenes platform even supports a concurrent CSI measurement from a heterogeneous array built with different hardware. 



\subsubsection{Evaluation of the maximum packet injection rate and CSI measurement speed}
Another promising feature of PicoScenes is its $>$10 kHz packet injection and CSI measurement.
We conducted a thorough evaluation to answer two simple questions:

\textit{How quickly can these devices inject packets?}

\textit{How quickly can these devices measure the CSI?}

In this evaluation, for each device and all its supported bandwidths, we used PicoScenes to inject 200k CSI Probing Frames as quickly as possible.
The evaluation encompassed the QCA9300, IWL5300, and SDR devices.
SDR is a special case for which we provide two working modes: a real-time mode and a replay mode. Both modes are based on PicoScenes on SDR. In the former, the baseband signals are generated and transmitted immediately, while in the latter, \textit{the baseband signals are generated, saved to a file, and then replayed in a second run}. In this way, by controlling the length of the inter-frame spacing (IFS), the replay mode can drive an SDR device to reach the theoretical limit of packet injection.

\begin{figure}[t] 
    \centering
    \includegraphics[width=1\columnwidth]{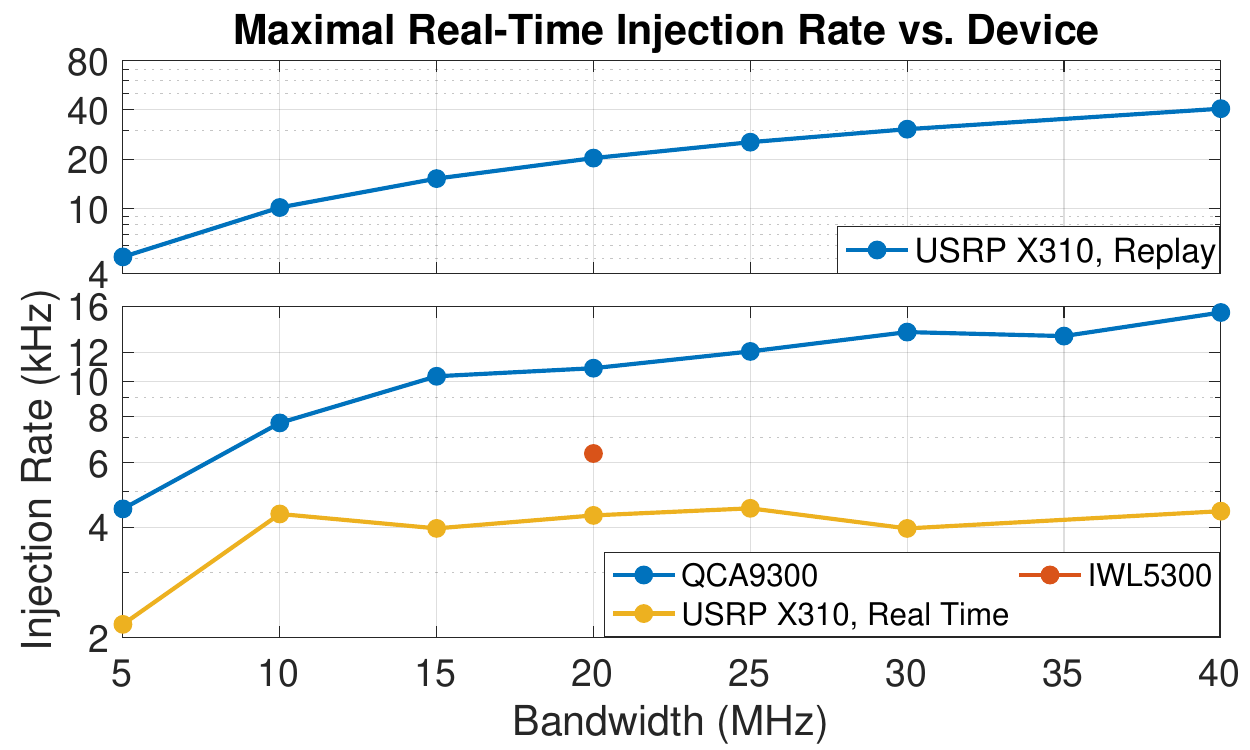}
    \caption{Maximal packet injection rate by device.}
    \label{fig:injection_rate}
\end{figure} 

\begin{figure}[t] 
    \centering
    \includegraphics[width=1\columnwidth]{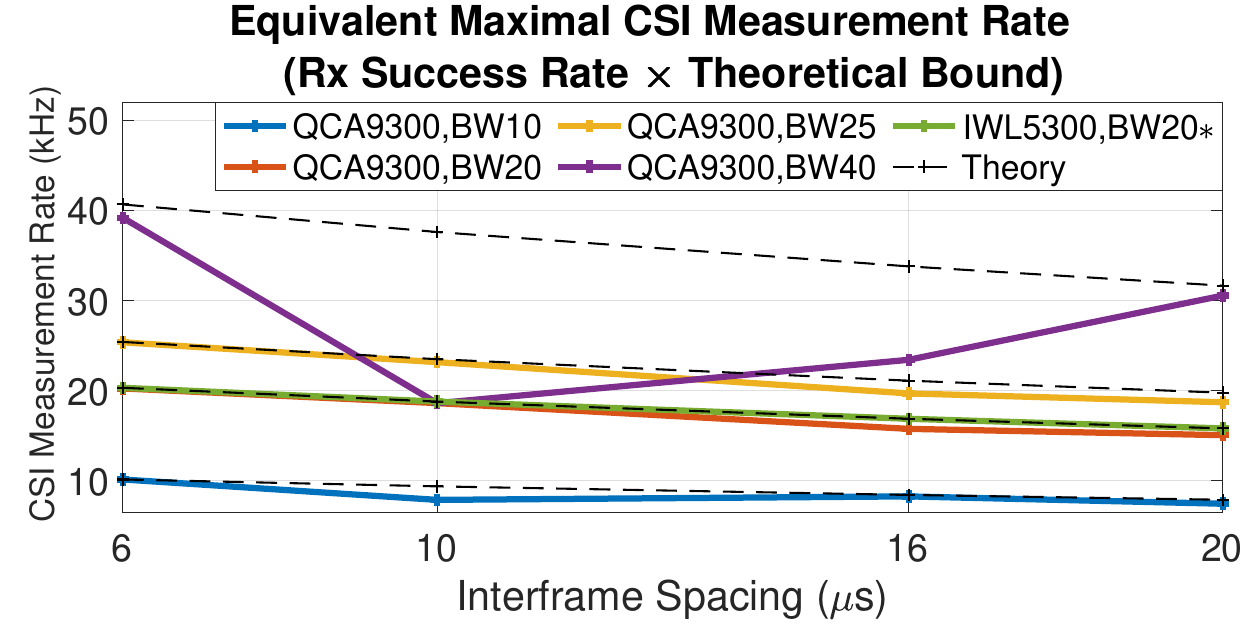}
    \caption{Maximum CSI measurement rate by device.}
    \label{fig:max_rx_rate}
\end{figure}

Fig.~\ref{fig:injection_rate} shows the results. The QCA9300 NIC yielded a rate of 10845 Hz, \ie, 92 $\mu$s per packet on average.
The IWL5300 yielded only a 6350 Hz maximal injection rate, \ie, 157 $\mu$s per packet on average.
For SDR in real-time mode, PicoScenes yielded 4304 Hz on the USRP X310, \ie, 232 $\mu$s per packet on average.
In regard to the software-based baseband encoding, we believe that this is a good result.
The injection rates achieved in replay mode are identical to the theoretical limit. Taking a 20 MHz bandwidth as an example, the theoretical limit of the injection rate was calculated to be $20M / (864 + 120)=20325$, where the 120 samples correspond to the 6 $\mu$s IFS for the 20 MHz bandwidth.

To measure the maximal CSI measurement rate,
we conducted another straightforward evaluation.
We used the replay mode of PicoScenes on SDR to pre-generate and replay 50000 CSI Probing Frames at the theoretical limit rate and logged the Rx rate. Then, we multiplied the Rx rate by the theoretical limit to obtain the equivalent maximal CSI measurement rate.


Fig.~\ref{fig:max_rx_rate} shows the results.
The most impressive result is that the QCA9300 attained a 40 kHz CSI measurement rate under a 40 MHz bandwidth and 6 $\mu$s IFS.
For the IWL5300, we see that its Rx performance was also close to the theoretical limit; however, during the evaluation, we experienced \textit{frequent firmware crashes} that happened shortly after the start of the signal burst.
We discuss in a later section the test of the real-time mode of PicoScenes on SDR.


\algrenewcommand\algorithmicindent{0.6em}%
\begin{algorithm}[t]
    \caption{Performance evaluation for PicoScenes on SDR}
    \label{list:list_baseband}
    \begin{algorithmic}[1]
        \State $FMT$ = $\{$11a/g, 11n, 11ac-SU, 11ax-SU$\}$
        \State $CBW$ = $\{20, 40, 80, 160\}$ MHz
        \State $N_{STS}$ = $\{1, 2, 3, 4\}$
        \State $MCS$ = $\{0 - 7\}$
        \State $CODING$ = $\{$BCC, LDPC$\}$
        \State $LEN$ = $\{250, 500, 1000, 2000, 4000, 8000\}$ MHz
        \ForEach {$fmt \in FMT $}
            \ForEach {$cbw \in CBW $}
                \ForEach {$n_{sts} \in N_{STS}$}
                    \ForEach {$mcs \in MCS $}
                        \ForEach {$code \in CODING $}
                            \ForEach{$len \in LEN$}
                                \If{is\_valid($fmt$, $cbw$, $n_{sts}$, $mcs$, $code$, $len$)}
                                \State Repeat the encoding $\rightarrow$ decoding loopback 
                                \State for 1000 times, and log the time consumption.
                                \EndIf
                            \EndFor
                        \EndFor
                    \EndFor
                \EndFor
            \EndFor
        \EndFor
    \end{algorithmic}
\end{algorithm}

\begin{figure}[t] 
    \centering
    \includegraphics[width=1\columnwidth]{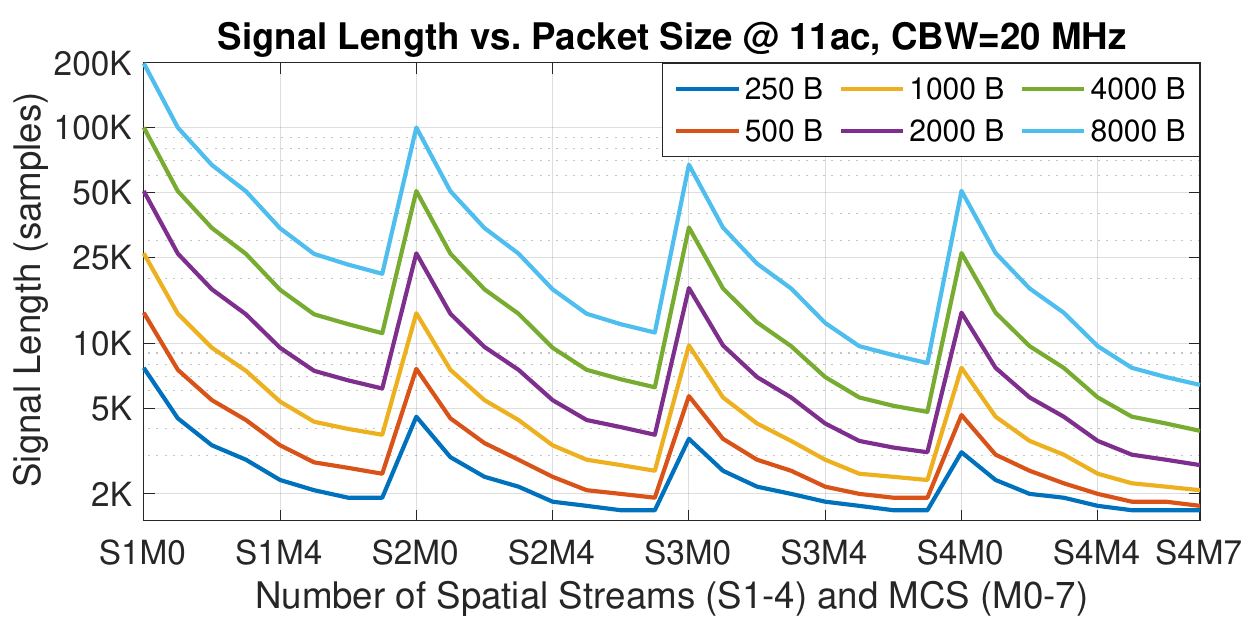}
    \caption{Baseband signal length under different packet lengths, $N_{STS}$ and MCSs. The measurement was performed with the 802.11ac protocol and 20 MHz CBW.}
    \label{fig:sdr_signallength}
\end{figure}





\subsection{Performance evaluation of PicoScenes on SDR}

We evaluated the performance of PicoScenes on SDR in two stages.
In the first stage, we utilized the internal loopback mechanism of PicoScenes to comprehensively measure the encoding and decoding performance for the Wi-Fi baseband signal \wrt the protocol, channel bandwidth (CBW)\footnote{The CBW refers to the parameter of the bandwidth used in encoding. In this paper, the CBW does not necessarily equal the baseband bandwidth. For example, as shown in Fig.~\ref{fig:qca9300_full_spectrum}, we transmit 802.11n HT20 (CBW=20 MHz) and HT40+/- (CBW=40 MHz) packets with many different bandwidths.}, $N_{STS}$, MCS, coding scheme and packet length.
In the second stage, we introduced PicoScenes into real-world CSI measurement scenarios and measured its real-time performance \wrt the injection rate, $N_{STS}$ and number of Rx antennas ($N_{Ant}$).

\begin{figure}[t] 
    \centering
    \includegraphics[width=1\columnwidth]{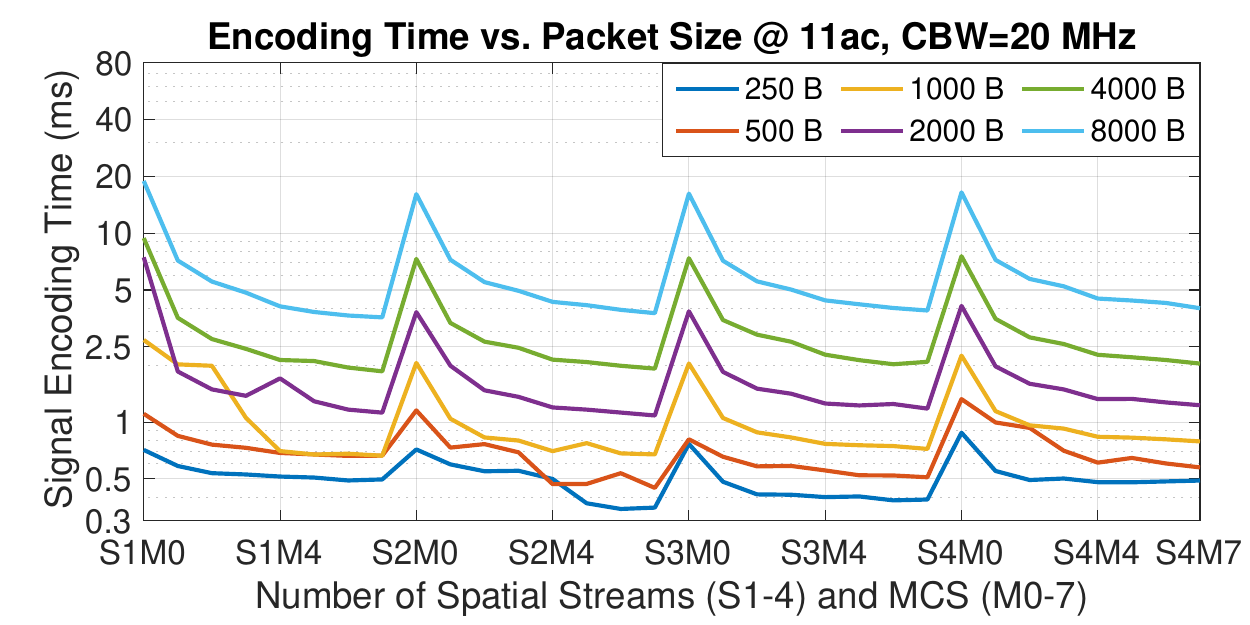}
    \caption{Baseband signal encoding times under different packet lengths, $N_{STS}$ and MCS settings. The measurements were performed with the 802.11ac protocol and a 20 MHz CBW.}
    \label{fig:sdr_walltime}
\end{figure}
\begin{figure}[t]
    \centering
    \includegraphics[width=1\columnwidth]{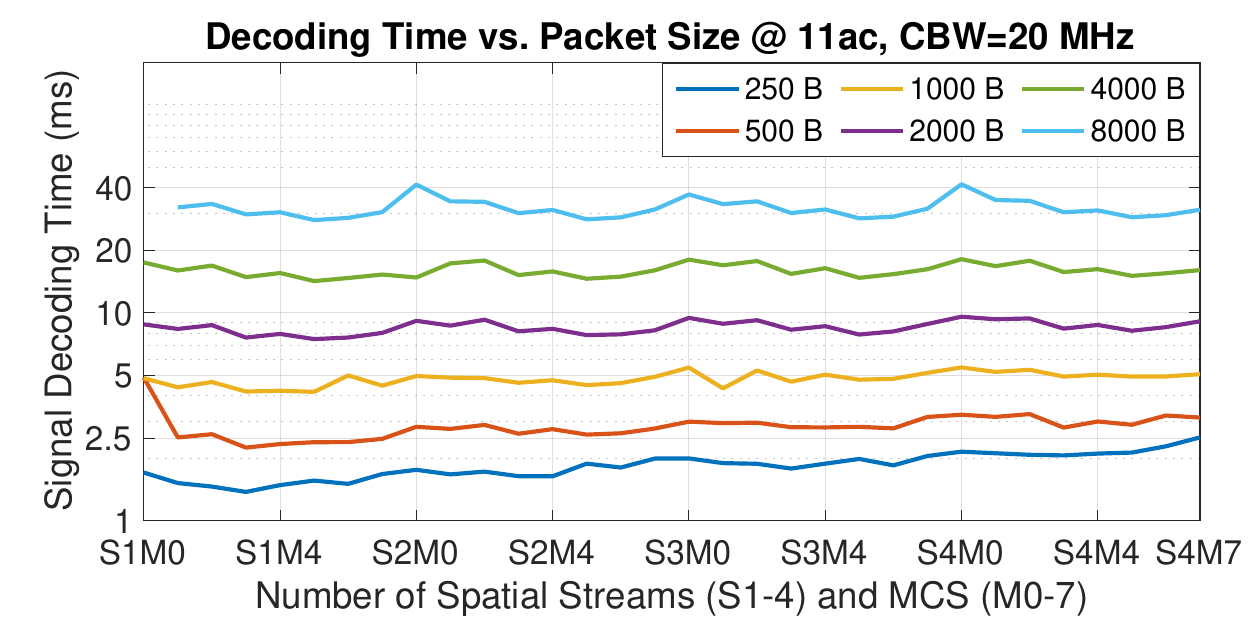}
    \caption{Baseband signal decoding times under different packet lengths, $N_{STS}$ and MCS settings. The measurements were performed with the 802.11ac protocol and a 20 MHz CBW.}
    \label{fig:sdr_rx_walltime}
\end{figure}

\begin{figure}[t] 
    \centering
    \includegraphics[width=1\columnwidth]{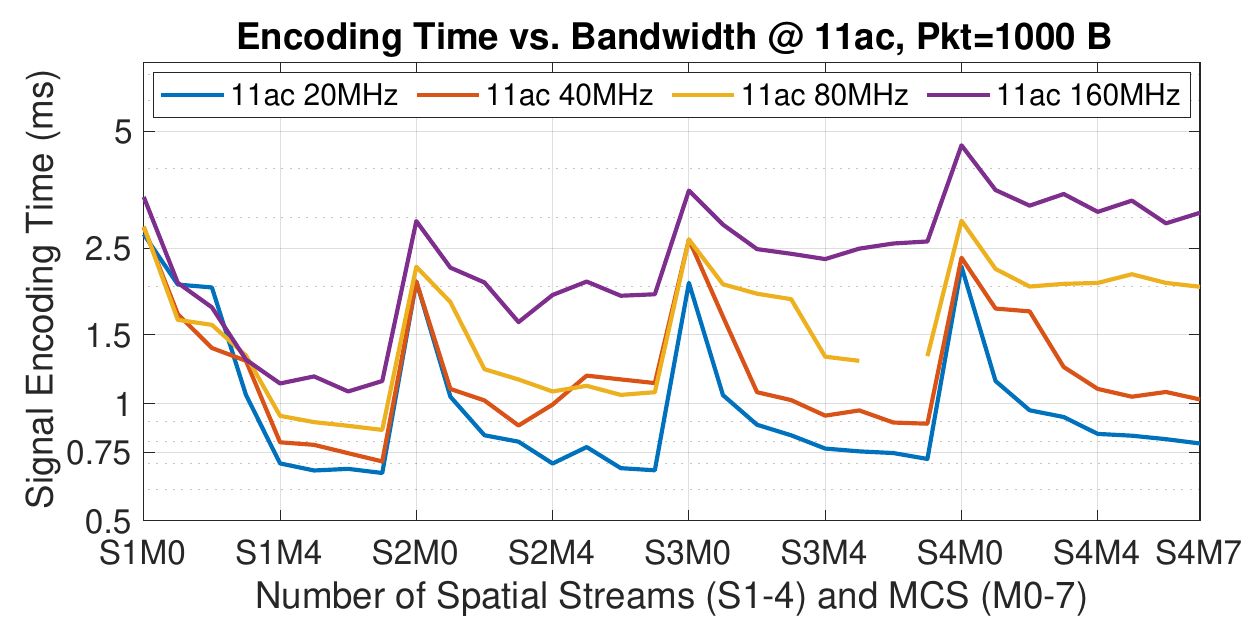}
    \caption{Baseband signal encoding times under different CBWs, $N_{STS}$ and MCS settings. In each configuration, the packet to be encoded was a 1000 byte-long 802.11ac single-user packet.}
    \label{fig:sdr_walltime_cbw}
\end{figure}
\begin{figure}[t] 
    \centering
    \includegraphics[width=1\columnwidth]{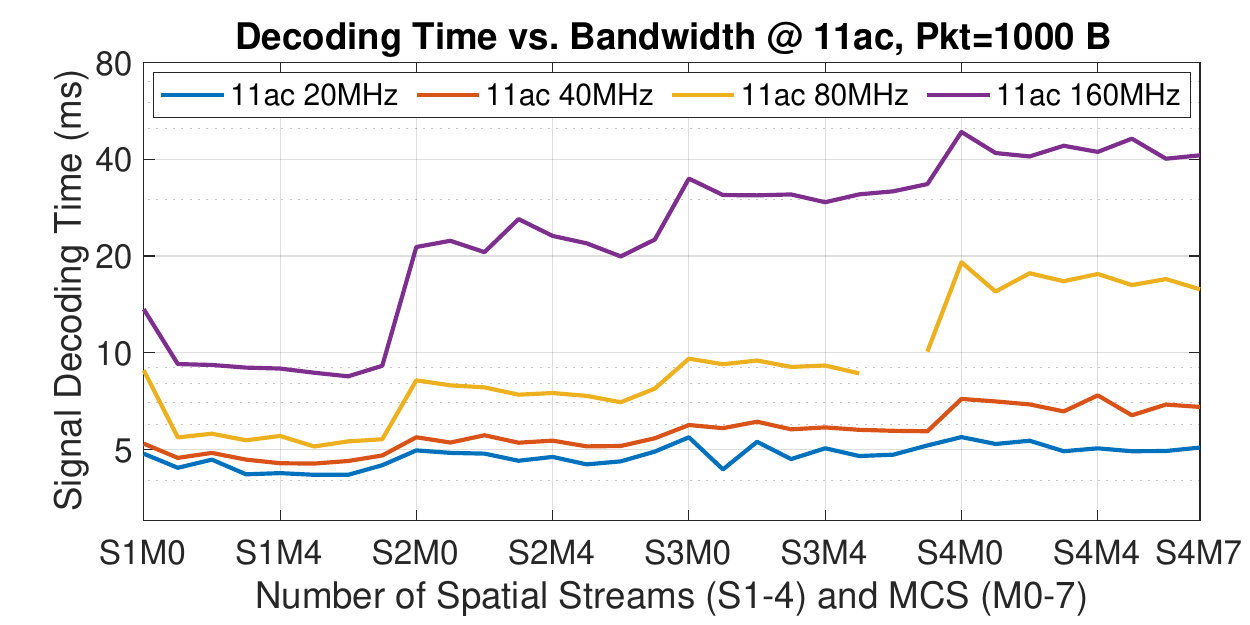}
    \caption{Baseband signal decoding times under different CBWs, $N_{STS}$ and MCS settings. In each configuration, the baseband signal to be decoded was a 1000 byte-long 802.11ac single-user packet.}
    \label{fig:sdr_rx_walltime_cbw}
\end{figure}

\subsubsection{Stage 1: performance evaluation of PicoScenes on SDR via loopback}
\begin{figure}[t] 
    \centering
    \includegraphics[width=1\columnwidth]{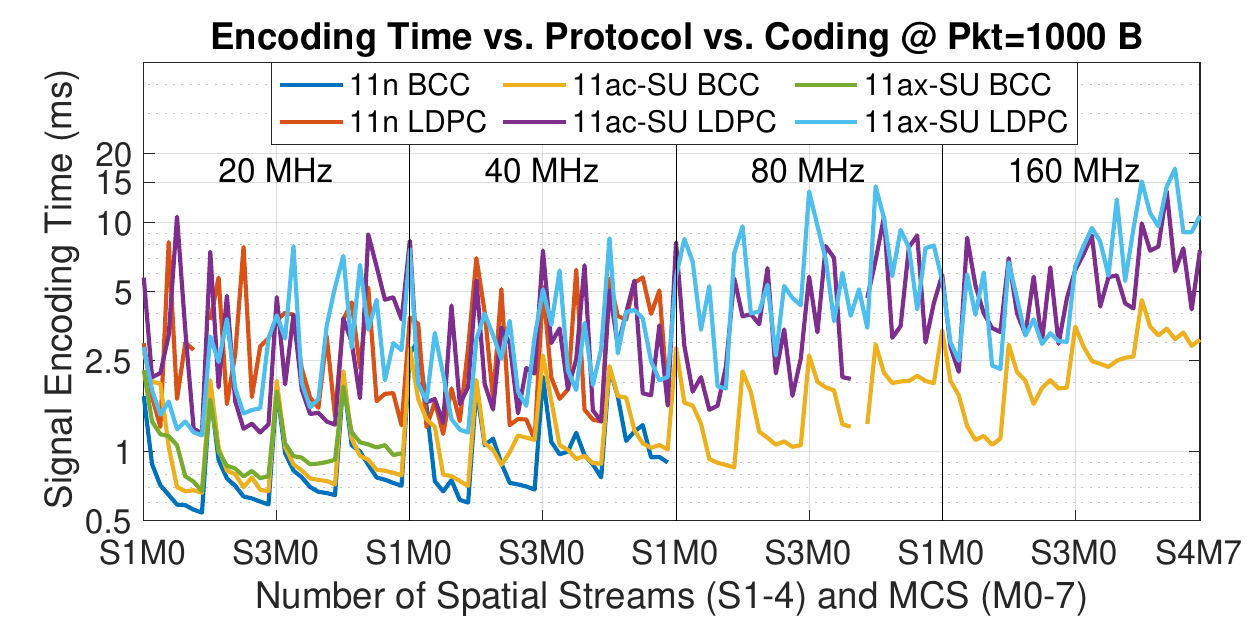}
    \caption{Baseband signal encoding times under different protocols, CBWs, and $N_{STS}$ and MCS settings. In each configuration, the packet to be encoded was a 1000 B single-user packet.}
    \label{fig:sdr_walltime_protocol}
\end{figure}

\begin{figure}[t] 
    \centering
    \includegraphics[width=1\columnwidth]{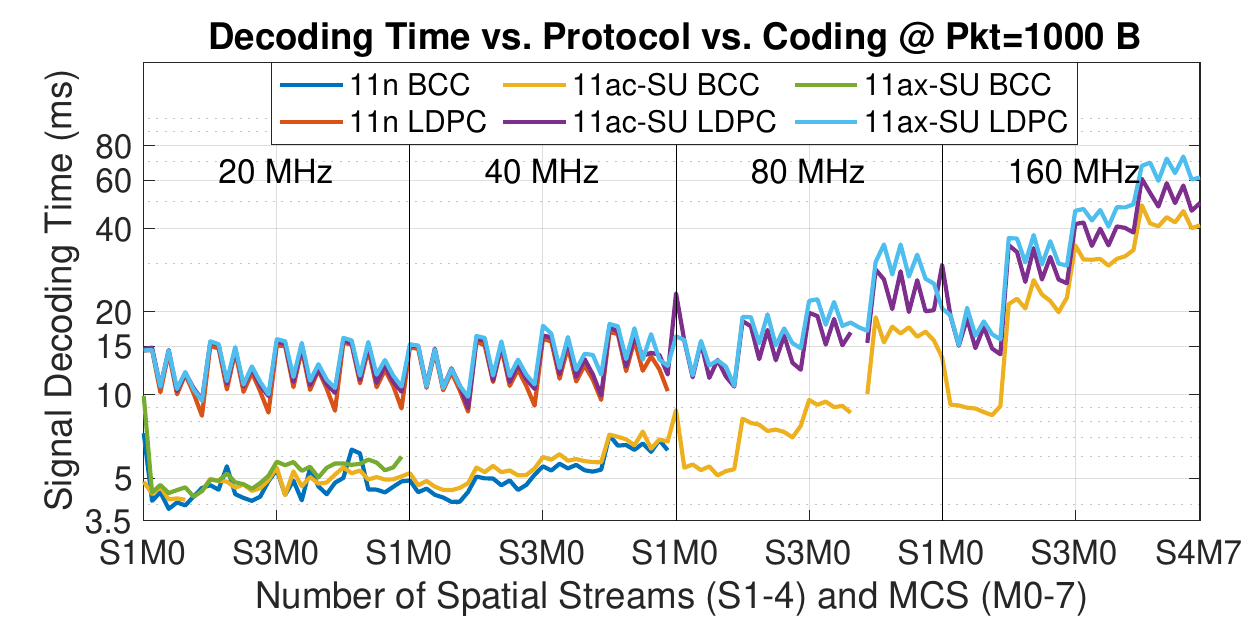}
    \caption{Baseband signal decoding times under different protocols, CBWs, and $N_{STS}$ and MCS settings. In each configuration, the baseband signal to be decoded was previously encoded from a 1000 B single-user packet.}
    \label{fig:sdr_rx_walltime_protocol}
\end{figure}

In this evaluation, we looped back the encoded Wi-Fi baseband signal directly into the decoder; in this manner, we obtained interference-free performance measurements for both the encoder and decoder.
As shown in Listing~\ref{list:list_baseband}, we enumerated $>$3000 valid channel configurations in total\footnote{Some of the configurations are invalid, such as a $>$20 MHz CBW for the 11a/g protocol, a $>$40 MHz CBW for the 11n protocol, BCC coding for a $>$40 MHz CBW in the 11ax protocol, and some special cases, such as the invalid configuration of $N_{STS}$=3 and MCS=6 for the 11ac protocol.}. In each configuration, we encoded/decoded 1000 packets and measured the encoding time, decoding time and signal length. The computer used for the evaluation was equipped with an i9-10850K CPU, 32 GB of RAM and a 512 GB SSD.
The results are shown in Figs.~\ref{fig:sdr_signallength} to \ref{fig:sdr_rx_walltime_protocol}.

Fig.~\ref{fig:sdr_signallength} presents the length of the encoded signal (in samples) \wrt different packet lengths (in bytes) and different $N_{STS}$ and MCS settings.
This figure shows that an increase in both the $N_{STS}$ and MCS values can dramatically reduce the signal length, and that a change in the MCS index leads to greater signal length reduction than a change in $N_{STS}$.
We chose the 802.11ac protocol with a 20 MHz CBW and BCC as the baseline protocol rather than the 802.11n protocol because, compared with 802.11n or the latest 802.11ax protocol, 802.11ac is more inclusive in its baseband configuration. It supports a CBW of up to 160 MHz and both BCC and LDPC coding in single-user (SU) or MU scenarios. This inclusiveness allows us to compare the performances across different CBWs, protocols and coding schemes.

Figs.~\ref{fig:sdr_walltime} and \ref{fig:sdr_rx_walltime} present the encoding and decoding times, respectively, \wrt the packet length.
These two figures reveal four interesting aspects of PicoScenes on SDR.
First, the signal length and the encoding time show similar trends. This similarity occurs because the $N_{STS}$ and MCS settings control the encoded length of a packet (in bits) and the time that the BCC encoder consumes is closely related to the encoded length.
Second, the decoding time seems independent of the signal length and is strongly proportional to the packet length. This is because the time consumed by the BCC decoder is proportional to the raw packet length (in bytes) regardless of the encoded length.
Third, we observe that the decoding time consumption is approximately double the encoding time, which can be attributed to the nature of the forward error correction (FEC) coding.
Finally, in terms of the specific time consumption, the encoder achieves a $<$1 ms encoding time for 250 B packets. Moreover, the decoder achieves a $<$2 ms time for 250 B packets. Both performances are approximately equivalent to a $>$1000 Hz packet encoding and $>$500 Hz packet decoding.
We expect these high performances accompanied by the availability of rich PHY-layer information to be attractive for Wi-Fi sensing research.

Figs.~\ref{fig:sdr_walltime_cbw} and \ref{fig:sdr_rx_walltime_cbw} present the encoding and decoding times, respectively, \wrt the CBW.
%
In contrast to
Figs.~\ref{fig:sdr_walltime} and \ref{fig:sdr_rx_walltime},
if we look carefully, Figs.~\ref{fig:sdr_walltime_cbw} and \ref{fig:sdr_rx_walltime_cbw} show not only an average time increase but also a stair-step time increase in both the encoding and decoding times.
%
%
The configurations that trigger the stair-step increase are (CBW=40 MHz, $N_{STS}=4$, MCS = 0), (CBW=80 MHz, $N_{STS}=2$, MCS = 0), (CBW=80 MHz, $N_{STS}=4$, MCS = 0), (CBW=160 MHz, $N_{STS}=2$, MCS = 0), (CBW=160 MHz, $N_{STS}=3$, MCS = 0), and (CBW=160 MHz, $N_{STS}=4$, MCS = 0).
Apparently, these configurations are not associated with specific $N_{STS}$ or CBW values.
Our investigation links these large time increases to jumps in the number of OFDM subcarriers $N_{SC}$.
More specifically, at these triggering configurations, $N_{SC}$ increases by multiples of 512, \ie, the $N_{SC}$ changes are $384\rightarrow 640$, $256\rightarrow 512$, $768\rightarrow 1024$, $512\rightarrow 1024$, $1024\rightarrow 1536$ and $1536\rightarrow 2048$.
We hypothesize that these time increases triggered by values of $N_{SC}=512n$ are related to certain low-level details of the compilation optimization, such as the capacity of the per-SIMD instructions. It is also one of our future goals to determine the reason for this observation and further optimize the baseband performance.

Finally, Figs.~\ref{fig:sdr_walltime_protocol} and \ref{fig:sdr_rx_walltime_protocol} compare the signal encoding and decoding times \wrt the Wi-Fi protocol and coding scheme.
We make two main observations based on the results.
First, we observe only small differences in time consumption among the different protocols. These differences are mainly caused by protocol overhead, such as a longer and more complex SIG field in the preamble.
Second,
in terms of the coding scheme, the LDPC coding is much more complex than the BCC.
In both figures, the LDPC coding takes approximately 0.5-2$\times$ more time than does the BCC. In addition, the LDPC coding exhibits a larger variance in the codec time.
Based on this evaluation, we recommend that PicoScenes users prioritize the BCC, considering both the performance and compatibility issues. The compatibility issue is that the IWL5300 does not support LDPC coding.

\subsubsection{Stage 2: real-time performance evaluation}

In this evaluation, we answer an important question:
\textit{how fast can PicoScenes on SDR measure CSI in real time?}
We evaluate the performance from three orthogonal perspectives:
first, the Rx rate \wrt the packet injection speed;
second, the Rx rate \wrt $N_{STS}$ and $N_{ANT}$;
and third, the Rx rate \wrt the bandwidth and CBW.

\subsubsection*{Performance evaluation setup}
In the first and second tests, we used the QCA9300 to inject packets at high frequency and used two USRP X310s as our SDR frontend.
In the third test, we used two USRP X310 devices as the Tx and Rx ends.
In these tests, each X310 was equipped with two UBX-160 daughterboards; both X310 devices were connected to the host computer via an Intel X710 Quad Port 10 Gb Ethernet adapter.



\begin{figure}[t] 
    \centering
    \includegraphics[width=1\columnwidth]{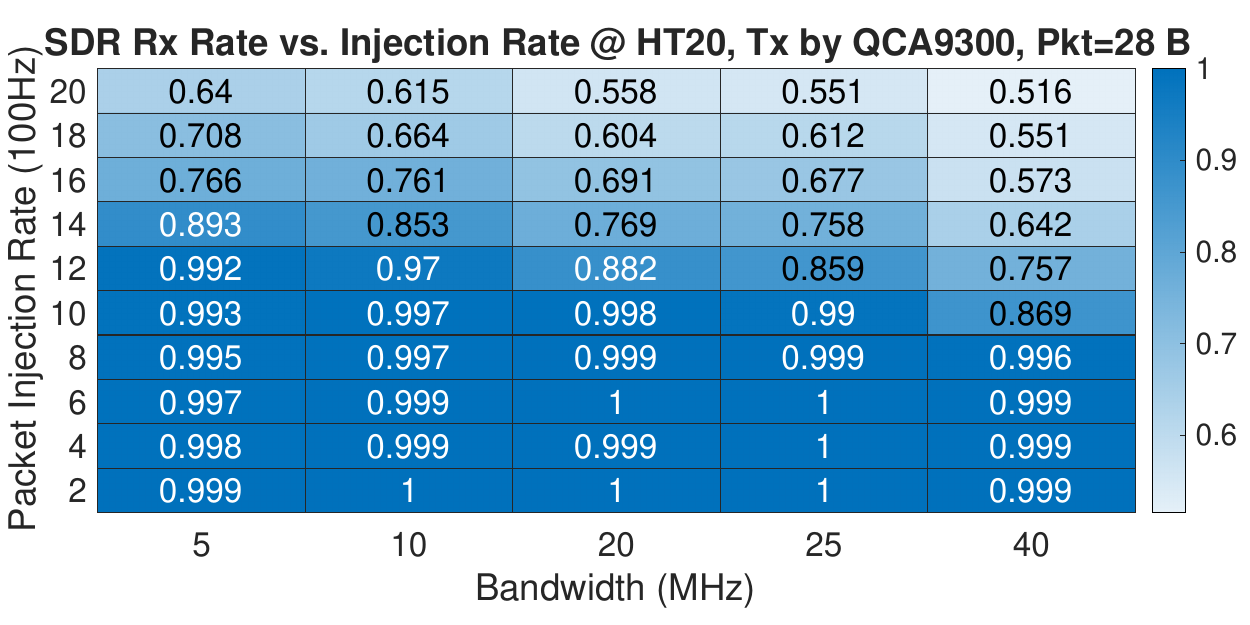}
    \caption{SDR real-time Rx rates under varying injection rates and bandwidths. The Rx end had 1 Rx antenna, and the injected packets were 32 B CSI Probing Frames with $N_{STS}=1$ and MCS=4.}
    \label{fig:sdr_rate_delay}
\end{figure}

\subsubsection*{T1: performance \wrt the packet injection rate}
In the first evaluation, both the Tx and Rx operated with one radio chain, \ie, $N_{STS}=1$ and $N_{ANT}=1$. The baseband bandwidths scanned from 5 to 40 MHz. At each bandwidth, the injection rate of the Tx was scanned from 200 to 2000 Hz, and at each injection rate, the Tx injected 10000 CSI Probing Frames, after which we logged the corresponding Rx rate.

Fig.~\ref{fig:sdr_rate_delay} shows the impressive results.
At the standard 20 MHz bandwidth, PicoScenes on SDR achieved a 99.8\% Rx rate at a 1000 Hz injection rate, and at the 2x faster 40 MHz bandwidth, it still achieved a 99.6\% Rx rate at an 800 Hz injection rate.
This high performance is groundbreaking. To the best of our knowledge, this is the first time that \textit{SDR devices have achieved CSI measurement performances comparable to those of COTS Wi-Fi NICs}.

In addition to the good figure, the results show predictable performance trends.
As the packet injection rate or the bandwidth increases, the baseband computational workload intensifies.
When the injection rate is above approximately 1100 Hz, the decoding speed can no longer keep up with the injection rate. To prevent buffer overflow, the baseband buffer drops signals, which eventually leads to a decline in the Rx rate.
Interestingly, as the bandwidth increases by 8$\times$ from 5 MHz to 40 MHz, the Rx rate shows only a 10-25\% decline.
This is due to the highly efficient packet detection implementation, which quickly filters out non-packet segments in the signals.

\begin{figure}[t] 
    \centering
    \includegraphics[width=1\columnwidth]{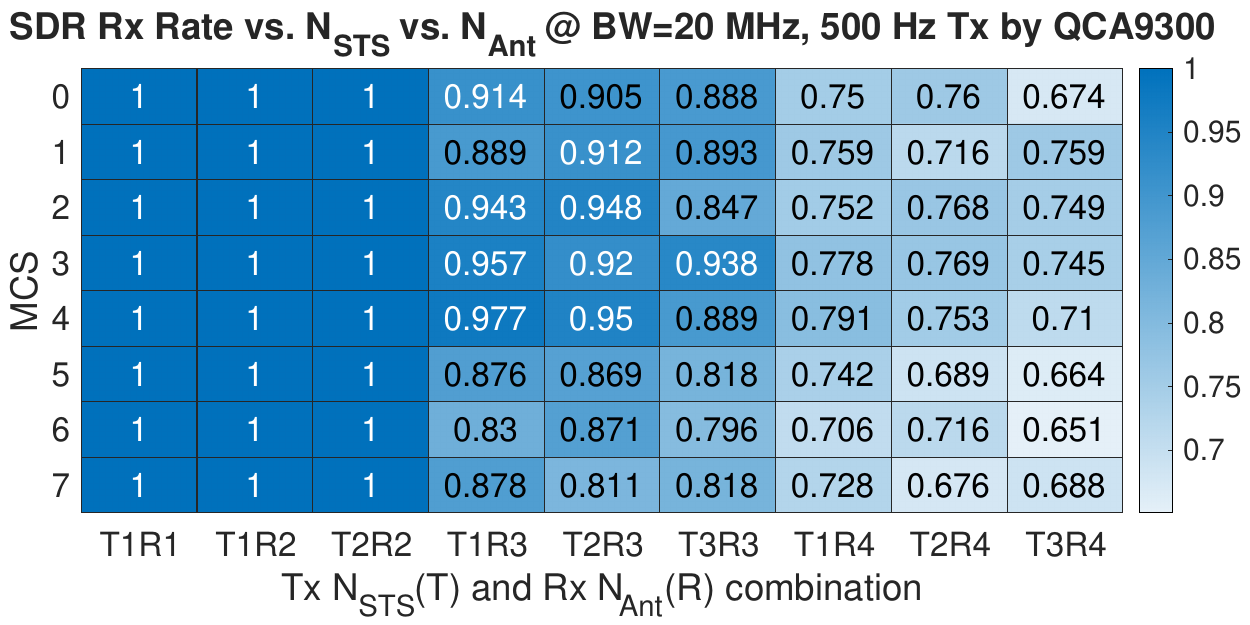}
    \caption{SDR real-time Rx rates under different Tx $N_{STS}$, MCS and Rx $N_{ANT}$ settings. The Tx end, a QCA9300 NIC, injected 32 B CSI Probing Frames at 500 Hz. Both Tx and Rx were operated at a 20 MHz bandwidth.}
    \label{fig:sdr_rate_sts_ant}
\end{figure}

\subsubsection*{T2: performance \wrt the modulation and number of antennas}

In this evaluation, we tested the real-time performance of PicoScenes on SDR \wrt the Tx $N_{STS}$, Rx $N_{ANT}$ and MCS settings.
We enumerated a total of 72 configurations corresponding to $N_{STS}\in\{1-3\}$, $N_{ANT}\in\{1-4\}$, and MCS$\in\{0-7\}$.
The Tx end, a QCA9300 NIC, injected 10000 CSI Probing Frames in each channel configuration. The injection rate was 500 Hz, which is acceptable for most Wi-Fi sensing research.
For the SDR Rx end, to provide up to 4 Rx radio chains,
we used the UHD \textsf{uhd::usrp::multi\_usrp} API to combine two USRP X310 devices into one virtual USRP device equipped with 4 independent radio chains.

Fig.~\ref{fig:sdr_rate_sts_ant} shows the Rx rates.
First, we see a 100\% Rx rate for the cases with $N_{ANT}\leq 2$ and $N_{STS}\leq 2$, \ie, 2$\times$2 MIMO with a 500 Hz injection rate. This result convinces us that SDR is now a competitive alternative to COTS NICs in Wi-Fi sensing research.
Second, compared to the $N_{ANT}\leq 2$ cases, we see an approximately 15\% to 25\% Rx rate drop in the $N_{ANT}=3$ and $N_{ANT}=4$ cases. This decline is mainly due to the 3$\times$ and 4$\times$ raw baseband signal input rates, which exceed the decoding speed and lead to signal dropping.
Third, regarding $N_{STS}$, we see that every increase in $N_{STS}$ by 1 brings about a 5\% loss in the Rx rate. This loss is due to the longer baseband processing required for MIMO decoding.
Finally, from the MCS perspective, we strangely find that for the MCS=0 cases, the most resilient MCS level, the performance is poor and is sometimes the worst. This outcome occurs because the encoder with MCS=0 produces the longest baseband signal, which is more prone to being dropped. In contrast, a higher MCS index results in shorter packet signals and therefore yields a better Rx rate.

\subsubsection*{T3: performance \wrt the bandwidth and CBW}
In this test,
to evaluate the performance throughout a wide bandwidth and CBW range, we used two USRP X310 devices as both the Tx and Rx ends.
We enumerated 56 configurations of the bandwidth, CBW and coding scheme, including bandwidths of up to 200 MHz\footnote{The maximum bandwidth supported by the USRP X310 is 200 MHz.}, CBWs of up to 160 MHz and both BCC and LDPC coding.
In each configuration, we used PicoScenes to inject 10000 CSI Probing Frames in 802.11ac format at a fixed 500 Hz injection rate and logged the Rx rate.

\begin{figure}[t] 
    \centering
    \includegraphics[width=1\columnwidth]{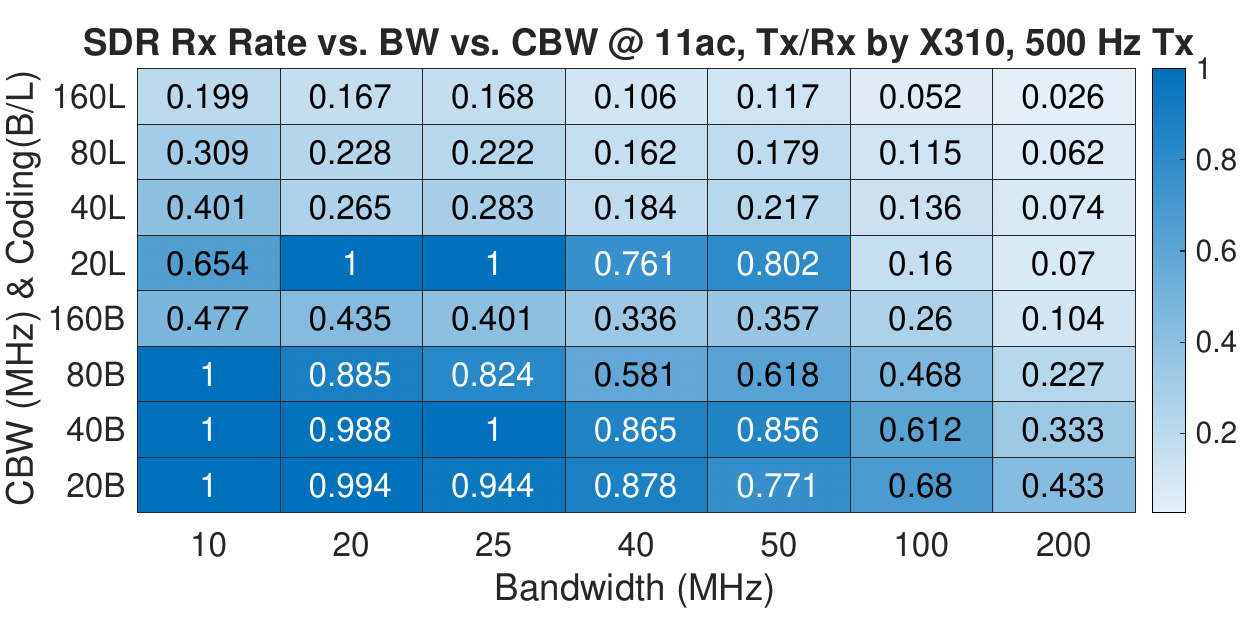}
    \caption{SDR real-time Rx rates under different bandwidths and CBWs. Both the Tx and Rx were USRP X310s with two UBX-160 daughterboards. The Tx end injected 32 B CSI Probing Frames in the 802.11ac BCC and LDPC formats. The injection rate was 500 Hz. 20B/20L and similar labels on the Y-axis denote a 20/40/80/160 MHz CBW with BCC/LDPC coding.}
    \label{fig:sdr_rate_bw_cbw}
\end{figure}

Fig.~\ref{fig:sdr_rate_bw_cbw} shows the results.
We first observe a gradual declining trend of the Rx rate with increasing bandwidth and CBW.
The reason for the CBW-related Rx rate decline is that a higher CBW leads to longer OFDM symbols and, inevitably, a larger computational overhead. This slows the processing speed and leads to signal dropping.
Second, the figure shows that the Rx rate with BCC is approximately 3$\times$ higher than the LDPC-based rate, which is consistent with the loopback test presented in Fig.~\ref{fig:sdr_rx_walltime_protocol}.
This result occurs because the 3$\times$ longer decoding time of the LDPC codec causes an approximately 3$\times$ higher signal drop rate.
Third, if we look at the Rx rates under high-bandwidth conditions, the results are remarkable.
For example, a 20 MHz CBW with BCC yields a 43.3\% Rx rate at the 200 MHz baseband, which can be equivalently interpreted as the maximal 200 Hz CSI sampling over the 200 MHz wide bandwidth.
We conclude that this high-bandwidth CSI measurement can improve the accuracy of Wi-Fi sensing.

\begin{figure}[t]
	\begin{center}
		\begin{tabular}{cc}
			\hspace{-0.13in}
			\includegraphics[width=0.5\columnwidth]{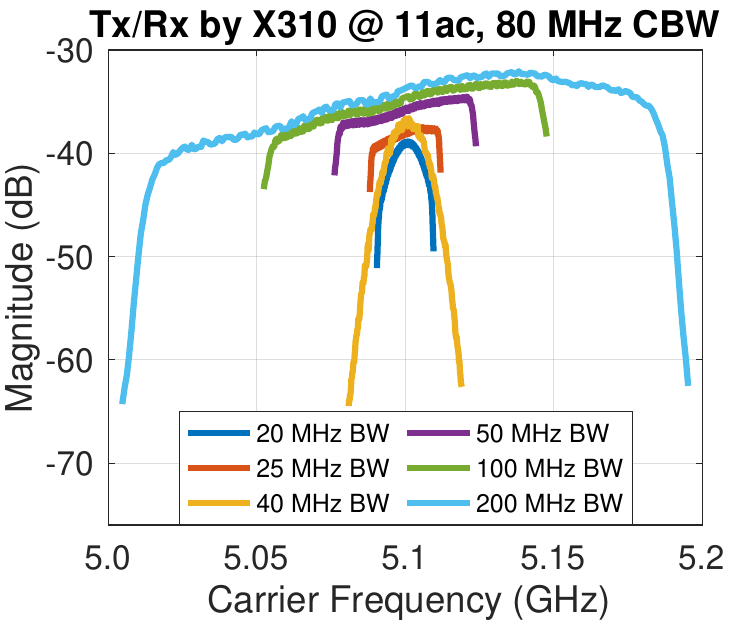} &
			\hspace{-0.2in}
			\includegraphics[width=0.5\columnwidth]{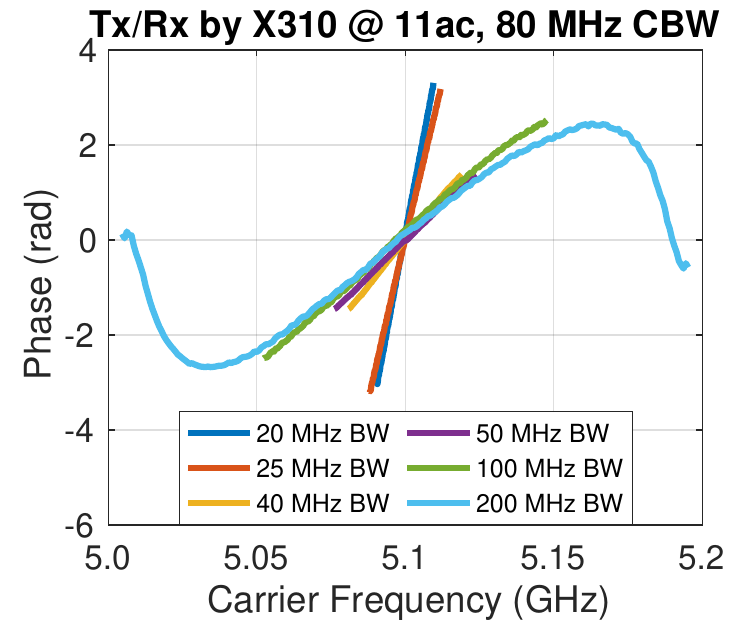} \\
            \hspace{-0.13in}
			\includegraphics[width=0.5\columnwidth]{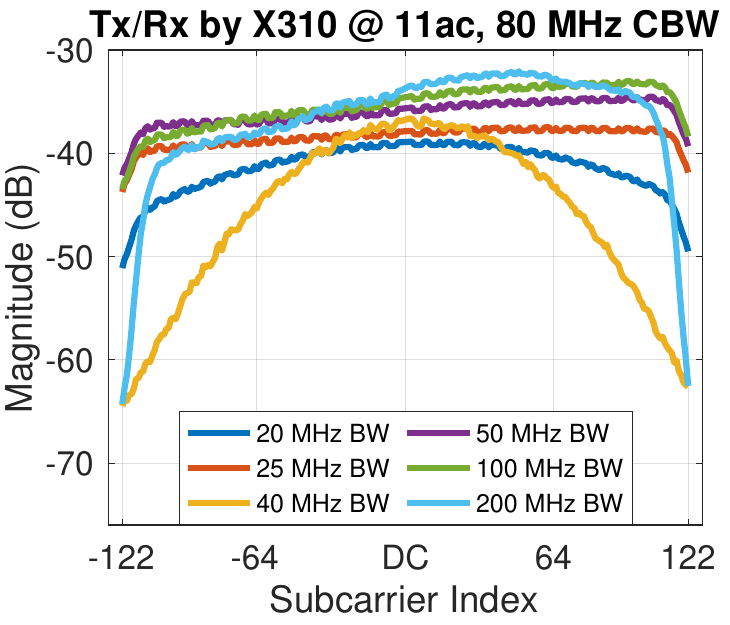} &
			\hspace{-0.2in}
			\includegraphics[width=0.5\columnwidth]{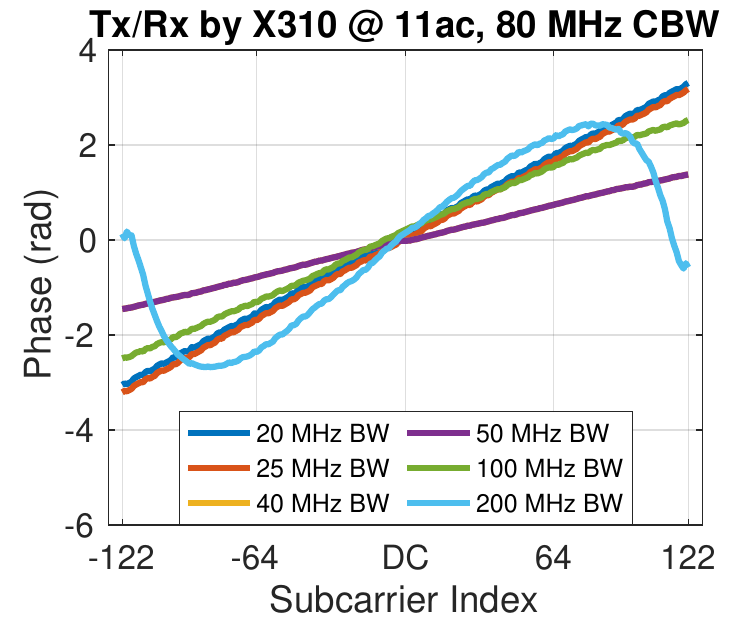} \\
		\end{tabular}
	\end{center}
    \caption{CSI (magnitude and phase response) measured by the USRP X310 under different bandwidths. Both the Tx and Rx were USRP X310s with two UBX-160 daughterboards. The Tx end injected 32 B CSI Probing Frames using the 802.11ac protocol with an 80 MHz CBW. At both ends, the bandwidth was scanned from 200 MHz. The top and bottom figures show the CSI in the spectrum view and the subcarrier view, respectively.}
	\label{fig:sdr_csi}                
\end{figure}

In addition to the performance evaluation, the CSI measurements themselves are also important.
Fig.~\ref{fig:sdr_csi} plots the CSI measurements from the 80 MHz CBW test group.
In the 80 MHz CBW configuration, the CSI includes a total of 245 subcarriers, comprising 234 data subcarriers, 8 pilot subcarriers and 3 interpolated subcarriers around the DC.
The most prominent observation is the CSI measured at a 200 MHz bandwidth. It shows the familiar horizontal S-shaped phase distortion, which was previously identified as Type-I distortion.
The phase distortion observed at this bandwidth corroborates our conjecture that the baseband filter causes the CSI distortion.
Specifically, the RF frontend we used in this evaluation, \ie, a UBX-160 daughterboard~\cite{ubx160} installed on a USRP X310, has a smaller bandwidth (160 MHz) than the maximal bandwidth (200 MHz) of the X310 motherboard.
Accordingly, for transmission at less than a 160 MHz bandwidth, the response of the baseband filter is flat; however, for bandwidths near or above 160 MHz, the filter shows significant influence.
Careful observation of the magnitude response at a 20/40 MHz bandwidth reveals a much stronger roll-off than the response at other bandwidths.
This is due to the cascaded integrated comb (CIC)
roll-off fading caused by odd-number downclocking. In our case, we use multiples of 5 (10 and 5) to downclock the master clock rate of 200 MHz to 20 and 40 MHz, respectively. Interestingly, this fading is another demonstration of the influence of baseband filters.

\subsection{Summary of evaluations}

Here, we briefly summarize the evaluations.
\begin{itemize}
\item The QCA9300 yielded superior link quality over a 2 GHz spectrum with a 3$\times$3 MIMO and 40 MHz bandwidth. The QCA9300 also achieved reliable SISO communication at all bandwidths from 2.5 to 70 MHz.
\item We presented the design of a 27-NIC Wi-Fi sensing array and evaluated its concurrent CSI measurement performance. It yielded a $>$95\% Rx success rate under an 8 kHz packet injection rate.
\item We explored the limit of the CSI measurement rate for a single NIC. The CSI measurement rate of the QCA9300 can reach the theoretical limits of 20 kHz and up to 40 kHz with a 40 MHz bandwidth.
\item We evaluated the in situ encoding and decoding performance of PicoScenes on SDR. It was shown to support most baseband configurations, such as the 802.11a/g/n/ac/ax protocols, 20/40/80/160 MHz CBWs, MCS$\leq10$, $N_{STS}\leq4$, $N_{ANT}\leq4$, and BCC and LDPC coding schemes.
\item We evaluated the CSI measurement performance of PicoScenes on SDR in real-world scenarios. It yielded 1 kHz and 800 Hz CSI measurements under bandwidths of 20 and 40 MHz, respectively. It supported up to 4 Rx antennas and Tx/Rx bandwidths of up to 200 MHz. We also demonstrated the first CSI measurement at an 80 MHz CBW and a 200 MHz bandwidth.
\end{itemize}

\section{Discussion \& Future Work} 
\label{sec:discussion}

In this work, we focus on the Wi-Fi baseband design, CSI distortion, and release of the PicoScenes software. There are several directions in which our work can be further extended.

\textit{Deeper exploration of COTS hardware and drivers.} Although we unlock some important features for the QCA9300, many aspects of the driver still have not been explored. We plan to continue to explore the QCA9300 hardware design and driver modification. We hope to add more powerful features for the QCA9300 and other CSI-available COTS Wi-Fi NICs.

\textit{Higher performance}: The decoding flow of our Wi-Fi baseband implementation is currently a single-threaded implementation; therefore, the computational resources cannot be fully utilized. We expect a 5$\times$ performance improvement on an 8-core CPU by parallelizing the decoding flow. 

\textit{More platform support}: Since the CSI-related driver or firmware modifications are all based on the open-source kernel driver, the existing CSI measurement tools, including PicoScenes, are all Linux-based, which is not a familiar OS for a large number of researchers. We are working to port PicoScenes on SDR to the Windows and macOS platforms. In this way, researchers can perform CSI measurements and data analysis on their favorite platforms.

\section{Related Work} 
\label{sec:related_work}




Remarkable advancements in Wi-Fi sensing~\cite{sensesurvey1} have been achieved over the last 10 years of development,
leading to many new areas of research related to fine-grained sensing, such as indoor localization~\cite{jiang2014communicating, xi2014electronic, Kotaru2015SpotFi, LiFS, Navi, mDTrack, RIM}, trajectory tracking~\cite{AnalysisofHand, TagRay, WiCapture}, material identification~\cite{WiMi}, hand gesture recognition~\cite{WiWrite} and location-based security~\cite{xiong2013securearray, jiang2013rejecting, xie2018genewave}. mDTrack~\cite{mDTrack} relies on a joint approach to simultaneously estimate the angle of arrival (AoA), time of flight (ToF) and Doppler effect, achieving decimeter-level resolution for indoor localization. Nopphon \etal~\cite{AnalysisofHand} extracted the Doppler frequency from CSI and used it to track a hand trajectory with centimeter-level accuracy. WiMi~\cite{WiMi} is a sophisticated system that can identify the type of a material regardless of its motion state. WiWrite~\cite{WiWrite} observes the different signal reflection patterns caused by different hand gestures and uses them to realize precise character recognition and word estimation. RIM~\cite{RIM} is a recently proposed system focusing on inertial measurement for tracking movement distance, heading direction and rotation angles. It leverages the phase difference between antennas to infer the movement direction and speed.


Regarding the hardware used in related research,
the IWL5300 is the most widely used NIC for CSI extraction~\cite{Kotaru2015SpotFi,WiCapture,RIM}. This NIC works with common laptops and allows users to easily carry out CSI measurements.
The second most popular NIC is the QCA9300~\cite{Xie:2019ji,XieSWAN, RIM}, which reports CSI with 10-bit resolution and uncompressed subcarriers. However, the Atheros CSI Tool prioritizes the development of router versions, making it less convenient for Wi-Fi sensing. The Nexmon Channel State Information Extractor~\cite{Nexmon, zhu2018tu} is the latest entry in the list of available CSI-ready devices. It is based on a BCM43xx series chip, such as those equipped on several Android smartphones, Raspberry Pi models, and Wi-Fi routers. The inability to access low-level hardware control is a barrier to more advanced Wi-Fi sensing research; however, SDR-based CSI extraction is more difficult~\cite{Xiong:2015dr, 2018BLoc, 246372}. USRP devices, the Wireless Open-Access Research Platform (WARP) and other SDR hardware devices are merely RF frontends, and the lack of a Wi-Fi software implementation is a serious problem.

\section{Conclusions} 
\label{sec:conclusion}

This paper accomplishes three tasks aiming to eliminate the key barriers to Wi-Fi sensing research.
First, we conduct an in-depth study of the baseband design of the QCA9300, which helps us characterize the CSI distortion. The lessons learned provide us with a paradigm from which we can speculate how other Wi-Fi NICs operate. We also propose a trivial distortion removal method.
Second, regarding hardware inadequacies,
we reintroduce both the QCA9300 NIC and SDR as two of the recommended hardware options for Wi-Fi sensing research. We enable over-GHz spectrum access on the QCA9300 and develop a high-performance software implementation of the Wi-Fi baseband.
Third, we release the PicoScenes software, which supports concurrent packet injection and CSI measurement using the QCA9300, IWL5300 and SDR hardware. The software is architecturally flexible, allowing users to prototype their advanced CSI measurement plugins easily.
Finally, extensive evaluations verify the performance of PicoScenes and yield many state-of-the-art results.


\bibliographystyle{IEEEtran}
\bibliography{reference}
\begin{IEEEbiography}[{\includegraphics[height=1.1in,clip,keepaspectratio]{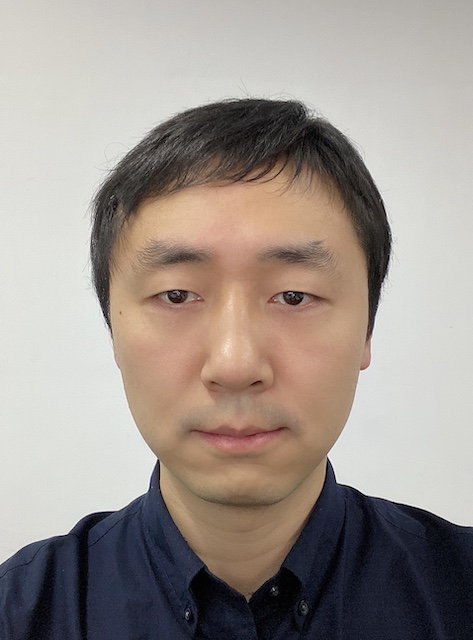}}]{Zhiping Jiang} received the PhD degree in computer science from Xi'an Jiaotong University in 2017. He is now an assistant professor at the School of Computer Science and Technology, Xidian University.
His research interests include wireless sensing, Wi-Fi/acoustic communication and mobile computing. For more information, please visit \href{https://zpj.io/}{https://zpj.io/}.
\end{IEEEbiography}
\begin{IEEEbiography}[{\includegraphics[height=1.1in,clip,keepaspectratio]{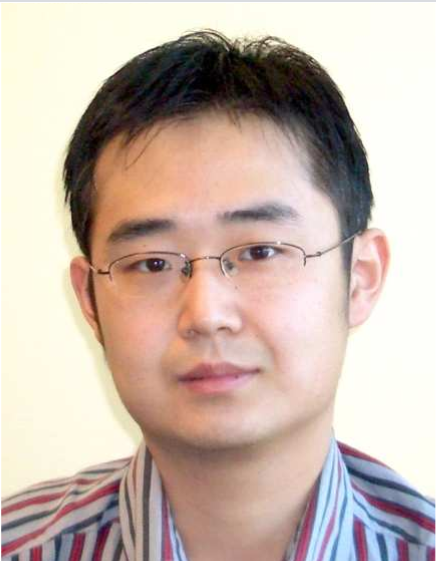}}]{Tom H. Luan}
    received the B.E. degree from the Xi'an Jiaotong University, China, in 2004, the Master degree from the Hong Kong University of Science and Technology, Hong Kong, in 2007, and the Ph.D. degree from the University of Waterloo, Canada, in 2012, all in Electrical and Computer Engineering. During 2013 to 2017, Dr. Luan was a Lecturer in Mobile and Apps at the Deakin University, Australia. Since 2017, he is with the School of Cyber Engineering in Xidian University, China, as a professor. His research mainly focuses on the content distribution and media streaming in vehicular ad hoc networks and peer-to-peer networking, and protocol design and performance evaluation of wireless cloud computing and edge computing. 
\end{IEEEbiography}
\begin{IEEEbiography}[{\includegraphics[height=1.1in,clip,keepaspectratio]{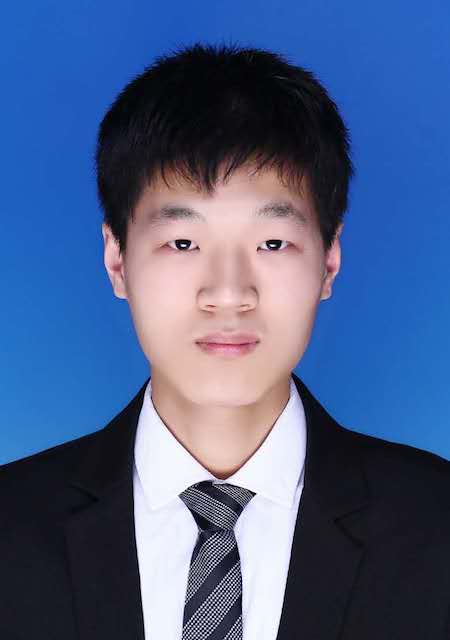}}]{Xincheng Ren}
    received the BS degree in oil-gas storage and transportation engineering from Changzhou University in 2019. He is currently a graduate student the School of Computer Science and Technology, Xidian University. His research interests include Wi-Fi sensing and agile baseband signal processing.
\end{IEEEbiography}
\begin{IEEEbiography}[{\includegraphics[height=1.1in,clip,keepaspectratio]{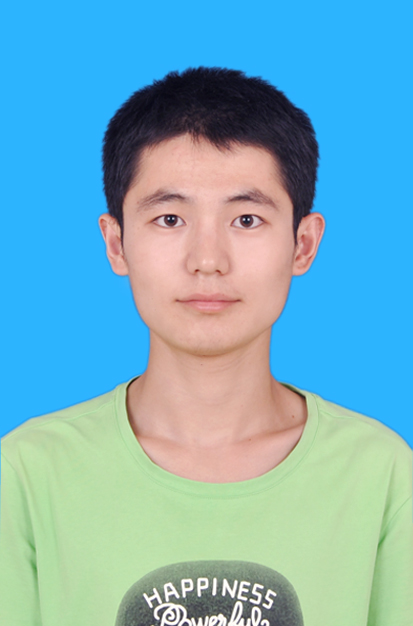}}]{Dongtao Lv}
    received the BS degree in communication engineering from Xi’an University of Science and Technology in 2019. He is currently a graduate student the School of Computer Science and Technology, Xidian University. His research interests include agile baseband signal processing and heterogenous DSP acceleration.
\end{IEEEbiography}
\begin{IEEEbiography}[{\includegraphics[height=1.1in,clip,keepaspectratio]{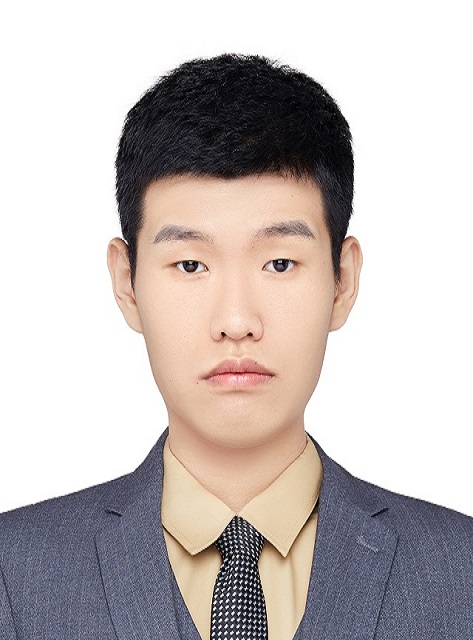}}]{Han Hao}
    received the BS degree in electrical engineering from North China Electric Power University in 2019. He is currently a graduate student with the School of Computer Science and Technology, Xi'an Jiaotong University. His research interests include cross-protocol communication, pervasive computing and wireless networks.
\end{IEEEbiography}
\begin{IEEEbiography}[{\includegraphics[height=1.1in,clip,keepaspectratio]{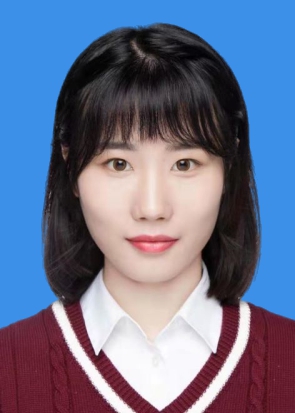}}]{Jing Wang}
    received the BS degree in telecommunication engineering from Shenzhen University in 2020. She is currently a graduate student the School of Computer Science and Technology, Xidian University. Her research interests include wireless sensing, wireless communication and mobile computing.
\end{IEEEbiography}
\begin{IEEEbiography}[{\includegraphics[height=1.1in,clip,keepaspectratio]{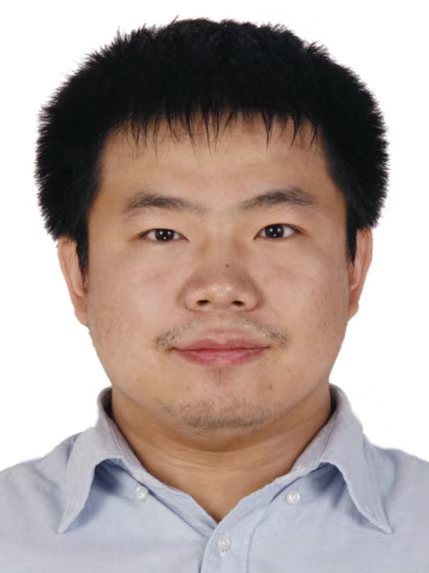}}]{Kun Zhao}
    received his Ph.D degree on Computer Science from Xi’an Jiaotong University in 2019.
    He is currently an assistant professor at the School of Computer Science and Technology, Xi’an Jiaotong University. His research interests include signal processing, wireless security, and federated learning.
\end{IEEEbiography}
\begin{IEEEbiography}[{\includegraphics[height=1.1in,clip,keepaspectratio]{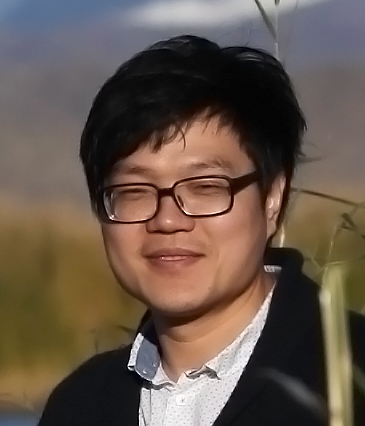}}]{Wei Xi}
    received his Ph.D degree on Computer Science from Xi’an Jiaotong University in 2014.
    He is currently an associate professor at the School of Computer Science and Technology, Xi’an Jiaotong University. His research interests include wireless networks, mobile computing, and AI.
\end{IEEEbiography}
\begin{IEEEbiography}[{\includegraphics[height=1.1in,clip,keepaspectratio]{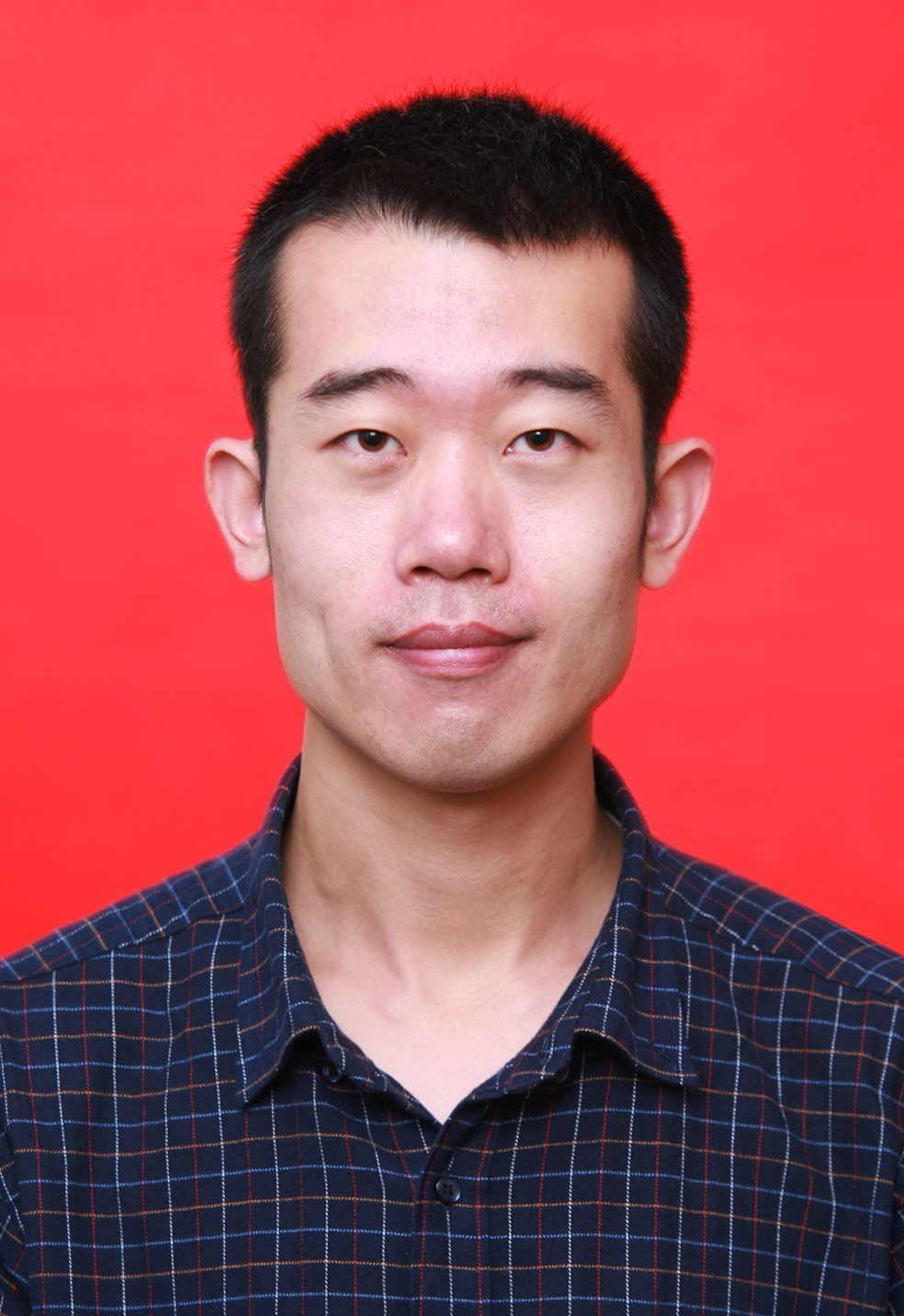}}]{Yueshen Xu}
    received the PhD degree from Zhejiang University, and was a co-trained Ph.D. student with the University of Illinois, Chicago. He is now an associate professor with the School of Computer Science and Technology, Xidian University. His research interests include recommender systems, mobile computing, and service computing.
\end{IEEEbiography}
\begin{IEEEbiography}[{\includegraphics[height=1.1in,clip,keepaspectratio]{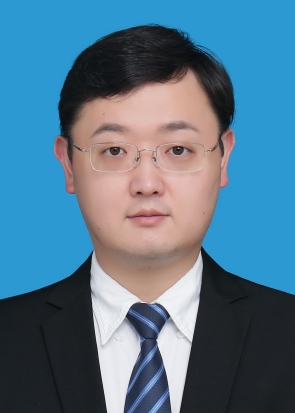}}]{Rui Li}
    received the PhD degree in computer science from Xi'an Jiaotong University in 2014. He is now an associate professor at the School of Computer Science and Technology, Xidian University, China.
His research interests include smart sensing, radar imaging and pervasive computing.
\end{IEEEbiography}


\end{document}